\documentclass[11pt,letterpaper]{article}

\usepackage{jheppub}
\usepackage{amsmath}
\usepackage{graphicx}
\usepackage{amssymb}
\usepackage[usenames,dvipsnames]{xcolor}
\usepackage[normalem]{ulem}

\newcommand{\beq}{\begin{equation}}
\newcommand{\eeq}{\end{equation}}
\newcommand{\nn }{\nonumber}
\newcommand{\HypF}{{}_2F_{1}}

\def\bnslash{\bar n\!\!\!\slash}

\newcommand{\mcdot}{\!\cdot\!}

\newcommand{\bra}[1]{\left\langle #1\right\rvert}
\newcommand{\ket}[1]{\left\lvert #1\right\rangle}

\newcommand{\as}{\alpha_s}
\newcommand{\minus}{\!-\!}
\newcommand{\plus}{\!+\!}

\newcommand{\cO}{\mathcal{O}}

\newcommand{\cP}{\mathcal{P}}

\DeclareMathOperator{\Tr}{Tr}

\newcommand{\eq}[1]{Eq.~\eqref{eq:#1}}
\newcommand{\eqs}[2]{Eqs.~\eqref{eq:#1} and \eqref{eq:#2}}

\newcommand{\fig}[1]{Fig.~\ref{fig:#1}}

\title{Probing light quark Yukawa couplings through angularity distributions in Higgs boson decay}
\author{Bin Yan$^{1,2}$, Christopher Lee$^{2}$}

\affiliation{$^1$Institute of High Energy Physics, Chinese Academy of Sciences, Beijing 100049, China}
\affiliation{$^2$Theoretical Division, Los Alamos National Laboratory, P.O. Box 1663, MS B283, \\ Los Alamos, NM 87545, USA}

\emailAdd{yanbin@ihep.ac.cn}
\emailAdd{clee@lanl.gov}

\abstract{
We propose to utilize  angularity distributions in Higgs boson decay to probe light quark Yukawa couplings at $e^+e^-$ colliders. Angularities $\tau_a$ are  a class of 2-jet event shapes with variable and tunable sensitivity to the distribution of radiation in hadronic jets in the final state. Using soft-collinear effective theory (SCET), we  present a prediction of angularity distributions from Higgs decaying to quark and gluon states at $e^+e^-$ colliders to ${\rm NNLL}+\mathcal{O}(\alpha_s)$ accuracy.  Due to the different color structures in quark and gluon jets, the angularity distributions from $H\to q\bar{q}$ and $H\to gg$ show different behaviors and can be used to constrain the light quark Yukawa couplings. 
We show that the upper limit of light quark Yukawa couplings could be probed to the level of $\sim15\%$ of the bottom quark Yukawa coupling in the Standard Model in a conservative analysis window far away from nonperturbative effects and other uncertainties; the limit can be pushed to $\lesssim 7-9\%$ with better control of the nonperturbative effects especially on gluon angularity distributions and/or with multiple angularities.
}


\begin{document}
\preprint{LA-UR-23-32929}
\maketitle
\section{Introduction}
After the discovery of the Higgs boson at the Large Hadron Collider (LHC)~\cite{Aad:2012tfa,Chatrchyan:2012ufa}, precision measurements of the properties of the  Higgs  and proving that the Higgs boson 
is indeed responsible for electroweak symmetry breaking and mass generation have become a forefront goal of high energy physics. 
Determining the Yukawa couplings of the Higgs boson to fermions is one of the avenues to verify the Standard Model (SM) of particle physics.
Since the Yukawa couplings are completely determined by the fermion mass in the SM, i.e. $y_q=\sqrt{2}m_q/v$ with $v=246 ~{\rm GeV}$, it is virtually impossible to probe the light quark Yukawa couplings directly due to the smallness of their mass. But these light quark Yukawa couplings may receive large modifications from new physics (NP) beyond the SM~\cite{Bar-Shalom:2018rjs}, and thus these NP effects could be probed by careful measurements of the Yukawa couplings.

Due to the large QCD backgrounds for the hadronic decay of Higgs boson at the LHC, a direct measurement of the light quark Yukawa couplings is challenging. 
Several approaches have been proposed to constrain the light quark Yukawa couplings indirectly.  For example, one can measure the light quark Yukawa coupling through {\it i}) rare modes of Higgs boson decays, e.g. $h\to J/\Psi\gamma  ~(\phi\gamma,~\rho\gamma,~\omega\gamma)$~\cite{Bodwin:2013gca,Kagan:2014ila,Han:2022rwq};
{\it ii}) Higgs production in association with a charm-tagged jet~\cite{Brivio:2015fxa};
iii) global analysis of the Higgs data~\cite{Perez:2015aoa,Zhou:2015wra,Perez:2015lra,deBlas:2019rxi}; {\it iv}) the transverse momentum ($p_T$) distribution of Higgs boson or jet in Higgs production processes~\cite{Bishara:2016jga,Soreq:2016rae,Bonner:2016sdg,Bizon:2018foh,Chen:2018pzu,Billis:2021ecs}; {\it v}) the  kinematic shapes of Higgs pair production~\cite{Alasfar:2022vqw}, off-shell Higgs production~\cite{Vignaroli:2022fqh,Balzani:2023jas} and triple heavy vector boson production~\cite{Falkowski:2020znk}; etc. The above proposals demand accurate calculations of light quark meson formation, charm tagging efficiency and faking rates of light quarks, or precise knowledge of the $p_T$ spectrum of Higgs boson.  
Compared to hadron colliders, $e^+e^-$ colliders provide direct access to all possible decay channels of the Higgs boson due to the clean environment. Several plans for future lepton colliders have been proposed, including the CEPC~\cite{CEPCStudyGroup:2018ghi}, ILC~\cite{Baer:2013cma}, CLIC~\cite{deBlas:2018mhx} and FCC-ee~\cite{Abada:2019zxq}. Through the Higgs and $Z$ boson associated production, the inclusive cross section could be measured to $0.5\%$ accuracy at $\sqrt{s}=250~{\rm GeV}$ with integrated luminosity of $5~{\rm ab}^{-1}$~\cite{CEPCStudyGroup:2018ghi}. Such high accuracy of the total cross section offers a possibility to measure the light Yukawa couplings directly at the $e^+e^-$ colliders. 

The main difficulty of measuring the light quark Yukawa couplings at $e^+e^-$ colliders is due to contamination from Higgs boson decays to gluons, obscuring jets initiated by light quarks.
To suppress the gluon background, it is necessary to use some form of quark and gluon jet  discrimination.  A wide variety of quark and gluon discriminants have
been proposed, e.g., in Refs.~\cite{Fodor:1989ir,Pumplin:1991kc,Gallicchio:2011xq,Gallicchio:2012ez,Larkoski:2014pca,Bhattacherjee:2015psa,Gras:2017jty,Chien:2019osu,Li:2023tcr,Wang:2023azz}. To first approximation, the main underlying feature is that the  initiating energetic quark radiates soft or collinear gluons in proportion to its color factor $C_F=4/3$, while initiating hard gluons radiate additional gluons proportional to the factor $C_A=3$, making gluon-initiated jets ``broader'' or more ``diffuse'' than quark-initated jets. Good discriminants tease out these and other more subtle differences between gluon and quark jets. Ideally, the discriminants have properties that can also be predicted reliably from first principles in QCD, though powerful methods also exist that are totally data-driven and require no input from QCD at all, e.g. \cite{Metodiev:2018ftz}.

One class of observables that can discriminate broad features of quark and gluon jets and can also be predicted to high accuracy in QCD are hadronic event shapes (e.g. \cite{Dasgupta:2003iq,Becher:2008cf,Abbate:2010xh,Monni:2011gb,Hoang:2014wka}).
Thus we explore in this paper their potential to constrain the light quark Yukawa couplings at $e^+e^-$ colliders.
Similar ideas have been discussed in  Ref.~\cite{Gao:2016jcm}, e.g. event shapes including thrust \cite{Farhi:1977sg}, heavy hemisphere mass \cite{Chandramohan:1980ry,Clavelli:1981yh}, $C$ parameter \cite{Parisi:1978eg,Donoghue:1979vi}, broadening \cite{Catani:1992jc} and Durham 2-to-3-jet transition parameter \cite{Catani:1991hj}, and jet energy profile \cite{Seymour:1997kj,Isaacson:2015fra,Li:2018qiy,Li:2011hy}. Ref.~\cite{Gao:2016jcm} showed that these event shapes can provide a much stronger sensitivity for the light quark Yukawa couplings compared to methods proposed for the LHC, e.g. $y_q<0.091 y_b$ by event shapes for $q=u,d,s$ at 95\% confidence level  at CEPC~\cite{Gao:2016jcm}, while it only can be constrained to $0.4\sim0.5 y_b$ by Higgs $p_T$ spectrum at the LHC~\cite{Bishara:2016jga,Soreq:2016rae}, where $y_b$ is the bottom quark Yukawa coupling in the SM. Recently, Ref.~\cite{Balzani:2023jas} studied the potential of off-shell Higgs production to further improve the Yukawa coupling limit, projecting improvements of 20\% or more compared to Higgs+jet production depending on assumptions about uncertainties on background.

Event shapes in $e^+e^-$ collisions to hadrons have already been computed to very high accuracy, up to N$^3$LL$'$ resummed accuracy matched to NNLO fixed-order calculations, e.g.~\cite{Becher:2008cf,Chien:2010kc,Abbate:2010xh,Hoang:2014wka}.
Recently, theoretical predictions of event shape observables in Higgs boson decays have also been computed to high accuracy, e.g.~the thrust distribution has been calculated to NLO accuracy plus NNLO singular terms~\cite{Gao:2019mlt}, the energy-energy correlation from Higgs decaying to gluon mode has been computed to NLO accuracy~\cite{Luo:2019nig}. Ref.~\cite{Alioli:2020fzf} also computed the 2-jettiness distribution from the decay of the Higgs boson to a $b\bar{b}$ quark pair and to gluons, to NNLL$'$+NNLO accuracy and even approximate N$^4$LL accuracy in~\cite{Ju:2023dfa}. In this work, we propose to use  a class of event shapes, \emph{angularities} \cite{Berger:2003iw}, to separate the $q\bar{q}$ channel from the $gg$ channel in Higgs decays and to improve measurements of the light quark Yukawa couplings at lepton colliders. (The thrust and total jet broadening used in Ref.~\cite{Gao:2019mlt} are special cases of the angularities.) We take advantage of recent results that make possible the prediction of angularity distributions to NNLL$'$ accuracy in resummed perturbation theory \cite{Bell:2018gce}. We also newly compute the LO fixed-order $\mathcal{O}(\as)$ corrections to general angularities in Higgs decay. We find that using any single angularity distribution allows determination of the light quark Yukawa couplings to similar or somewhat better precision as other individual event shapes studied in Ref.~\cite{Gao:2019mlt}, depending on whether we use a conservative analysis window at larger $\tau_a$ away from large nonperturbative effects and other corrections, yielding a potential bound of $y_q\lesssim 0.15 y_b$, or a more aggressive window that enters further into the nonperturbative region but thus takes advantage of the very different peaks for the quark and gluon angularity distributions, yielding $y_q\lesssim 0.7-0.9 y_b$.
We also perform a very preliminary study using Pythia of the ability of double differential distributions of angularities to improve this reach further, attempting to utilize their additional power over single angularity distributions to distinguish quark and gluon jets. We find that, at the present time, realizing large additional improvements is difficult due to the backgrounds from $H\to b\bar{b},c\bar{c}$ and $Zq\bar{q}$. However, if those backgrounds could be further suppressed, say by a factor 10, both single- and multi-angularity distributions could yield even better limits on $y_q$.

The paper is organized as follows: In Section~\ref{sec:fac}, we obtain the factorization of angularity distributions from Higgs boson decay in the formalism of SCET, allowing large logs in the distributions to be resummed through renormalization group evolution.  Both the decay processes $H\to q\bar{q}$ and $H\to gg$ are calculated to ${\rm NNLL}+\mathcal{O}(\alpha_s)$ accuracy. Although higher orders are now possible (e.g. \cite{Zhu:2023oka}), for our illustrative study here, we do not include higher-order corrections. The numerical predictions are given in Section~\ref{sec:num}. We then show the precision with which light quark Yukawa couplings could be measured using angularity distributions in Section~\ref{sec:yukawa}. Finally, we conclude in Section~\ref{sec:sum}. The one-loop angularity distributions in Higgs boson decay and some technical details of our analysis are discussed in the Appendix.

\section{Factorization  and resummation of event shapes}
\label{sec:fac}

\subsection{Factorization of cross section in SCET}

The angularities are defined as~\cite{Berger:2003iw,Berger:2003pk},
\beq
\tau_a=\frac{1}{Q}\sum_i |p_T^i|e^{-|\eta_i|(1-a)},
\eeq
where, for us, we will take $Q=m_H=125~{\rm GeV}$, the sum is over all final state particles $i$.  The pseudorapidity $\eta_i$ and transverse momentum $p_T^i$ of each particle $i$ is measured with respect to the thrust axis \cite{Brandt:1964sa,Farhi:1977sg} in the rest frame of the Higgs boson. The  parameter
$a$ defining each angularity $\tau_a$ is a continuous parameter $a<2$ for infrared safety, though we will focus on $a<0.5$ in this work, since soft recoil effects which complicate the resummation become important as $a\to 1$~\cite{Dokshitzer:1998kz,Berger:2003iw,Becher:2011pf,Chiu:2011qc,Chiu:2012ir,Budhraja:2019mcz}. 
The angularities have found a wide range of applications in  $e^+e^-$ collisions producing hadrons (e.g. Refs.~\cite{Bauer:2008dt, Hornig:2009vb, Ellis:2010rwa, Larkoski:2014tva, Procura:2018zpn,Bell:2018gce}), and their distributions have been calculated to ${\rm NNLL}^\prime$ resummed and $\mathcal{O}(\alpha_s^2)$ matched accuracy~\cite{Bell:2018gce}. 

To describe the decay of  Higgs boson, we will use the following effective Lagrangian,
\beq
\label{eq:Leff}
\mathcal{L}_{\rm eff}=\frac{\alpha_s(\mu)}{12\pi v}HG^{\mu\nu,b}G_{\mu\nu}^b+\sum_q \frac{y_q(\mu)}{\sqrt{2}}H\bar{\Psi}_q\Psi_q,
\eeq
where $\mu$ is the renormalization scale, $b$ is the color index (summed over) of the gluon field with field strength tensor $G_{\mu\nu}^b$. 
In our calculation, we ignore the masses of the light quarks, but keep the Yukawa coupling $y_q$ itself. The coupling of the Higgs to the gluon fields comes from integrating out the top quark.

For small values of $\tau_a\ll 1$, the degrees of freedom in the final state at leading power are collinear quarks and gluons in the two back-to-back directions in the Higgs rest frame, and (ultra)soft gluons radiated between them. We can predict the cross section in this regime by matching the operators in \eq{Leff} onto operators in SCET:
\beq
\label{eq:hardmatching}
\bar{\Psi}_q\Psi_q \to C_2^q(m_H,\mu)\bar{\chi}_n\chi_{\bar{n}},\quad
G^{\mu\nu,b}G_{\mu\nu}^b\to C_2^g(m_H,\mu)m_H^2g_{\mu\nu}B_{n\perp}^{\mu,b}B_{\bar{n}\perp}^{\nu,b}.
\eeq
Here $\chi_{n,\bar{n}}$ and  $B_{n\perp,\bar{n}\perp}$ are the SCET gauge invariant collinear quark and gluon fields respectively  \cite{Bauer:2000ew,Bauer:2000yr,Bauer:2001ct,Bauer:2001yt}. 
Computing the cross section in this regime, the event shape distributions in $e^+e^-$ collisions can be factorized  into hard, jet and soft functions~\cite{Berger:2003iw,Bauer:2008dt,Hornig:2009vb,Almeida:2014uva,Bell:2018gce},
\beq
\label{eq:spectrum}
\frac{d\Gamma^i_H}{\Gamma_{H0}^id\tau_a}=H^i(m_H,\mu)\int d\tau_a^n d\tau_a^{\bar{n}}d\tau_a^s\delta\Bigl(\tau_a-\frac{t_a^n + t_a^{\bar n}}{m_H^{2-a}}-\frac{k_s}{m_H}\Bigr)J_n^i(t_a^n,\mu)J_{\bar{n}}^i(t_a^{\bar{n}},\mu)S^i(k_s,\mu),
\eeq
where $i=q, g$ corresponds to $H\to q\bar{q}, gg$, $t_{a}^{n,\bar n}$ are the natural arguments of the jet functions of dimension $2-a$ \cite{Hornig:2009vb,Almeida:2014uva}, and $k_s$ is the natural dimension-1 argument of the soft function. The decay spectra in \eq{spectrum} are normalized to the leading-order partial decay widths,
\beq
\Gamma^q_{H0}=\frac{y_q^2(m_H)m_HC_A}{16\pi},\quad\quad
\Gamma^g_{H0}=\frac{\alpha_s^2(m_H)m_H^3}{72\pi^3v^2}.
\eeq
$H^i(m_H,\mu)$ is the hard coefficient of Higgs boson decaying to quarks or gluons, and is given by the square of the matching coefficients in \eq{hardmatching},
\beq
H^i(m_H,\mu)=|C_2^i(m_H,\mu)|^2.
\eeq
These encode virtual fluctuations at scales $\mu\sim m_H$ that give quantum corrections to the decay operators on the left-hand sides of \eq{hardmatching} that are integrated out of the lower-scale effective theory of collinear and soft excitations.
At leading order in QCD factorization, soft wide-angle radiation can be shown to factor for the dynamics inside collinear jets \cite{Korchemsky:1993uz}, such that the sum of all soft emissions couple only to Wilson lines encoding the light-cone directions $n,\bar n$ of jets and their color representations. In SCET this is encapsulated in a field redefinition of collinear fields with soft Wilson lines so that soft gluons no longer couple directly to any collinear particles \cite{Bauer:2001yt}.
All collinear radiation and splitting inside jets are described by the jet functions,
$J_{n,\bar{n}}^i(t_a^{n,\bar{n}},\mu)$, 
defined here by~\cite{Ellis:2010rwa,Almeida:2014uva}
\begin{align}
J_n^q(t_a^n,\mu) & = \frac{1}{2N_C}\!\int\! \frac{dl^+}{2\pi}  \!\Tr \!\int\! d^4x \, e^{il\cdot x}\! \bra{0}\frac{\bnslash}{2}\chi_n(x)\delta(t_a^n - Q^{2-a}\hat \tau_a^n)\delta(Q+\bar n\mcdot\cP)\delta^2(\cP_\perp)\bar\chi_n(0)\ket{0}, \nn \\
J_n^g(t_a^n,\mu) &= -\frac{Q}{2(N_C^2-1)}\int\!\frac{dl^+}{2\pi}\!\Tr \!\int\! d^4x \, e^{il\cdot x}\nn\\
&\qquad\times \bra{0}B_{n\perp}^{\mu A}(x)\delta(t_a^n - Q^{2-a}\hat \tau_a^n)\delta(Q-\bar n\mcdot\cP)\delta^2(\cP_\perp)B_{n\perp}^{\nu B}(0) \ket{0}
\label{eq:Jdef}
\end{align}
where the traces are over color and Dirac (Lorentz) indices, the ``label'' momentum operators $\cP^\mu$ fix the large label momentum components of the SCET collinear quark (gluon) fields $\chi_n$ ($B_{n\perp}$) \cite{Bauer:2001ct} (here, $Q=m_H$), and $\hat\tau_a^n$ is an operator that fixes the angularity of final states produced by the collinear quark (gluon) fields \cite{Bauer:2008dt}.
Meanwhile, the soft functions $S^i(k^s,\mu)$ are defined by
\begin{align}
\label{eq:Sdef}
S^q(k_s,\mu)&=\frac{1}{N_C} \Tr\langle 0|\overline{T}[{Y}_{\bar{n}}Y_n^\dagger](0)\delta(k_s-Q\hat{\tau}_a^s)T[Y_nY^\dag_{\bar{n}}](0)|0\rangle,\nn\\
S^g(k_s,\mu)&=\frac{1}{N_C^2-1}\Tr\langle 0|\overline{T}[{\mathcal{Y}}_{\bar{n}}\mathcal{Y}_n^\dagger](0)\delta(k_s-Q\hat{\tau}_a^s)T[\mathcal{Y}_n\mathcal{Y}_{\bar{n}}](0)|0\rangle,
\end{align}
where $Y_n (\mathcal{Y}_n)$ are soft Wilson lines in the fundamental (adjoint) representation along  the direction of $n_\mu$,
\beq
Y_n(x)=P{\rm exp}\left[ig_s\int_0^\infty\!ds\, n\mcdot A_s(ns+x)\right],
\eeq
with $A_s$ in the fundamental representation, and $\mathcal{Y}_n$ is similar but in the adjoint representation.

In \eqs{Jdef}{Sdef}, the operators $\hat{\tau}_a^{n,s}$ acting on a collinear or soft final state returns the contribution to the angularity $\tau_a$ of that state, defined by its action on the collinear (soft) states $X_{n,s}$,
\beq
\hat{\tau}_a^{n,s}|X_{n,s}\rangle=\frac{1}{Q}\sum_{i\in X_{n,s}}|p_\perp^i|e^{-|\eta_i|(1-a)}|X_{n,s}\rangle\,,
\eeq
and can also be constructed in terms of the energy-momentum tensor in QCD~\cite{Sveshnikov:1995vi,Cherzor:1997ak,Belitsky:2001ij,Bauer:2008dt}.

In the next subsection we will review perturbative calculations of the above functions appearing in the factorized decay spectra and use them to evolve each piece and sum large logarithms in the full decay spectra.

\subsection{RG Evolution and Resummation}

The prediction for the decay spectrum \eq{spectrum} in fixed order QCD perturbation theory contains logs of $\tau_a$ at every order in $\as$, which become large for $\tau_a\ll 1$ and need to be resummed. This can be achieved by RG evolution of each piece of the factorized cross section (see, e.g. \cite{Contopanagos:1996nh,Almeida:2014uva}).

The Yukawa coupling $y_q(\mu)$ and strong coupling $\alpha_s(\mu)$ obey the following renormalization-group  (RG) equations,
\beq
\mu\frac{d}{d\mu}y_q(\mu)=\gamma_y[\alpha_s(\mu)]y_q(\mu),\qquad \mu\frac{d}{d\mu}\alpha_s(\mu)=\beta[\alpha_s(\mu)].
\eeq
The anomalous dimension  $\gamma_y$ and the $\beta$ function have expansions in $\as$ that we express:
\beq
\label{eq:yqas_expansion}
\gamma_y[\alpha_s(\mu)]=\sum_{n=0}^\infty\left(\frac{\alpha_s}{4\pi}\right)^{n+1}\gamma_y^n,\quad \beta[\alpha_s(\mu)]=-2\alpha_s\sum_{n=0}^\infty\left(\frac{\alpha_s}{4\pi}\right)^{n+1}\beta_n.
\eeq
where the coefficients $\gamma_y^n,\beta_n$ are given in \eqs{betan}{gammayn}.
The one-loop hard functions are~\cite{Berger:2010xi},
\begin{align}
H^q(m_H,\mu)&=1-\frac{\alpha_s(\mu) C_F}{2\pi}\left[\ln^2\frac{\mu^2}{m_H^2}+3\ln\frac{\mu^2}{m_H^2}-2+\frac{7\pi^2}{6}\right], \nn \\
H^g(m_H,\mu)&=1-\frac{\alpha_s(\mu)}{2\pi}\left[C_A\ln^2\frac{\mu^2}{m_H^2}+\beta_0\ln\frac{\mu^2}{m_H^2}-\left(5+\frac{7\pi^2}{6}\right)C_A+3C_F\right].
\end{align}

The soft functions for quark and gluon jets  are known to one~\cite{Fleming:2007xt} and two loops~\cite{Kelley:2011ng,Monni:2011gb} for $a=0$. For generic values of $a$, the soft function was computed to  one loop order in~\cite{Hornig:2009vb}: 
The one-loop result in Laplace space is,
\beq
\widetilde{S}^i(\nu_a,\mu)=1-\frac{\alpha_s(\mu)}{4\pi}\frac{\pi^2}{1-a}C_i-\frac{\alpha_s(\mu)}{4\pi}\frac{8C_i}{1-a}\ln^2\frac{\mu e^{\gamma_E}\nu_a}{m_H}
\eeq
Here the color factor $C_i=C_{F,A}$ for quark and gluon respectively and $\nu_a$ is the Laplace-conjugate variable to the angularity $\tau_a$.
The two-loop soft functions for $a\neq 0$ have recently become computable using the program \texttt{SoftSERVE}~\cite{Bell:2015lsf,Bell:2018vaa,Bell:2018oqa}, which were used for predictions of $e^+e^-$ angularities to NNLL$'$ accuracy in \cite{Bell:2018gce}. To at least two-loop order, whether the directions of the $n,\bar n$ Wilson lines in \eq{Sdef} are incoming or outgoing does not affect the perturbative results for the soft functions \cite{Kang:2015moa}, so we can also use the results of \cite{Bell:2018vaa,Bell:2018oqa} here. We give the results for the non-cusp anomalous dimensions in \eq{gammaS1a}. The 2-loop constant terms for the quark channel can be found in \cite{Bell:2018gce,Bell:2018oqa}, though we will not use them in this paper, where we go only to NNLL$+\mathcal{O}(\as)$ accuracy.

The jet functions for massless quark and gluon jets are also known to one~\cite{Bauer:2003pi,Becher:2009th}, two~\cite{Becher:2006qw,Becher:2010pd} and even three loops~\cite{Bruser:2018rad,Banerjee:2018ozf} for $a=0$. For  generic values of $a$, the one-loop jet functions for quarks were computed in \cite{Hornig:2009vb}, and for gluons in~\cite{Ellis:2010rwa,Hornig:2016ahz}. The result can be expressed in Laplace space,
\begin{align}
\widetilde{J}^i(\nu_a,\mu)=1&+\frac{\alpha_s(\mu)}{4\pi}\left[f_i(a)+\frac{2\pi^2C_i}{3(1-a)(2-a)}\right]+\frac{\alpha_s(\mu)}{4\pi}\frac{4\gamma_i}{2-a}\ln\frac{\mu^{2-a}e^{\gamma_E}\nu_a}{m_H^{2-a}}\nn\\
&+\frac{\alpha_s(\mu)}{4\pi}\frac{4C_i}{(1-a)(2-a)}\ln^2\frac{\mu^{2-a}e^{\gamma_E}\nu_a}{m_H^{2-a}},
\end{align}
where the non-cusp anomalous dimension coefficient $\gamma_i$ is given by,
\beq
\gamma_q=\frac{3}{2}C_F,\quad \gamma_g=\frac{\beta_0}{2}.
\eeq
The coefficient of the constant term is a function of $a$, $f_i(a)$, which is defined as,
\begin{align}
\label{eq:fqfg}
f_q(a)&=\frac{4C_F}{1\minus a/2}\left\{\frac{7 \minus 13a/2}{4}-\frac{\pi^2}{12}\frac{3-5a+9a^2/4}{1-a}-\int_0^1 \! dx \frac{1\minus x\plus x^2/2}{x}\ln\left[(1\minus x)^{1-a}+x^{1-a}\right]\right\}, \nn \\
f_g(a)&=\frac{2}{1-a/2}\biggl\{ C_A\biggl[(1-a)\left(\frac{67}{18}-\frac{\pi^2}{3}\right)-\frac{\pi^2}{6}\frac{(1-a/2)^2}{1-a}\biggr] -T_F n_f \frac{20-23a}{18}  \\
&\qquad \qquad -\int_0^1dx  \biggl[C_A\frac{(1-x+x^2)^2}{x(1-x)} + T_F n_f (1-2x+2x^2) \biggr] \ln\left[(1-x)^{1-a}+x^{1-a}\right]\Bigr] \biggr\}\nn
\end{align}
Here $T_F=\frac{1}{2}$, and $n_f=5$ is the number of active quark flavors. The remaining integrals in the definitions of $f_{q,g}$ are easily evaluated numerically.

The hard, soft and jet functions $H,\widetilde{S},\widetilde{J}$ obey the renormalization group equations (RGEs),
\beq
\mu\frac{d}{d\mu}F^i(\mu)=\gamma_F^i(\mu)F^i(\mu),
\eeq
where $F=H,\widetilde{S},\widetilde{J}$. The anomalous dimension $\gamma_F^i$ is given by,
\beq
\gamma_F^i(\mu)=-\kappa_F \Gamma^i_{\rm cusp}[\alpha_s]\ln\frac{\mu^{j_F}e^{\gamma_E}\nu_a}{m_H^{j_F}}+\gamma_F^i[\alpha_s].
\eeq
Here $\kappa_H=4$, $\kappa_S=\frac{4}{2-a}$, $\kappa_J=-\frac{2}{1-a}$, $j_{H,S}=1$ and $j_J=2-a$.
Both the cusp and non-cusp anomalous dimensions can be expanded as,
\beq
\label{eq:gammaexpansion}
\Gamma^i_{\rm cusp}[\alpha_s]=\sum_{n=0}^\infty\left(\frac{\alpha_s}{4\pi}\right)^{n+1}\Gamma_n^i,\quad
\gamma^i_{F}[\alpha_s]=\sum_{n=0}^\infty\left(\frac{\alpha_s}{4\pi}\right)^{n+1}\gamma_{Fn}^i.
\eeq
The coefficients $\Gamma_n^i$ and $\gamma_{Fn}^i$ are given in App.~\ref{app:NNLL}.

For the two-loop jet functions for generic $a$, the cusp parts of the anomalous dimensions are known, while the non-cusp anomalous dimension for quark jets can be obtained from the hard and soft anomalous dimensions, by RG consistency,
\begin{equation}
\gamma_H + 2\gamma_J(a) + \gamma_S(a) = 0\,.
\end{equation}
The two-loop constant terms for $a\neq 0$ for quark jets were obtained in \cite{Bell:2018gce} from numerical computations of the QCD singular cross section using EVENT2 \cite{Catani:1996jh,Catani:1996vz} together with the numerical results for soft function constants in \cite{Bell:2018oqa}. Newer, highly precise results for $\gamma_J(a)$ and 2-loop constants for $a\neq 0$ for quark jet functions have been presented in \cite{Bell:2021dpb}. Similar computations could be done for the gluon jet function for arbitrary $a$ but lie outside the scope of this paper.

It is convenient to present the resummed results for the cumulative distribution,
\beq
\Gamma_{Hc}^i=\int_0^{\tau_a} d\tau_a^\prime\frac{d\Gamma^i_H}{d\tau_a^\prime},
\eeq
whose resummed form in momentum ($\tau_a$) space is conveniently expressed in terms of the Laplace-transformed jet and soft functions acting with derivative operators on a resummation kernel \cite{Becher:2006mr,Becher:2006nr,Almeida:2014uva}:
\begin{align}
\frac{\Gamma_{Hc}^i}{\Gamma_{H0}^i}&=e^{\widetilde{K}^i(\mu_H,\mu_J,\mu_S,m_H)+K_\gamma^i(\mu_H,\mu_J,\mu_S)}\left(\frac{1}{\tau_a}\right)^{\Omega^i(\mu_J,\mu_S)}H^i(m_H,\mu_H)\nn\\
&\times \widetilde{J}^i\left(\partial_{\Omega^i}+\ln\frac{\mu_J^{2-a}}{m_H^{2-a}\tau_a},\mu_J\right)^2\widetilde{S}^i\left(\partial_{\Omega^i}+\ln\frac{\mu_S}{m_H\tau_a},\mu_S\right)\frac{e^{\gamma_E\Omega^i}}{\Gamma(1-\Omega^i)}.
\label{eq:res}
\end{align}
Note that we use a form of the cusp evolution kernel proposed in~\cite{Bell:2018gce} that keeps the resummed cross section explicitly independent of the factorization scale $\mu$ appearing in the original factorized cross section \eq{spectrum}, at every order in resummed perturbation theory,
\beq
\widetilde{K}_{\Gamma}^i(\mu,\mu_F,Q)=\int_{\mu_F}^\mu \frac{d\mu^\prime}{\mu^\prime}\Gamma_{\rm cusp}^i[\alpha_s(\mu^\prime)]\ln\frac{\mu^\prime}{Q}.
\eeq 
While the other evolution kernels are defined as follows,
\begin{align}
\eta_{\Gamma}^i(\mu,\mu_F) &=\int_{\mu_F}^\mu\frac{d\mu^\prime}{\mu^\prime}\Gamma_{\rm cusp}^i[\alpha_s(\mu^\prime)], &
K_{\gamma_F}^i(\mu,\mu_F)&=\int_{\mu_F}^\mu\frac{d\mu^\prime}{\mu^\prime}\gamma_{F}^i[\alpha_s(\mu^\prime)]. 
\end{align}
Explicit expansions order by order for these kernels are given in~\cite{Bell:2018gce}.
The sums of individual hard, jet, and soft evolution kernels $\Omega^i$, $\widetilde{K}^i$ and $K_\gamma^i$ used in \eq{res} are defined,
\begin{align}
\Omega^i\equiv\Omega^i(\mu_J,\mu_S)&=-2\kappa_J\eta_\Gamma^i(\mu,\mu_J)-\kappa_S\eta_\Gamma^i(\mu,\mu_S), \nn\\
K_\gamma^i(\mu_H,\mu_J,\mu_S)&\equiv K_{\gamma_H}^i(\mu,\mu_H)+2K_{\gamma_J}^i(\mu,\mu_J)+K_{\gamma_S}^i(\mu,\mu_S),\nn\\
\widetilde{K}^i(\mu_H,\mu_J,\mu_S,Q) &\equiv -4\widetilde{K}_{\Gamma}^i(\mu,\mu_H,Q)-2j_J \kappa_J\widetilde{K}_{\Gamma}^i(\mu,\mu_J,Q)-\kappa_S\widetilde{K}_{\Gamma}^i(\mu,\mu_S,Q).
\end{align}

\subsection{Fixed-order matching}

In order to obtain a reliable prediction for large values of $\tau_a$, we also need to match our calculation to the full QCD fixed-order distribution. To $\mathcal{O}(\alpha_s)$, the full QCD distribution is,
\beq
\frac{1}{\Gamma_{H0}^i}\frac{d\Gamma_H^i}{d\tau_a}\bigg \vert_{\rm QCD}=\delta(\tau_a)+\left(\frac{\alpha_s}{2\pi}\right)A_a^i(\tau_a)+\mathcal{O}(\alpha_s^2).
\eeq
We follow the method in Ref.~\cite{Hornig:2009vb} to calculate the coefficient $A_a^i(\tau_a)$ numerically. The detail of the calculation can be found in the appendix.  Analytical results are only known for $a=0$, found, e.g. in Ref.~\cite{Ellis:1996mzs,Dasgupta:2003iq,Gao:2019mlt}.

\begin{figure}
\centering
\includegraphics[width=0.32\textwidth]{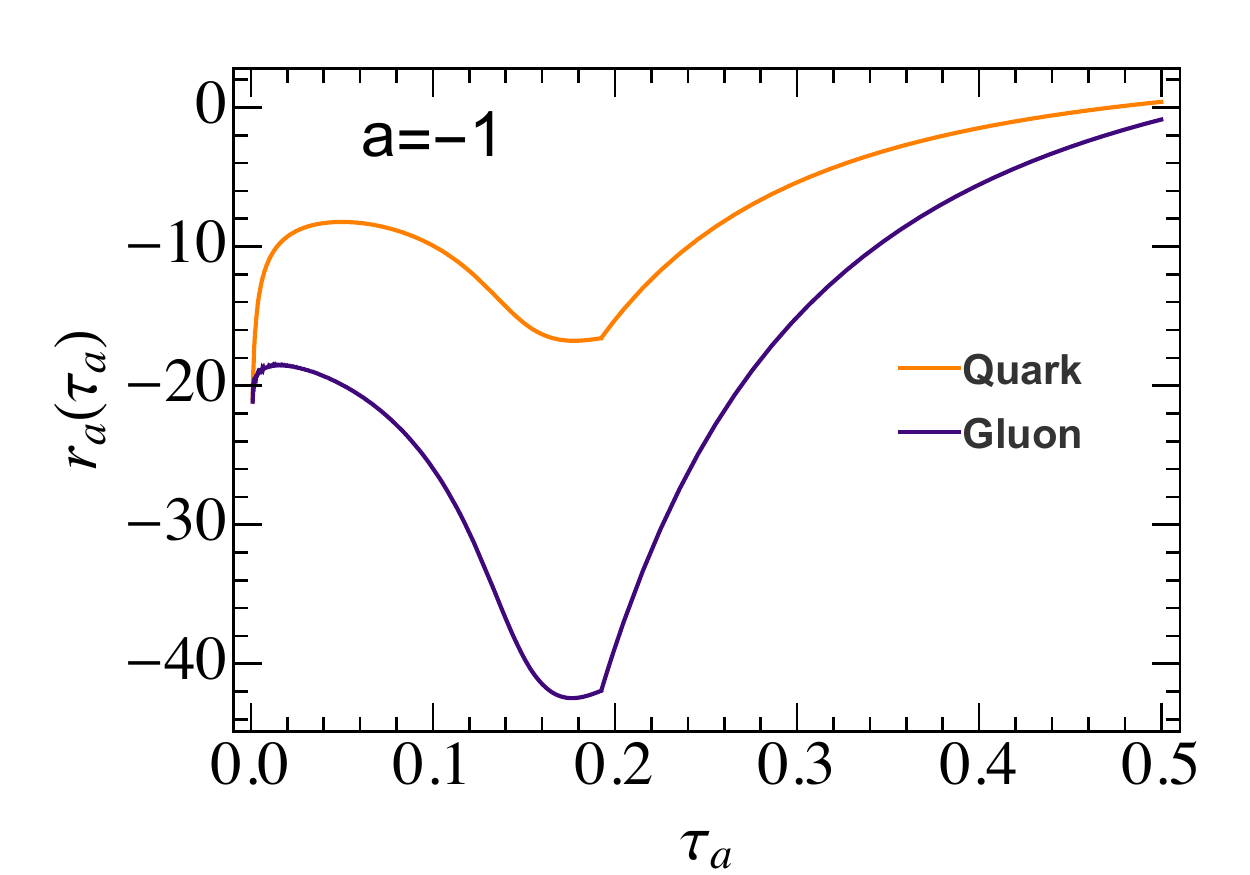}
\includegraphics[width=0.32\textwidth]{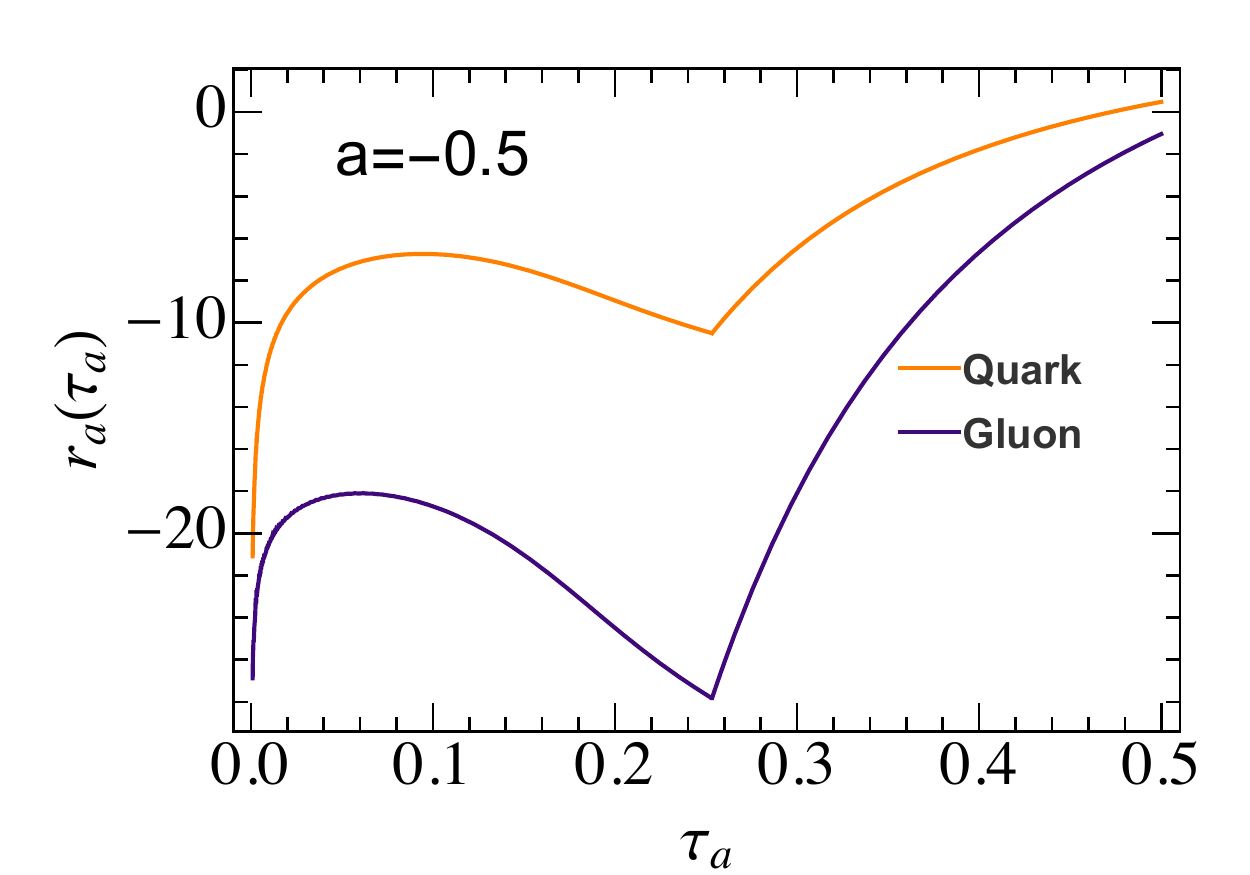}
\includegraphics[width=0.32\textwidth]{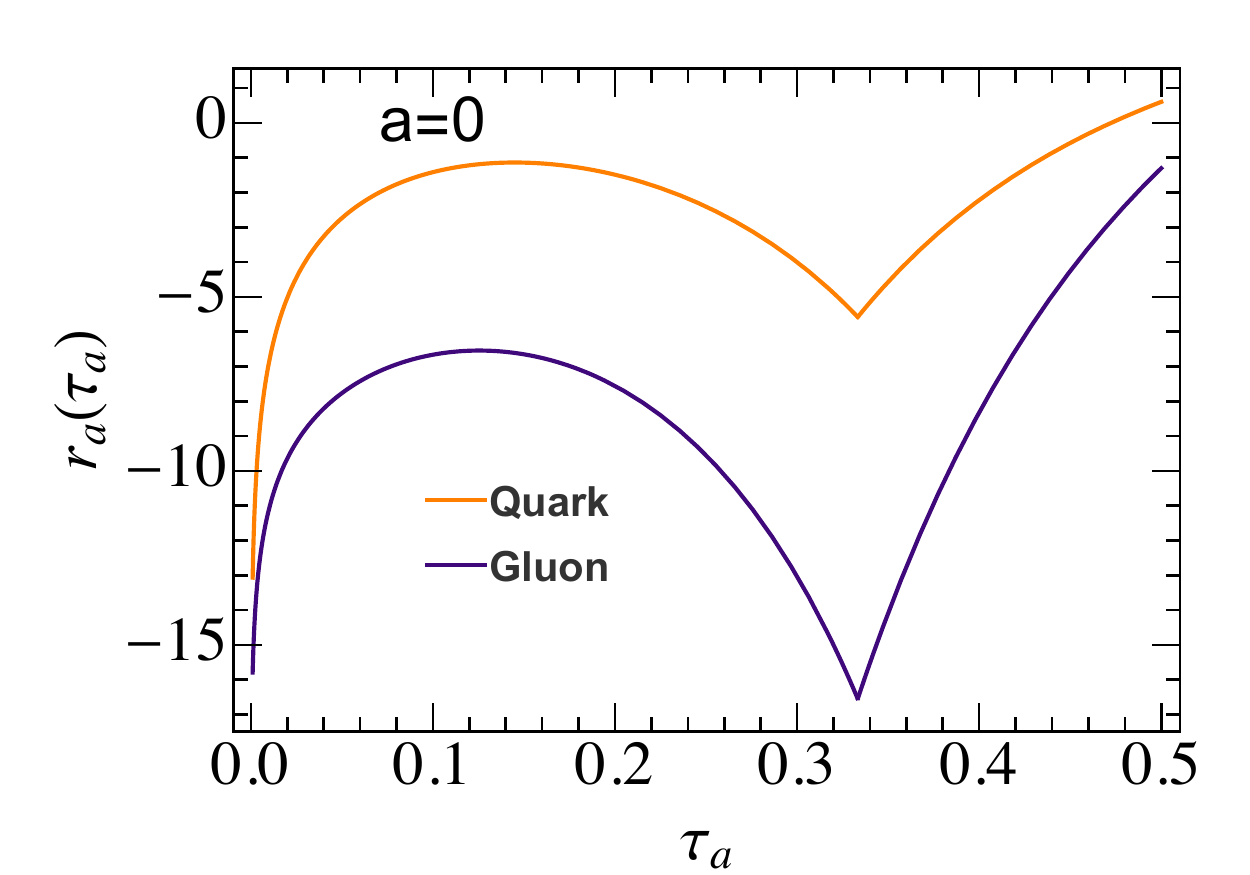}
\includegraphics[width=0.32\textwidth]{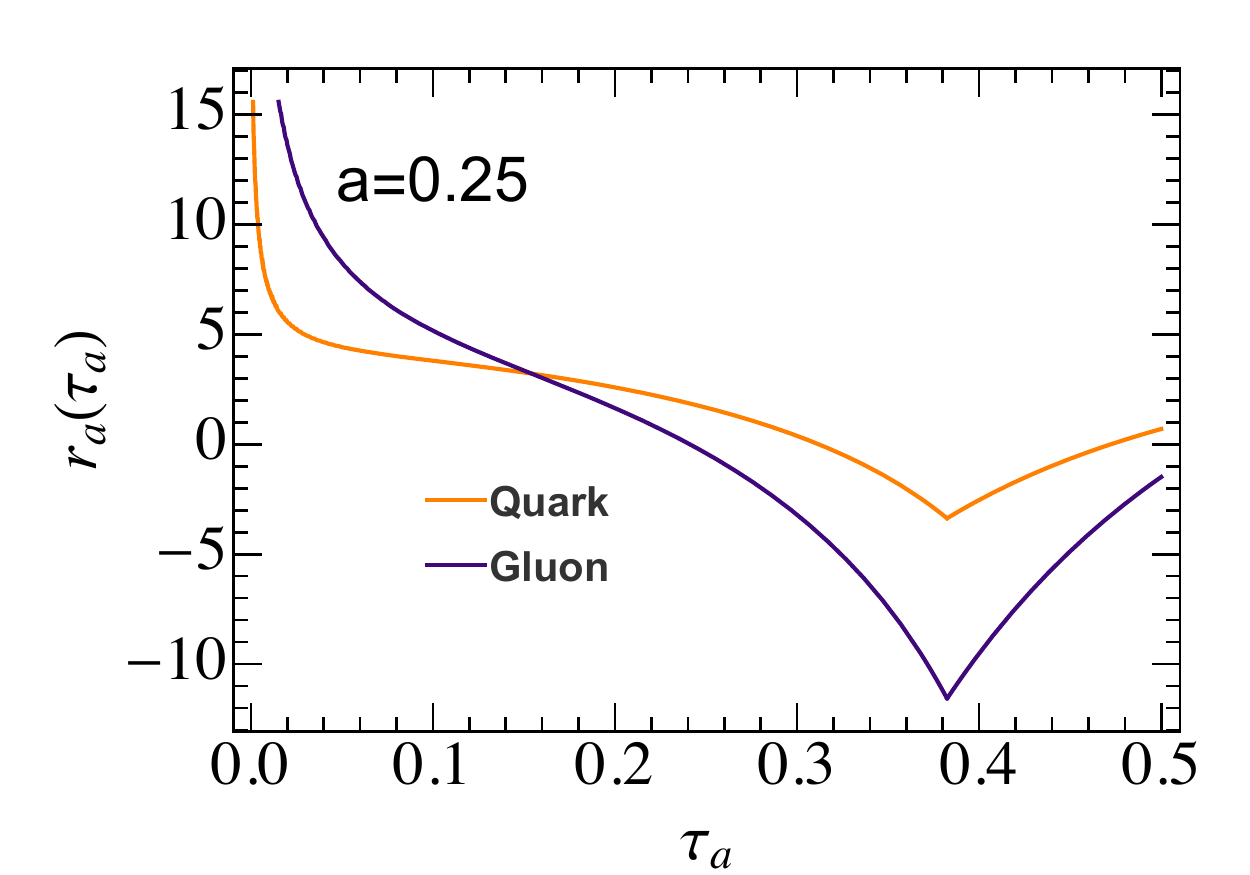}
\includegraphics[width=0.32\textwidth]{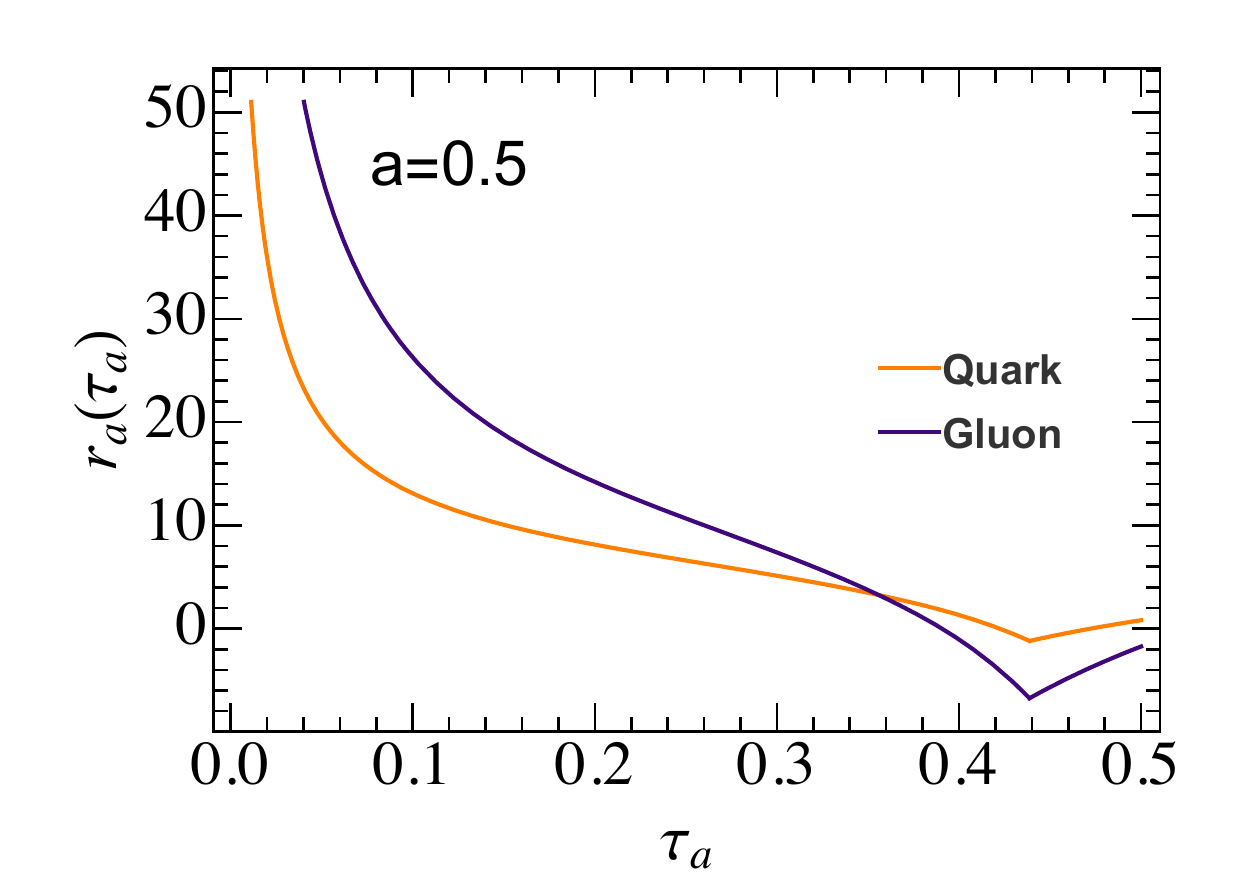}
\vspace{-1em}
\caption{The $\cO(\as)$ nonsingular remainder functions for Higgs decaying to quarks (orange) and gluons (purple). Their relative size compared to the singular and total fixed-order cross sections is shown in Fig.~\ref{fig:cross}.}
\label{fig:ra}
\end{figure}

The fixed-order angularity in SCET at $\mathcal{O}(\alpha_s)$ is given by,
\beq
\label{eq:singulardist}
\frac{1}{\Gamma_{H0}^i}\frac{d\Gamma_H^i}{d\tau_a}\bigg \vert_{\rm SCET}=\delta(\tau_a)D_a^{\delta i}+\left(\frac{\alpha_s}{2\pi}\right)[D_a^i(\tau_a)]_+.
\eeq
For quark final state,
\begin{align}
&D_a^{\delta q}=1-\frac{\alpha_s}{2\pi}\frac{C_F}{2-a}\Bigl\lbrace 2+5a-\frac{\pi^2}{3}\left(2+a\right)+6\left(a-2\right) \nn \\
&\qquad\qquad\qquad +4\int_0^1dx \frac{x^2-2x+2}{x}\ln\left[x^{1-a}+(1-x)^{1-a}\right] \Bigr\rbrace,\nn\\
&D_a^q(\tau_a)=-\frac{2C_F}{2-a}\frac{\theta(\tau_a)(3+4\ln\tau_a)}{\tau_a}.
\end{align}
For gluon final state,
\begin{align}
&D_a^{\delta g}=1+\frac{\alpha_s}{2\pi}\left\lbrace \left(5+\frac{7\pi^2}{6}\right)C_A-3C_F+\frac{C_A}{1-a}\frac{\pi^2}{6}+2f_g(a)\right\rbrace,\nn\\
&D_a^g(\tau_a)=-\frac{2}{2-a}\frac{\theta(\tau_a)\left(\beta_0+4C_A \ln \tau_a\right)}{\tau_a}\,,
\end{align}
where the function $f_g$ was given in \eq{fqfg}.
We define the difference away from $\tau_a=0$ between full QCD and SCET as,
\beq
\label{eq:remainder}
r_a^i(\tau_a)=A_a^i(\tau_a)-D_a^i(\tau_a).
\eeq
Note that the coefficient $\alpha_s/(2\pi)$ is not included in the above definition. We need to add this remainder function to the resummed distribution in order to obtain the $\rm{NLL}^\prime+\mathcal{O}(\alpha_s)$ and ${\rm NNLL}+\mathcal{O}(\alpha_s)$ accuracy.  The numerical results of $r_a^i(\tau_a)$ for values of $a$ equal to $\{-1.0,-0.5,0.0,0.25,0.5\}$ are shown in Fig.~\ref{fig:ra}. 
\begin{figure}
\centering
\includegraphics[width=0.32\textwidth]{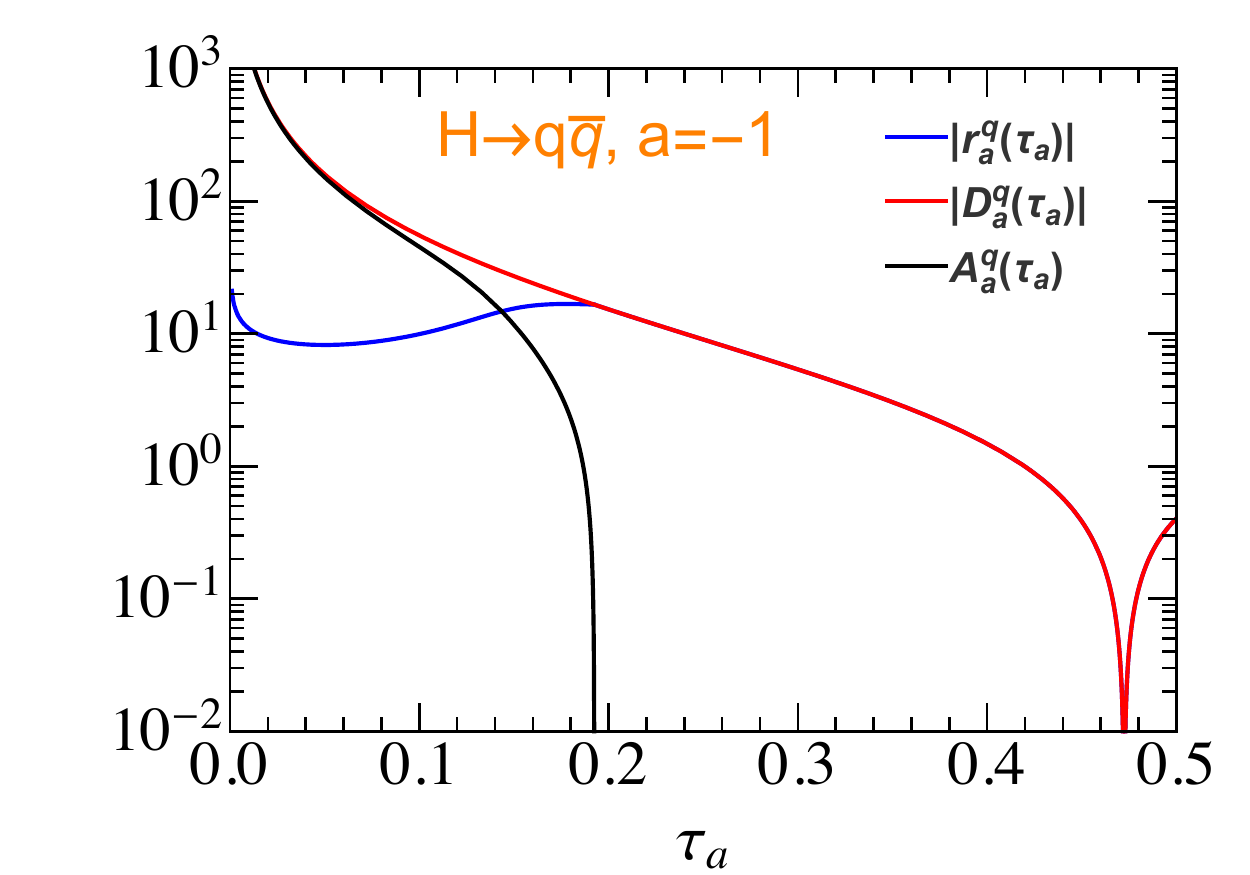}
\includegraphics[width=0.32\textwidth]{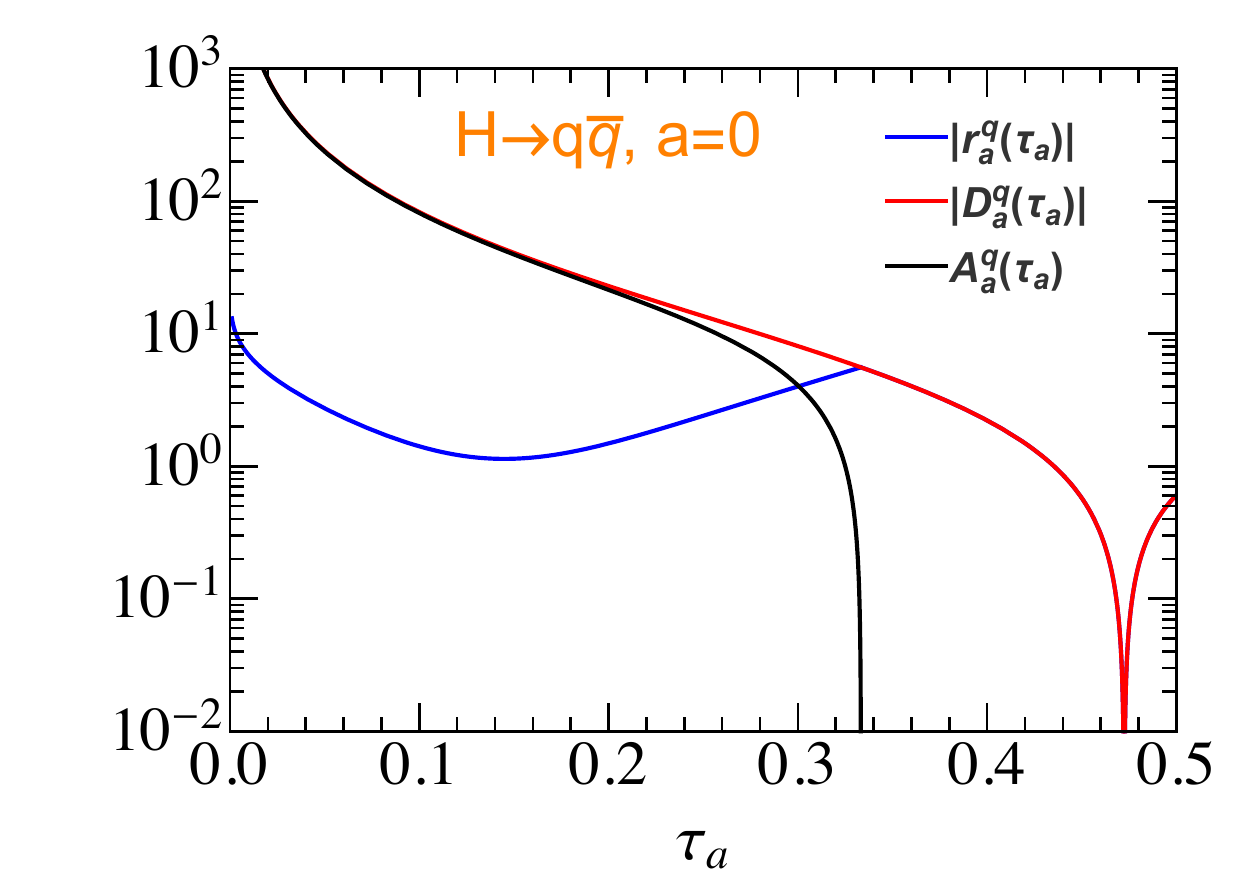}
\includegraphics[width=0.32\textwidth]{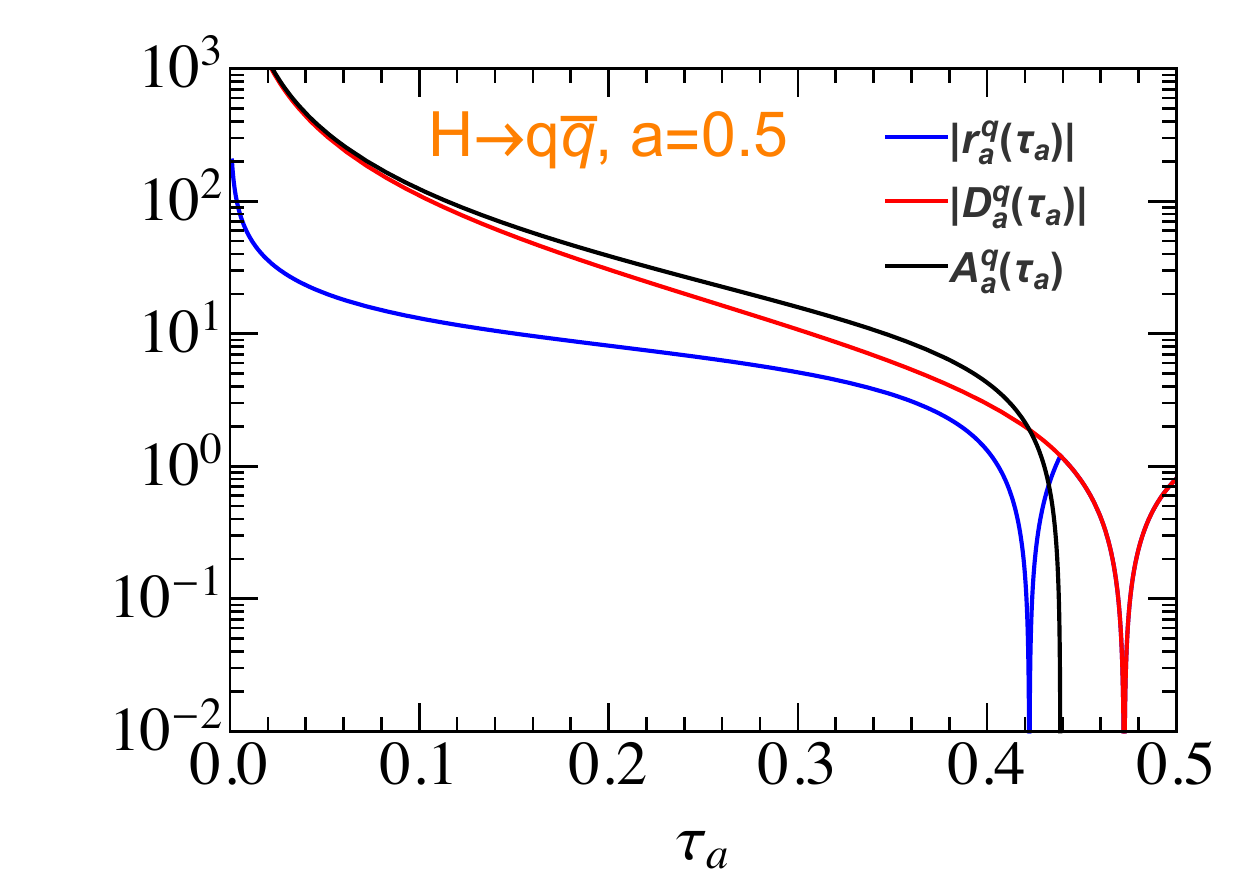}
\includegraphics[width=0.32\textwidth]{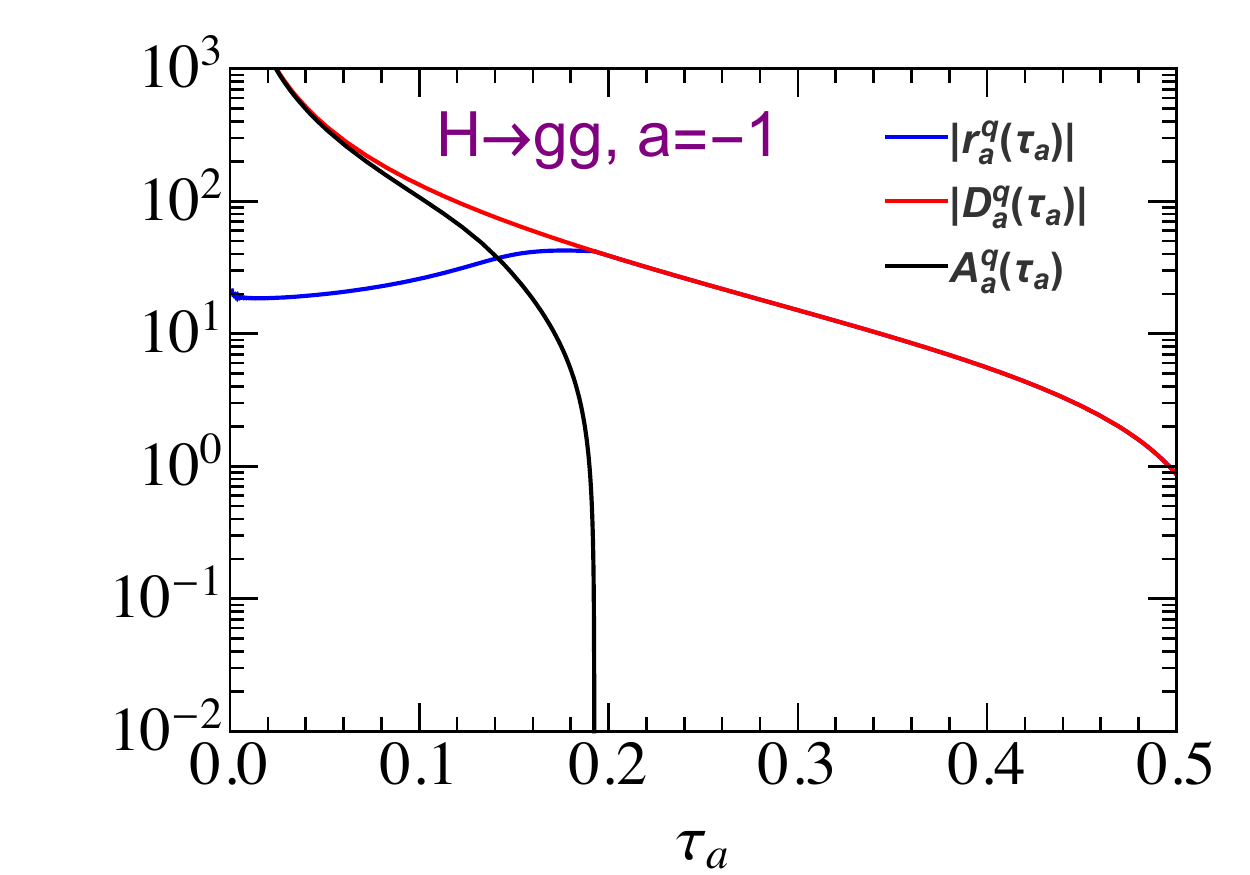}
\includegraphics[width=0.32\textwidth]{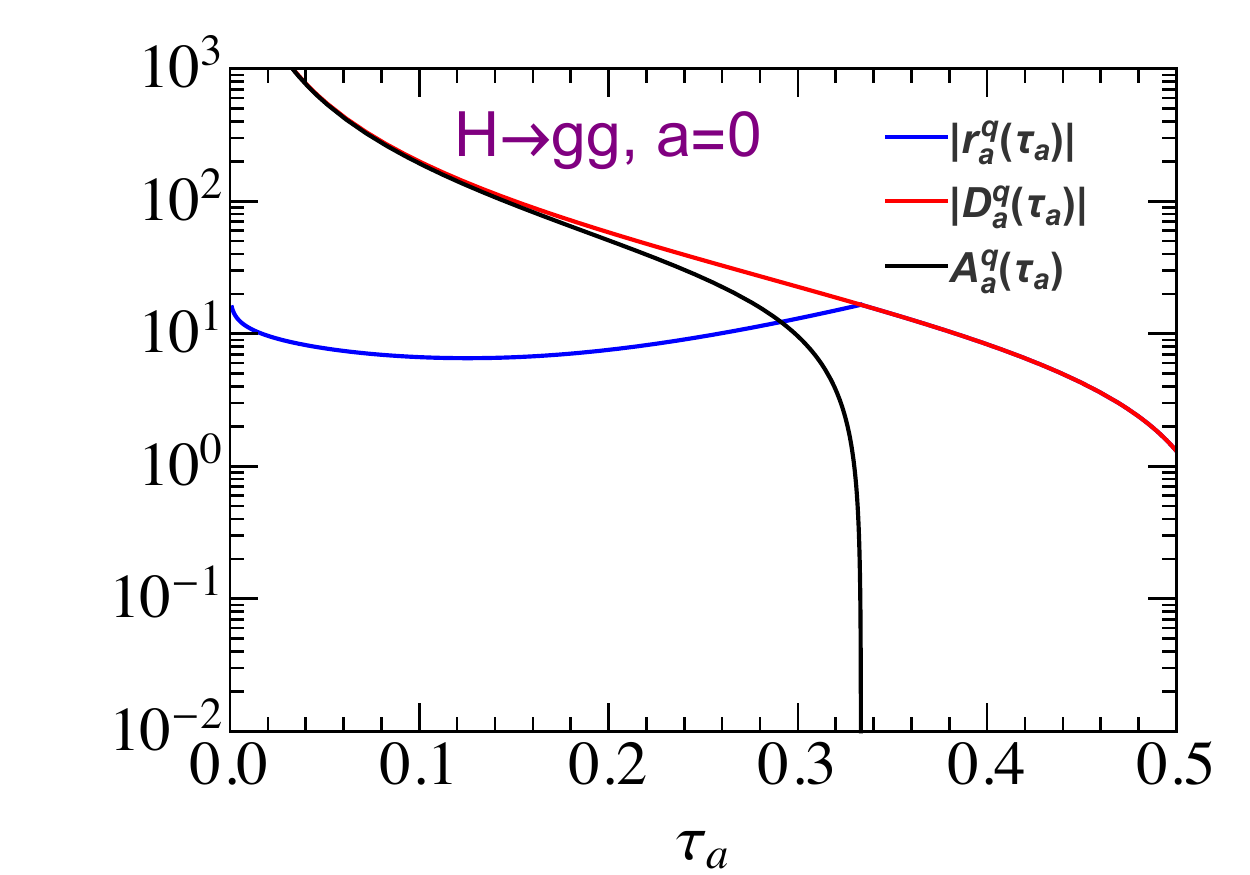}
\includegraphics[width=0.32\textwidth]{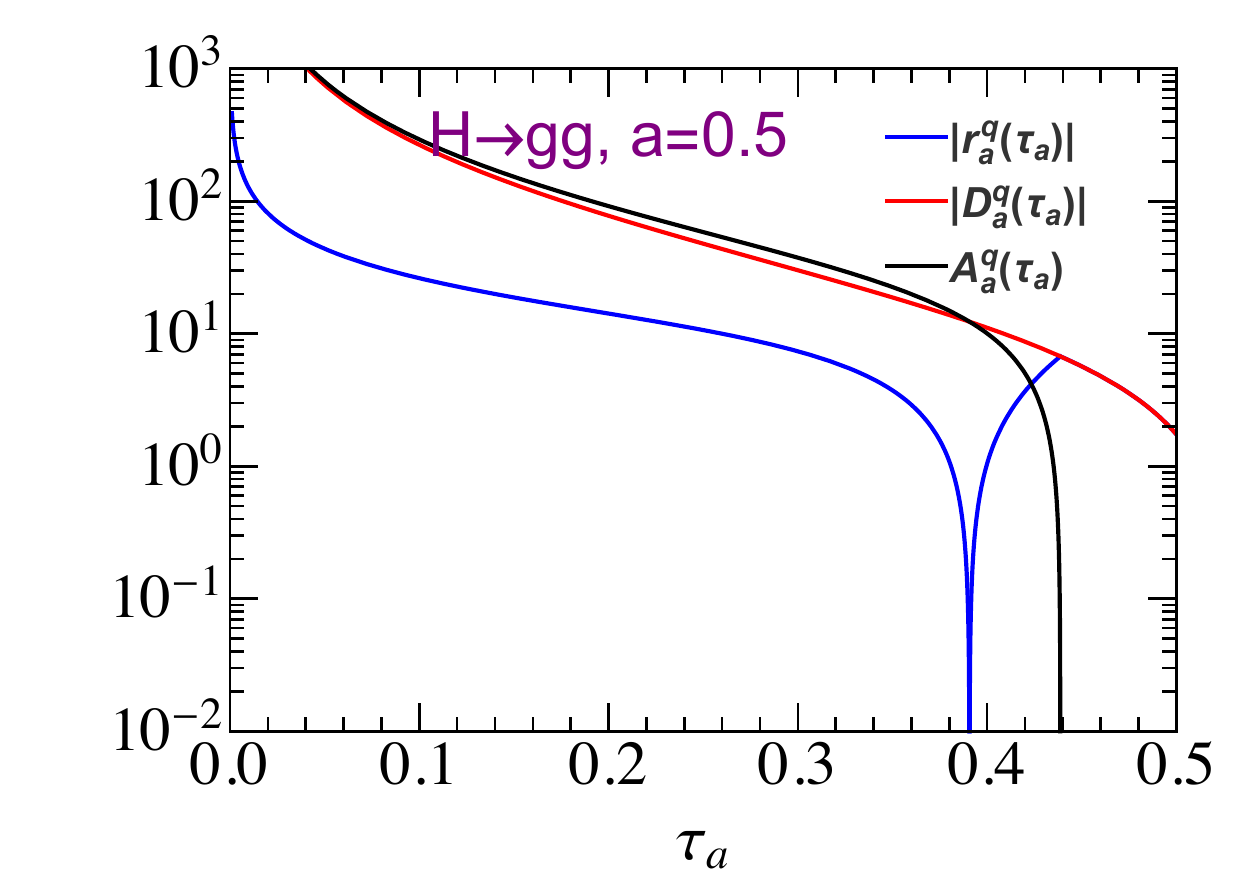}
\vspace{-1em}
\caption{The singular and nonsingular distributions for $a=(-1.0,0.0,0.5)$. The value of $\tau_a$ where the singular (black) and nonsingular (blue) cross determines 
the transition point $\tau_a=t_2(a)$ between resummation and fixed-order regions that will be used in the scale profile functions later.}
\label{fig:cross}
\end{figure}
The kink in Fig.~\ref{fig:ra} at $\tau_a^{\rm max}=\frac{1}{3^{1-a/2}}$ is due to the fact that the full QCD distribution vanishes above the maximum kinematic value of $\tau_a$ and  only the singular part will give a contribution above this value.
As an important consistency check of this matching technique, we integrate over  $\tau_a$ to get the fixed-order correction for the Higgs partial decay width. 
The total partial decay width can be written as,
\beq
\label{eq:GammaH}
\Gamma_{\rm H tot}^i=\Gamma_{H0}^i\left(1+\frac{\alpha_s(m_H)}{2\pi}\Gamma_{H1}^i\right).
\eeq
The numerical results of $\Gamma_{H1}^i$ are summarized in Table.~\ref{tbl:NLO}.
\begin{table}
\begin{center}
\begin{tabular}{c|c|c|c|c|c|c}
\hline 
$a$ & -1.0 & -0.5  & 0.0 & 0.25 & 0.5 & NLO~\cite{Djouadi:2005gi} \\ 
\hline 
$\Gamma_{H1}^q$ & 11.33 &  11.33 & 11.33 & 11.33 & 11.30 & 11.34 \\ 
\hline 
$\Gamma_{H1}^g$ & 35.83 & 35.83 & 35.83 & 35.83 & 35.77 & 35.83 \\ 
\hline 
\end{tabular} 
\end{center}
\vspace{-1em}
\caption{NLO QCD correction for the Higgs partial decay width for selected values of $a$.}
\label{tbl:NLO}
\end{table}
It shows our results are in agreement with  Ref.~\cite{Djouadi:2005gi} for all the values of $a$ of we are interested in.

\subsection{Nonperturbative shape function}

The soft function in the factorization theorem will also receive corrections from nonperturbative hadronization effects. It can be parameterized into a soft shape function $f_{\rm mod}(k)$ as~\cite{Korchemsky:1998ev,Korchemsky:1999kt,Hoang:2007vb},
\beq
\label{eq:softmodel}
S^i(k,\mu)=\int dk^\prime S_{\rm PT}^i(k-k^\prime,\mu)f_{\rm mod}^i(k^\prime-2\bar{\Delta}^i_a).
\eeq
Here $S_{\rm PT}^i$ is the perturbative soft function and $\bar{\Delta}^i_a=\frac{\bar{\Delta}^i}{1-a}$ is a gap parameter with $\bar{\Delta}^i\sim \Lambda_{\rm QCD}$ and $\bar{\Delta}^i$ is an $a$-independent parameter. This scaling of $\bar\Delta_a^i$ tracks the known scaling $1/(1-a)$ of the first moment of the shape function \cite{Berger:2003pk,Lee:2006nr}.  The shape function can be expanded in a complete set of orthonormal basis functions~\cite{Ligeti:2008ac},
\beq
f_{\rm mod}^i(k)=\frac{1}{\lambda_i}\left[\sum_{n=0}^\infty b_n f_n\left(\frac{k}{\lambda_i}\right)\right]^2,
\eeq
where 
\begin{align}
f_n(x)&=8\sqrt{\frac{2x^3(2n+1)}{3}}e^{-2x}P_n(g(x)), & 
g(x)&=\frac{2}{3}\left[3-e^{-4x}\left(3+12x+24x^2+32x^3\right)\right]-1.
\end{align}
Here $P_n$  are Legendre polynomials. In our calculation, we will only keep $b_0=1$ and set $b_n=0$ for $n>0$. The parameter $\lambda_i$ corresponds to the first moment of the $f_\text{mod}$. 
However, it has been shown in Ref.~\cite{Hoang:2007vb} that the gap parameter $\bar{\Delta}_a$ has a renormalon ambiguity. To ensure good perturbative convergence of the soft function \eq{softmodel} and resulting cross section, as well as a stable definition of $\bar\Delta_a^i$, it is necessary to subtract/add a series removing this ambiguity from both $S_{\rm PT}$ and $\bar \Delta_a^i$. In order to cancel the ambiguity, we can split the gap parameter as,
\beq
\label{eq:gapsubtraction}
\bar{\Delta}_a^i=\Delta_a^i(\mu_S,R)+\delta_a^i(\mu_S,R),
\eeq 
where $\delta_a^i$ can be perturbative expansion with a same renormalon ambiguity as $S_{\rm PT}^i$ and $\Delta_a^i$ is a renormalon free parameter. The $R$ is the subtraction scale,
which is defined through,
\beq
Re^{\gamma_E}\frac{d}{d\ln\nu}\left[\ln\left(e^{-2\nu\delta_a^i(\mu)}\widetilde{S}^i_{\rm PT}(\nu,\mu)\right)\right]_{\nu=1/(Re^{\gamma_E})}=0,
\eeq
where $\nu=\nu_a/m_H$, the same scheme adopted in, e.g. \cite{Hoang:2008fs,Abbate:2010xh,Bell:2018gce}. There are multiple other schemes to define the series $\delta_a^i$, see e.g. \cite{Bachu:2020nqn}, and one could study the quantitative effects of varying schemes (cf. \cite{Bell:2023dqs}), though this lies outside the scope of this paper.
The shift \eq{gapsubtraction} results in the renormalon-free soft function,
\beq
\label{eq:softfree}
S^i(k,\mu)=\int dk^\prime S_{\rm PT}^i(k-k^\prime-2\delta_a^i(\mu,R),\mu)f_{\rm mod}^i(k^\prime-2\Delta^i_a).
\eeq
The non-perturbative effects will shift the perturbative cross section. The shift parameter $\overline{\Omega}^i(\mu,R)$ is given by,
\beq
\frac{2\overline\Omega^i(\mu,R)}{1-a}=2\Delta^i_a(\mu,R)+\int dk\, k f^i_{\rm mod}(k).
\eeq
In our calculation, we choose $\overline\Omega^q(R_{\Delta},R_{\Delta})=0.4~{\rm GeV}$ and $\Delta_a^i(R_\Delta,R_\Delta)=\frac{0.1~{\rm GeV}}{1-a}$ with the reference scale $R_{\Delta}=1.5~{\rm GeV}$.  We also assume the Casimir scaling for the parameter $\overline\Omega^g(R_{\Delta},R_{\Delta})=\overline\Omega^q(R_{\Delta},R_{\Delta}) C_A/C_F$. This is purely an assumption, though probably fairly good, that we make to simplify our illustrative analysis below. A definitive study should measure $\overline\Omega^{q,g}$ separately.  The gap parameter $\Delta_a^i$ can be evolved to any other subtraction scale $R$ and soft scale $\mu_S$, using the formalism in \cite{Hoang:2008yj,Hoang:2009yr}, formulas also summarized in Ref.~\cite{Bell:2018gce}.

\subsection{Scales in resummation}

From the arguments of the logarithms in the hard, jet and soft functions, the canonical scales should be,
\begin{align}
\mu_H&=m_H, & \mu_J&=m_H\tau_a^{1/(2-a)}, & \mu_S=m_H\tau_a.
\end{align}
However, the canonical scales do not properly take into account the transition from the resummation region into the fixed-order region where $\tau_a$ is not small or into the nonperturbative region for $\tau_a\leq \Lambda_{\rm QCD}/m_H$. The use of profile scales has been  proposed to smooth the transition between those different scale regions~\cite{Ligeti:2008ac,Abbate:2010xh} and various specific forms for these profile functions are possible, e.g. \cite{Abbate:2010xh,Kang:2014qba,Hornig:2016ahz,Bell:2018gce}. We use:
\begin{align}
\mu_H &=e_H m_H, \nn\\
\mu_S(\tau_a)&=\left[1+e_S\theta(t_3-\tau_a)\left(1-\frac{\tau_a}{t_3}\right)^2\right]\mu_{\rm run}(\tau_a),\nn\\
\mu_J(\tau_a)&=\left[1+e_J\theta(t_3-\tau_a)\left(1-\frac{\tau_a}{t_3}\right)^2\right]\mu_H^{\frac{1-a}{2-a}}\mu_{\rm run}(\tau_a)^{\frac{1}{2-a}},
\end{align}
where $e_H$ simply controls the normalization of the hard scale, while $e_{J,S}$ control variations of the shape of the jet and soft scales as a function of $\tau_a$ above and below their default, canonical shapes. The ranges over which we vary them are given in Table~\ref{tbl:parameter}. (Actually we keep $e_S=0$ as variations of other parameters below will make varying $e_S$ redundant.) The running scale is defined as,
\begin{equation}
\mu_{\rm run}=
\begin{cases}
\mu_0,&\tau_a\leq\tau_0,\\
\zeta\left(\tau_a;\lbrace t_0,\mu_0,0\rbrace,\lbrace t_1,0,\frac{r}{\tau_a^{\rm sph}\mu_H}\rbrace \right), & t_0\leq\tau_a\leq t_1,\\
\frac{r}{\tau_a^{\rm sph}}\mu_H\tau_a, & t_1\leq\tau_a\leq t_2,\\
\zeta\left(\tau_a;\lbrace t_2,0,\frac{r}{\tau_a^{\rm sph}\mu_H}\rbrace, \lbrace t_3,\mu_H,0\rbrace \right), & t_2\leq\tau_a\leq t_3,\\
\mu_H, &\tau_a\geq t_3.
\end{cases}
\end{equation}
The function $\zeta$ smoothly transitions between regions, and is defined as
\begin{equation}
\zeta\left(\tau_a;\lbrace t_0,y_0,r_0\rbrace,\lbrace t_1,y_1,r_1\rbrace \right)=
\begin{cases}
a+r_0(\tau_a-t_0)+c(\tau_a-t_0)^2, & \tau_a\leq \frac{\tau_0+\tau_1}{2}\\
A+r_1(\tau_a-t_1)+C(\tau_a-t_1)^2, & \tau_a\geq \frac{\tau_0+\tau_1}{2}
\end{cases},
\end{equation}
where the coefficients are 
\begin{align}
a&=y_0+r_0 t_0, & A &=y_1+r_1t_1, & c&=2\frac{A-a}{(t_0-t_1)^2}+\frac{3r_0+r_1}{2(t_0-t_1)}, & C=2\frac{a-A}{(t_0-t_1)^2}+\frac{3r_1+r_0}{2(t_1-t_0)}.
\end{align}
\begin{figure}
\centering
\includegraphics[width=0.6\textwidth]{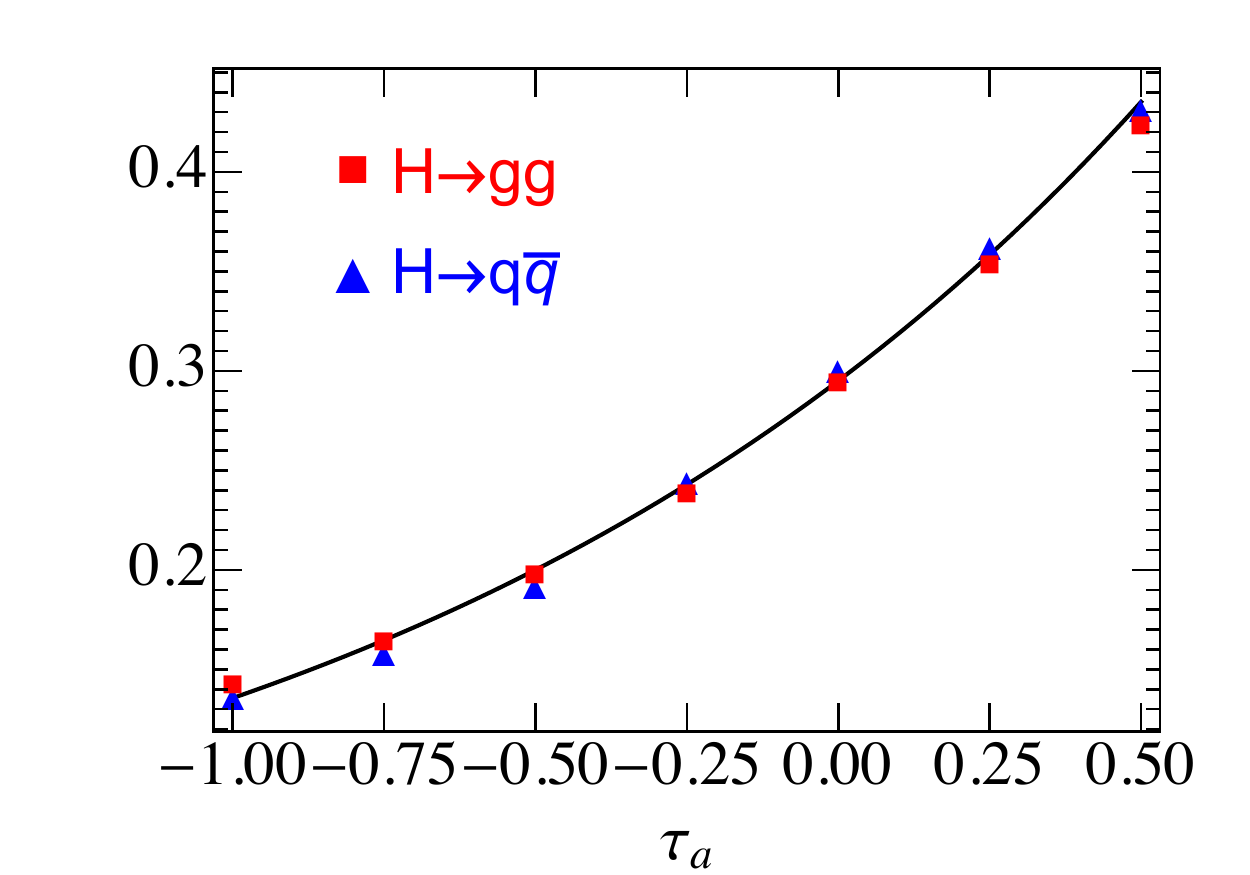}
\vspace{-1em}
\caption{The value of $t_2(a)$ with $a=(-1.0,-0.75,-0.5,-0.25,0.0,0.25,0.5)$, determined by the crossing of singular and nonsingular parts of the cross section in Fig.~\ref{fig:cross}. The blue triangle is from $H\to q\bar{q}$, while the red box is for $H\to gg$. The black line is the fitting function $t_2=0.295^{1-0.637a}$. }
\label{fig:t2}
\end{figure}
The parameters $t_i$ are used to control the transition of different regions and we parameterized them as,
\begin{align}
\label{eq:t0123}
t_0&=\frac{n_0}{m_H}3^a, & t_1&=\frac{n_1}{m_H}3^a,& t_2&=n_2\times 0.295^{1-0.637a}, & t_3&=n_3\tau_a^{\rm sph}.
\end{align}
Here $\tau_a^{\rm sph}=\frac{1}{2-\frac{a}{2}}\;\HypF\Bigl(1,-\frac{a}{2};3-\frac{a}{2};-1\Bigr) $ is the angularity of the spherically symmetric configuration.   The parameters $t_0$ and $t_1$ control the transition between the non-perturbative and resummation regions. In the nonperturbative region, we freeze the scale at $\mu_\text{run}\sim 1-3$ GeV to allow the a shape function to smoothly describe this region and avoid the Landau pole in $\as(\mu)$. 
The parameter $t_2$ was determined by the point where singular and nonsingular contribution  become comparable. As an example, we show the value of $t_2$ for $a=(-1.0,0.0,0.5)$ in Fig.~\ref{fig:cross}, and it is determined by the crossing points of blue and black lines.
It shows the $t_2$ is almost same for $H\to q\bar{q}$ and $H\to gg$, see Fig.~\ref{fig:t2}. The particular functional form in \eq{t0123} is the same as that found for 1-jettiness in DIS in \cite{Kang:2014qba}, and the same form happens to fit the transition points plotted in \fig{t2} as well. Therefore, we will use the same  $t_2$  formula  for both 
Higgs decaying to quark and gluon states. Now, the exact values of $t_{0,1,2,3}$ are somewhat arbitrary, thus we vary them using the parameters $n_{0,1,2,3}$ to estimate theoretical uncertainties, across ranges given in Table~\ref{tbl:parameter}.

For the renormalon subtraction scale, we choose $R(\tau_a)=\mu_S(\tau_a)$ with $\mu_0\to R_0$ and initial value of $R_0<\mu_0$ for $\tau_a<t_0$. The remainder function depends on another scale $\mu_{\rm ns}$ which is used to estimate the higher-order effects from fixed order calculation \cite{Hoang:2014wka},
\begin{equation}
\mu_{\rm ns}(\tau_a)=
\begin{cases}
\frac{1}{2}(\mu_H+\mu_J(\tau_a)), &n_s=1\nn\\
\mu_H, & n_s=0\nn\\
\frac{1}{2}(3\mu_H-\mu_J(\tau_a)), & n_s=-1
\end{cases}.
\end{equation}
The values we pick for $R_0$ for quark and gluon $\tau_a$ distributions are also given in Table~\ref{tbl:parameter}.

\section{Numerical results}
\label{sec:num}

\begin{figure}
\centering
\includegraphics[width=0.32\textwidth]{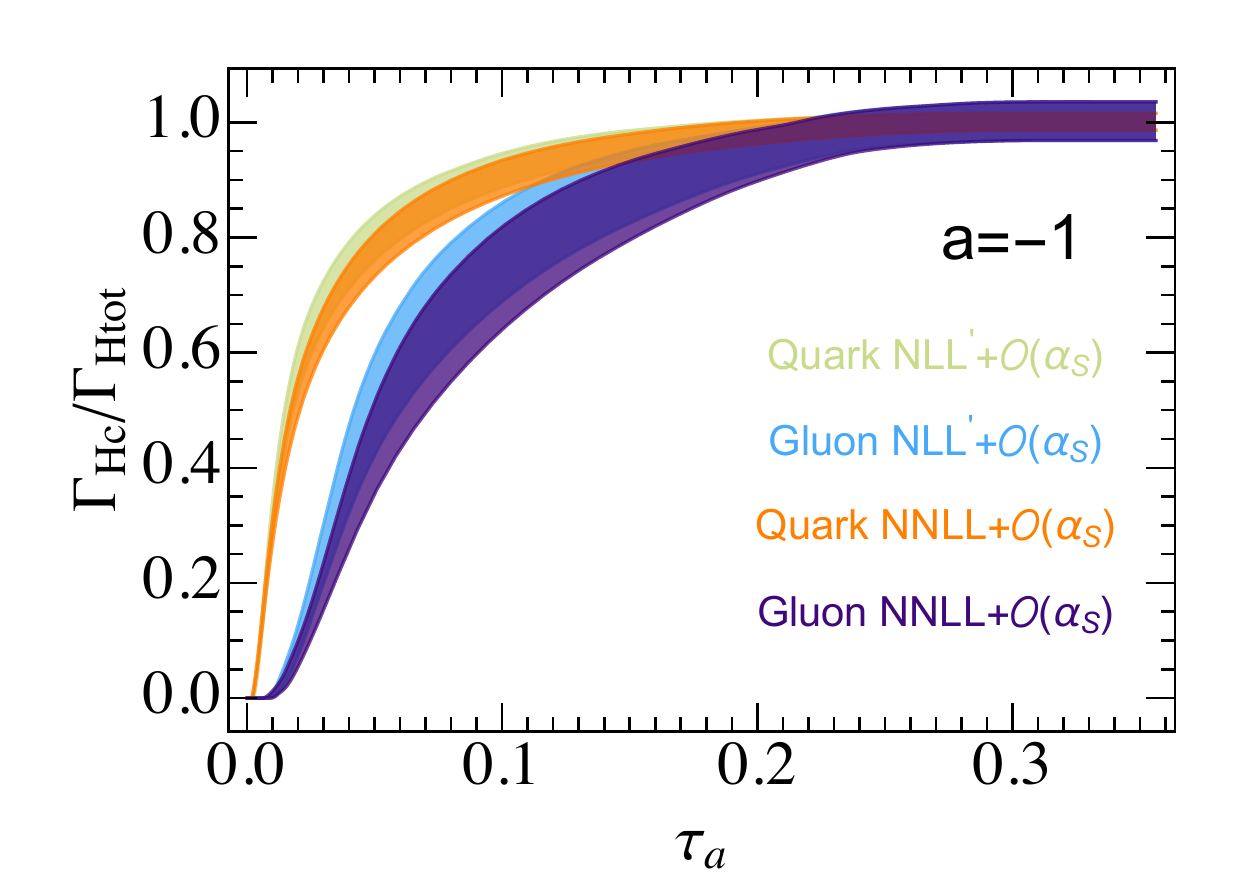}
\includegraphics[width=0.32\textwidth]{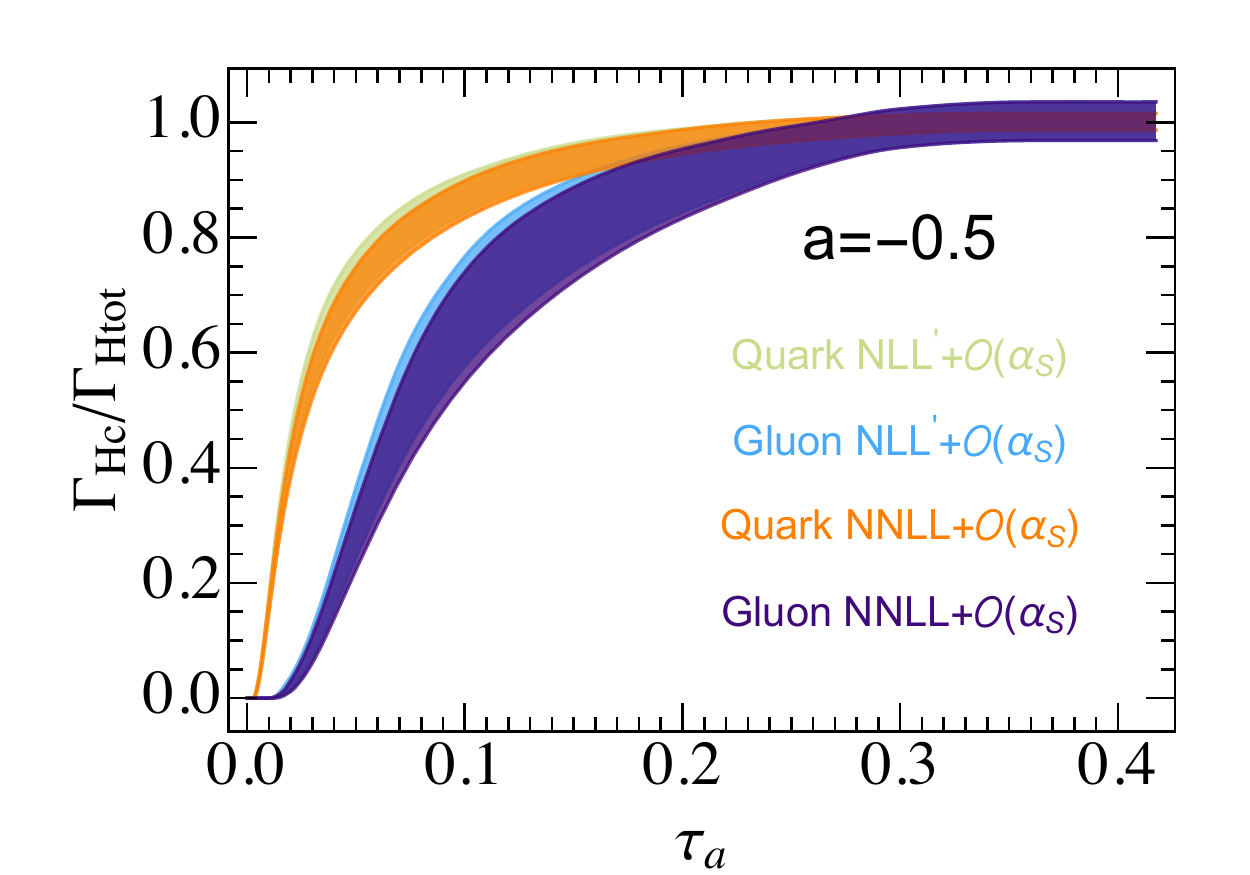}
\includegraphics[width=0.32\textwidth]{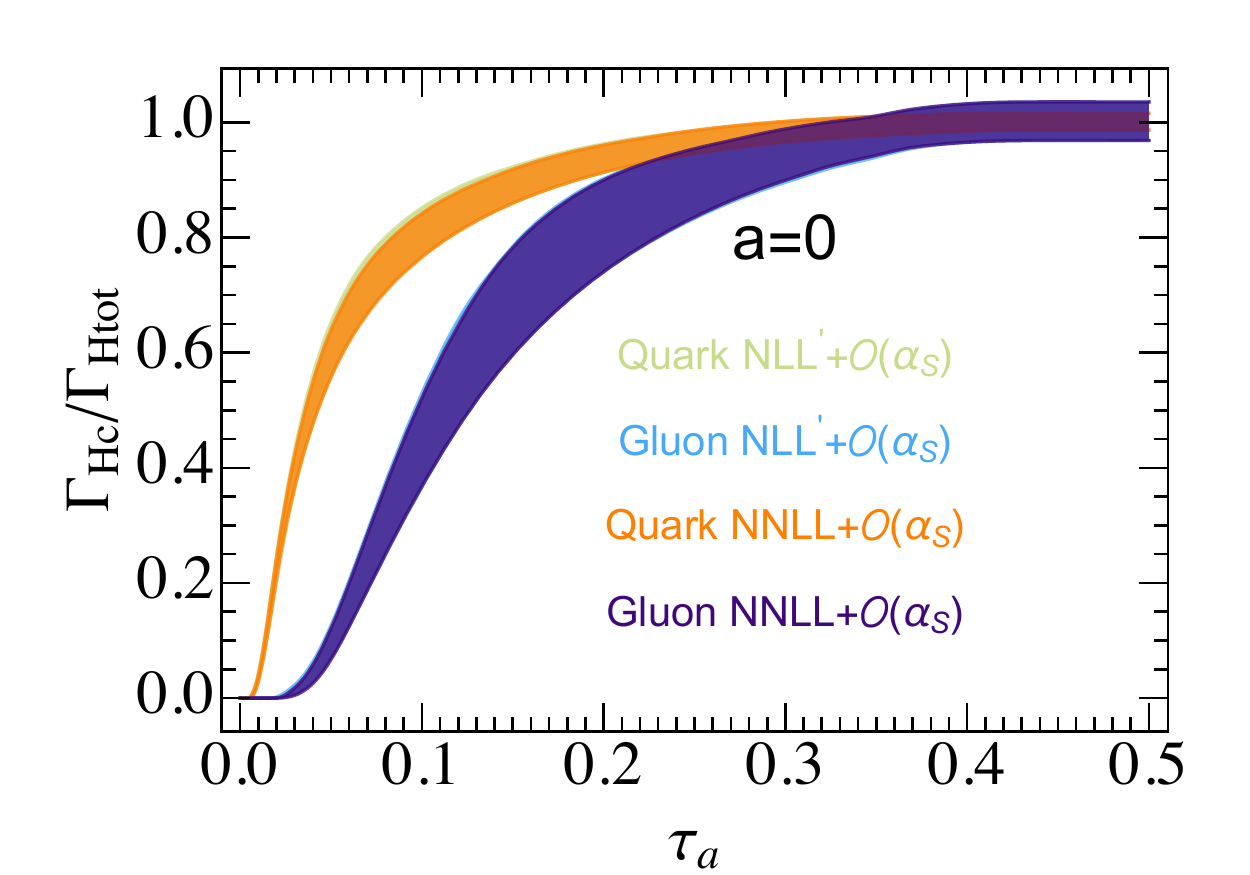}
\includegraphics[width=0.32\textwidth]{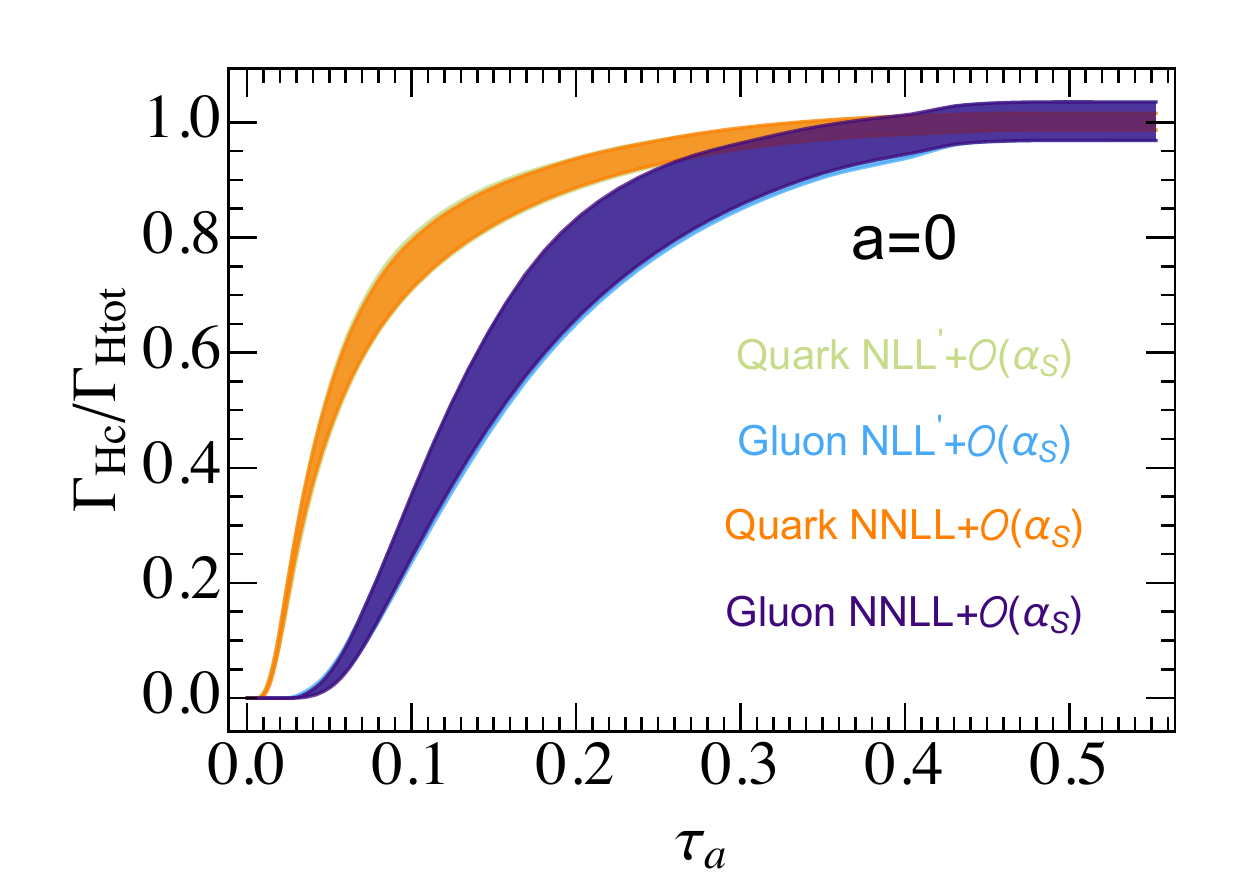}
\includegraphics[width=0.32\textwidth]{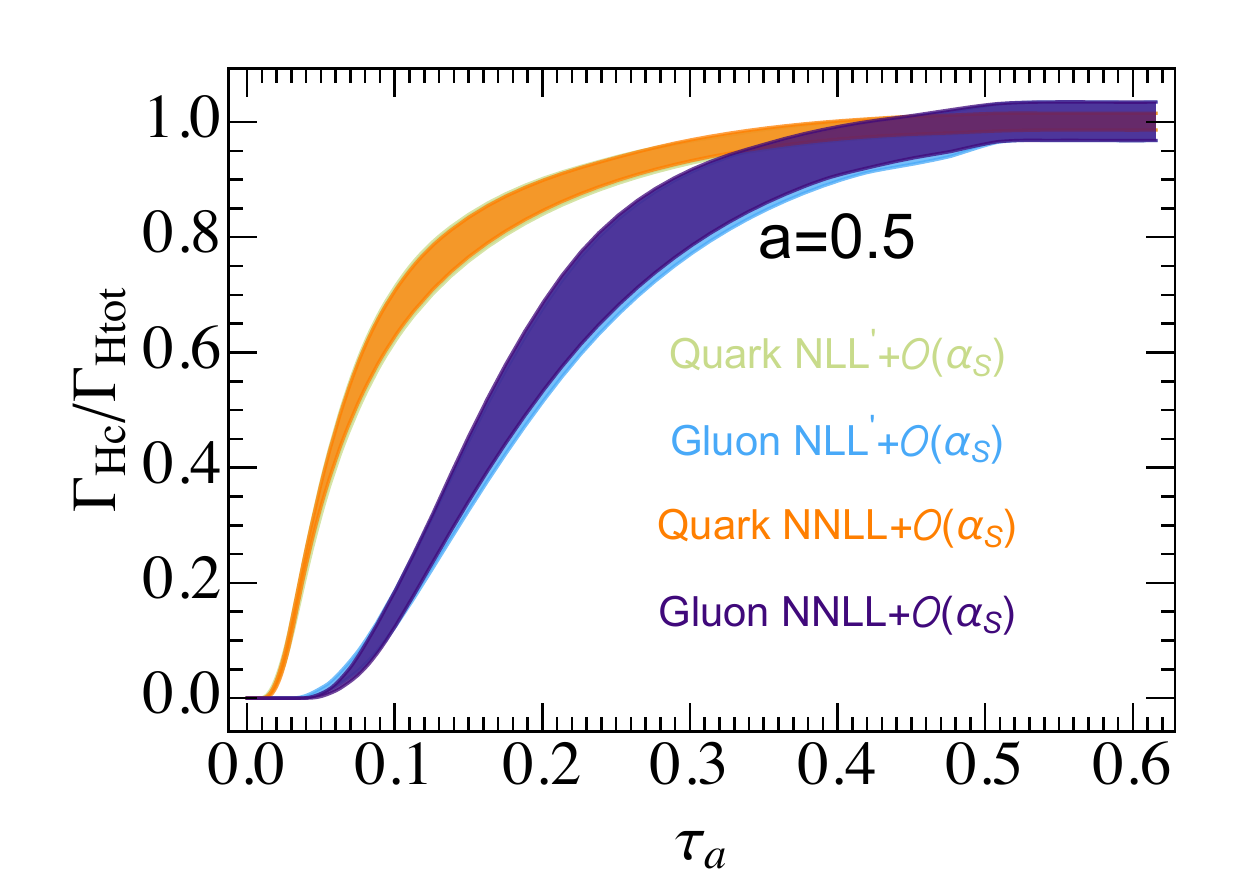}
\vspace{-1em}
\caption{Integrated angularity distributions for values of  $a=(-1.0,-0.5,0.0,0.25,0.5)$ at $\rm{NLL}^\prime+\mathcal{O}(\alpha_s)$ (green for quark and blue for gluon) and $\rm{NNLL}+\mathcal{O}(\alpha_s)$ (orange for quark and purple for gluon), including shape function effects. The theoretical uncertainties have been estimated with the scan method.}
\label{fig:sigc}
\end{figure}

Below, we present the numerical results of angularity distributions from Higgs boson decaying to quarks and gluons at ${\rm NLL}^\prime+\mathcal{O}(\alpha_s)$ and ${\rm NNLL}+\mathcal{O}(\alpha_s)$ accuracy. In order to obtain a comprehensive theory uncertainty, we should consider all the scales and parameters from the profile function
in our calculation.  For a conservative estimation, we use the scan method to calculate the uncertainty band~\cite{Abbate:2010xh}, and take 64 random selections of profile and scale parameters within the ranges shown in Table~\ref{tbl:parameter}. Note that the soft scale parameter $e_S$ has been fixed to zero and the variation of soft scale $\mu_S$ is from other parameters, e.g. $n_0,n_1,\mu_0,r$.
\begin{table}[b!]
\begin{center}
\begin{tabular}{c|c|c|c|c}
\hline
$e_H$ & $e_J$& $e_S$ & $n_0(q)$ &  $n_0(g)$  \\
\hline
 $0.5\leftrightarrow 2$ & $-0.5\leftrightarrow 0.5$ &0 & $1\leftrightarrow 2 ~{\rm GeV}$  & $2.8\leftrightarrow 3.5 ~{\rm GeV}$\\
 \hline
 $n_1(q)$ & $n_1(g)$ & $n_2$ & $n_3$ & $\mu_0(q)$\\
 \hline
$8.5\leftrightarrow 11.5 ~{\rm GeV}$&$25\leftrightarrow 28 ~{\rm GeV}$& $0.9\leftrightarrow 1.1$ & $0.8\leftrightarrow 0.9$ & $1.0\leftrightarrow 1.2 ~{\rm GeV}$ \\
\hline
$\mu_0(g)$& $R_0(q)$ & $R_0(g)$&$r$ &\\
 \hline
$2.2\leftrightarrow 3.0 ~{\rm GeV}$& $\mu_0(q)-0.4~{\rm GeV}$ &$\mu_0(g)-1.8~{\rm GeV}$  &$0.75 \leftrightarrow 1.33$  & \\
\hline
\end{tabular}
\end{center}
\vspace{-1em}
\caption{The parameter ranges for our theoretical uncertainty estimation~\cite{Bell:2018gce}. Note that we use the  different parameters $n_{0,1}$, $\mu_0$ and $R_0$ for quark and gluon final states.}
\label{tbl:parameter}
\end{table}
In Fig.~\ref{fig:sigc}, we show the integrated distribution $\Gamma_{Hc}^i(\tau_a)$ for 
five values of angularity parameter $a=(-1.0,-0.5,0.0,0.25,0.5)$ to ${\rm NLL^\prime}+\mathcal{O}(\alpha_s)$ and ${\rm NNLL}+\mathcal{O}(\alpha_s)$ accuracy.  
\begin{figure}
	\centering
	\includegraphics[width=0.32\textwidth]{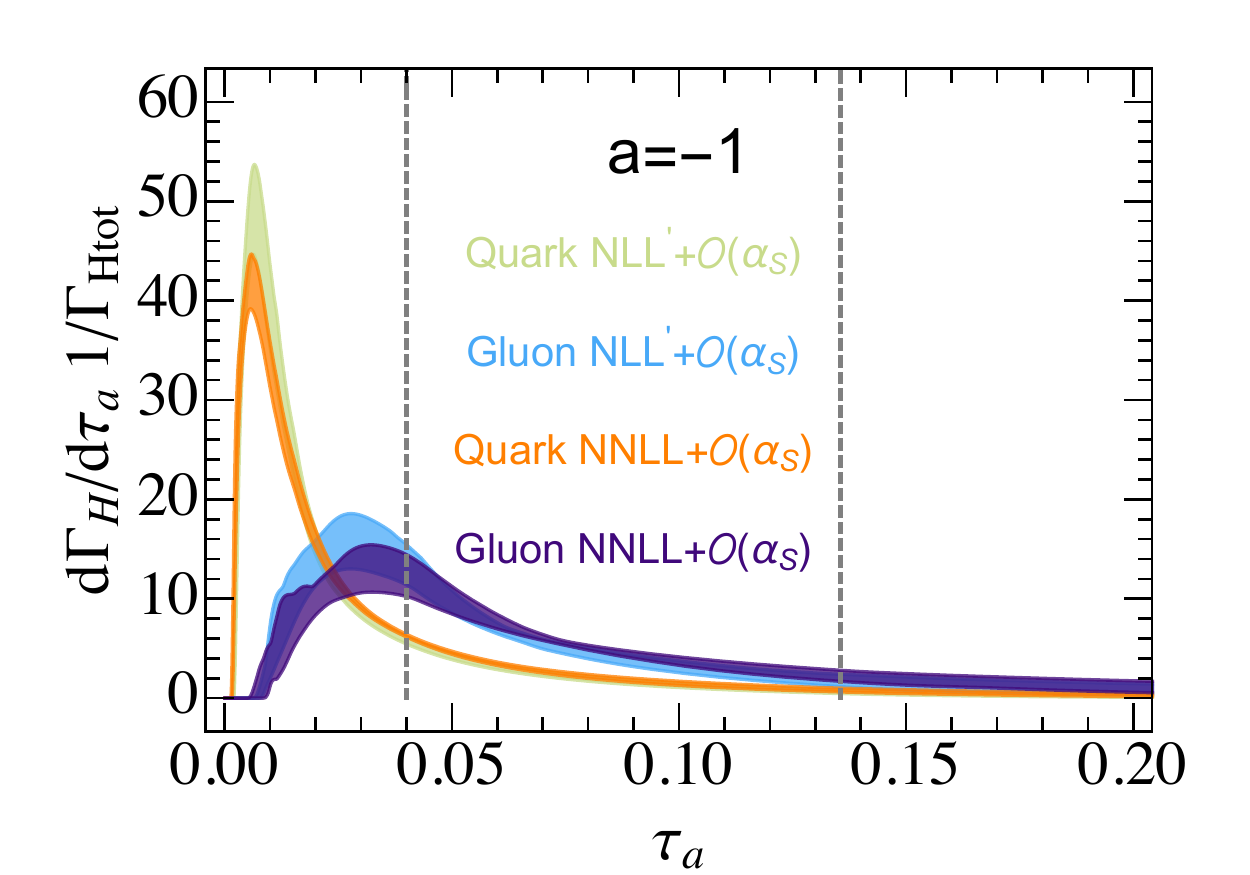}
	\includegraphics[width=0.32\textwidth]{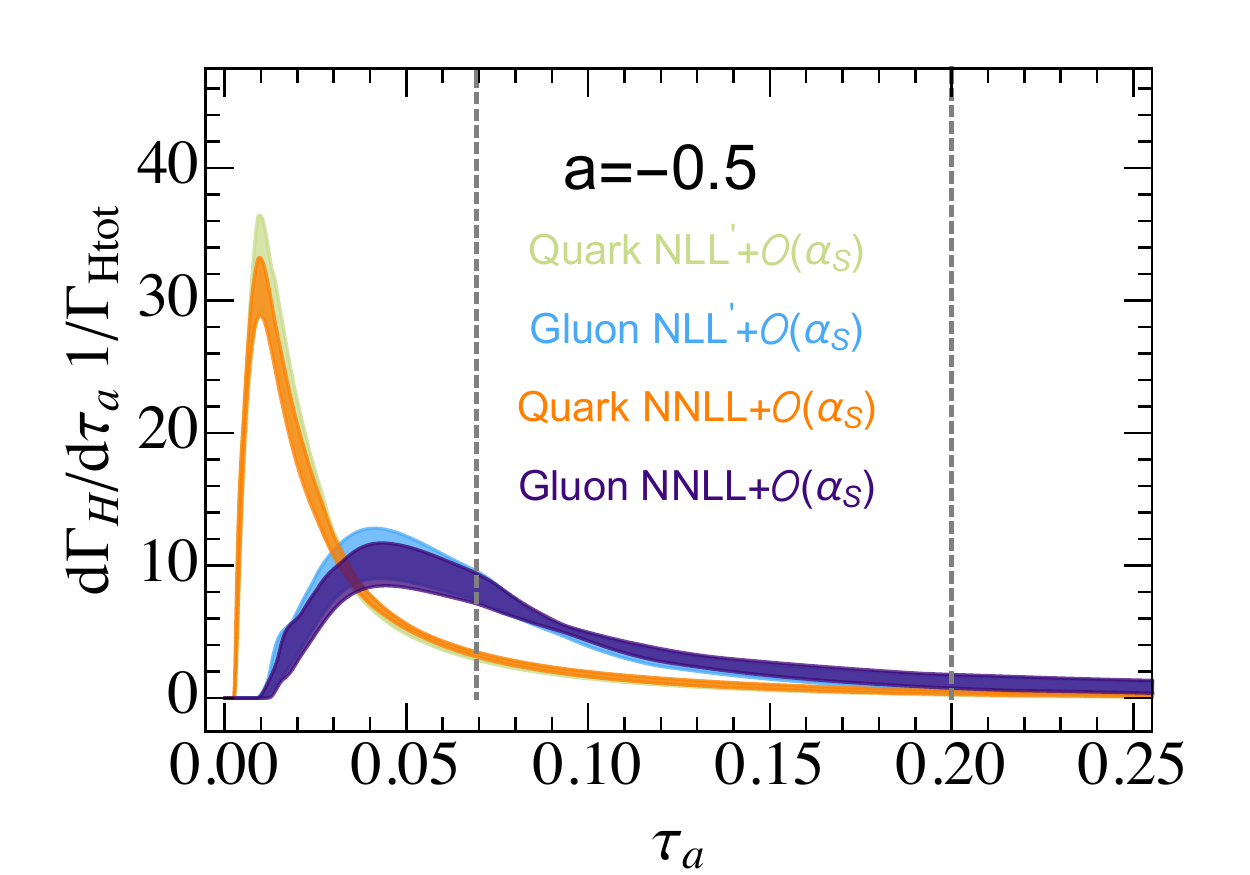}
	\includegraphics[width=0.32\textwidth]{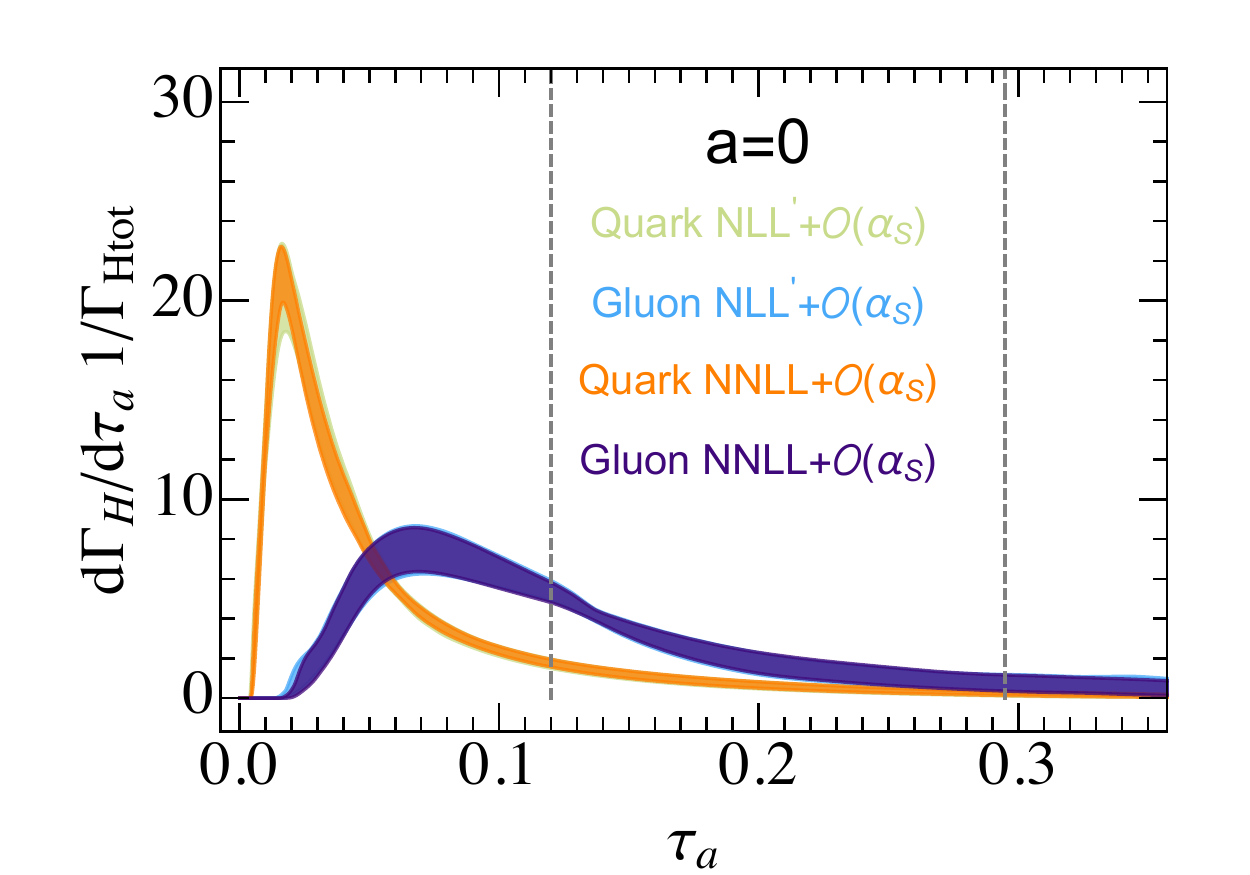}
	\includegraphics[width=0.32\textwidth]{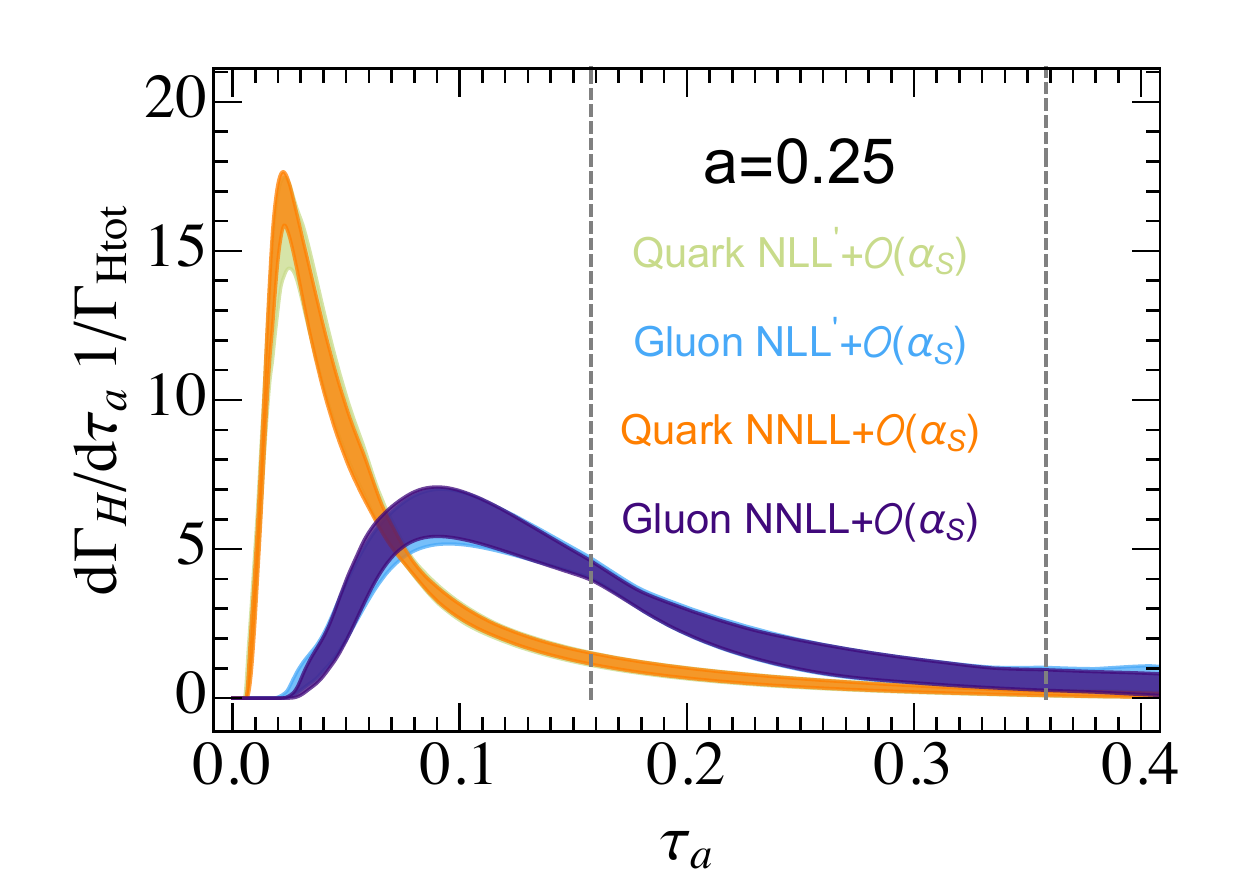}
	\includegraphics[width=0.32\textwidth]{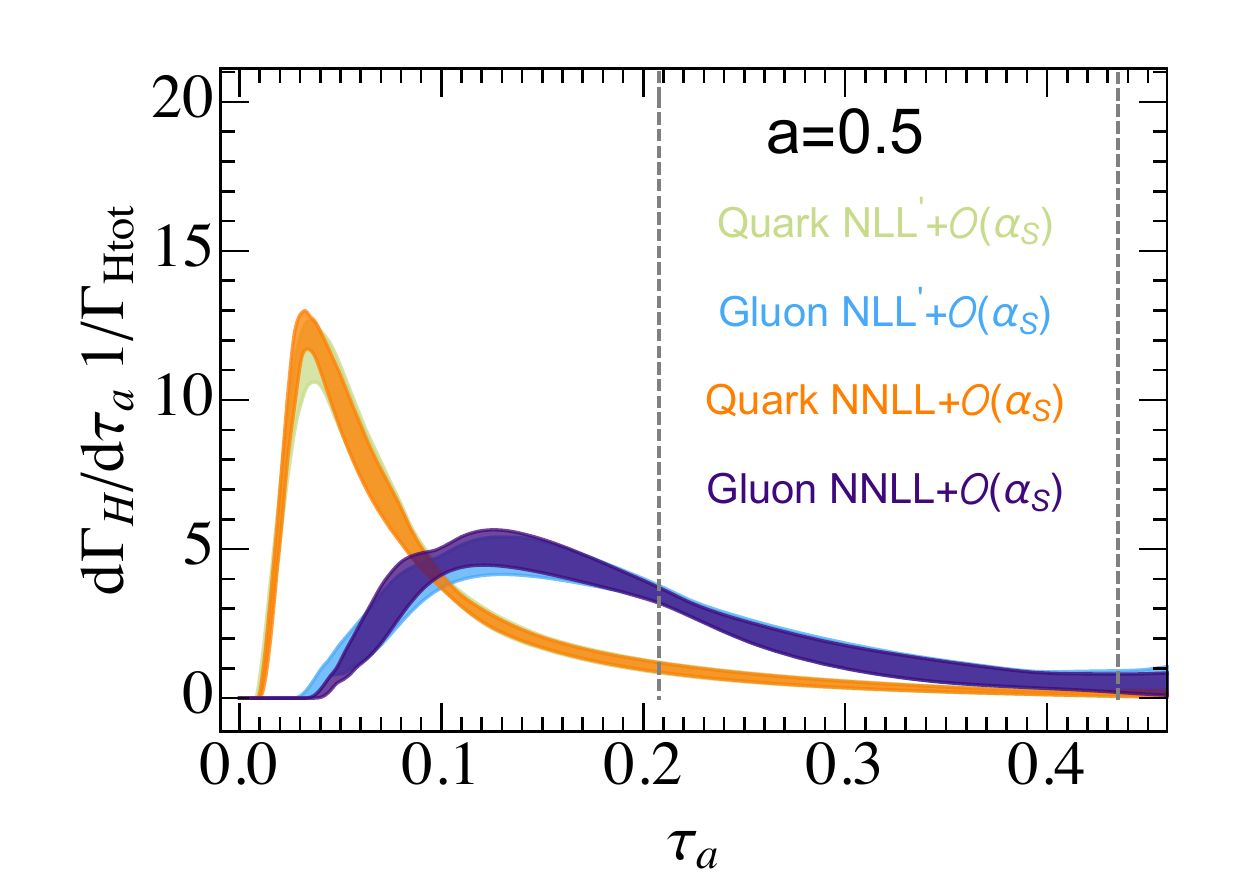}
 \vspace{-1em}
	\caption{Differential angularity distributions for values of  $a=(-1.0,-0.5,0.0,0.25,0.5)$ at $\rm {NLL}^\prime+\mathcal{O}(\alpha_s)$ (green for quark and blue for gluon) and $\rm {NNLL}+\mathcal{O}(\alpha_s)$ (orange for quark and purple for gluon) for Higgs decay, including shape function effects. The theoretical uncertainties have been estimated with the scan method. The perpendicular dashed lines denote the analysis regions [$\tau_{\rm min}(a),\tau_{\rm max}(a)$] for probing light quark Yukawa couplings.}
	\label{fig:sig}
\end{figure}
Note that the total partial decay width $\Gamma_{\rm Htot}^i$ is defined with the renormalization scale $\mu=m_H$.  From ${\rm NLL^\prime}+\mathcal{O}(\alpha_s)$ to ${\rm NNLL}+\mathcal{O}(\alpha_s)$, the scale uncertainties can reduce a little and the correction is sizable for small $a$, e.g. $a=-1$.  It is also clear  that the  quark and gluon decay modes from Higgs boson show different shapes in the small $\tau_a$ region since the different color structure of quark and gluon.
Therefore, the precise study of the angularity event shapes offers the possibility to distinguish the quark and gluon jets from Higgs boson decay and further can be used to constrain the light quark Yukawa couplings.

\begin{figure}
	\centering
	\includegraphics[width=0.45\textwidth]{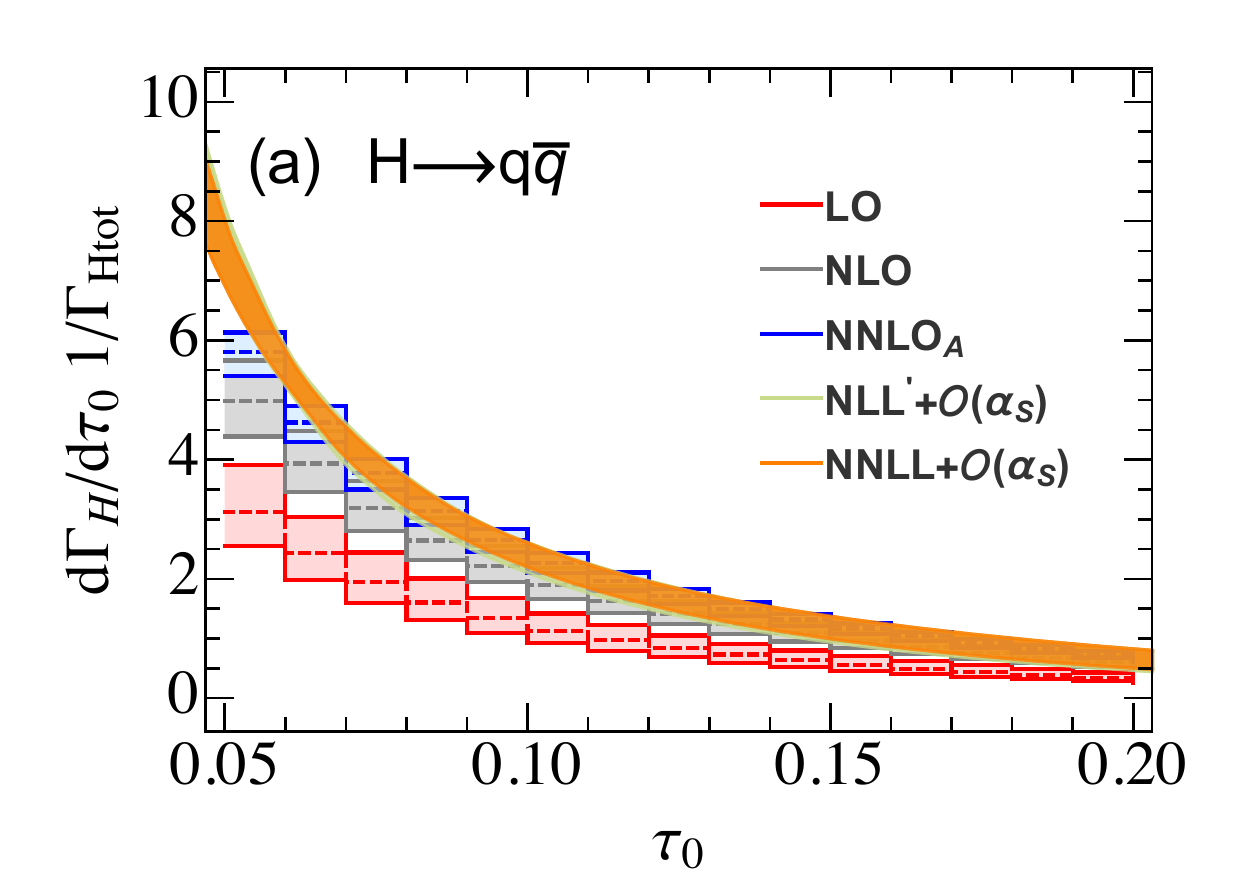}
	\includegraphics[width=0.45\textwidth]{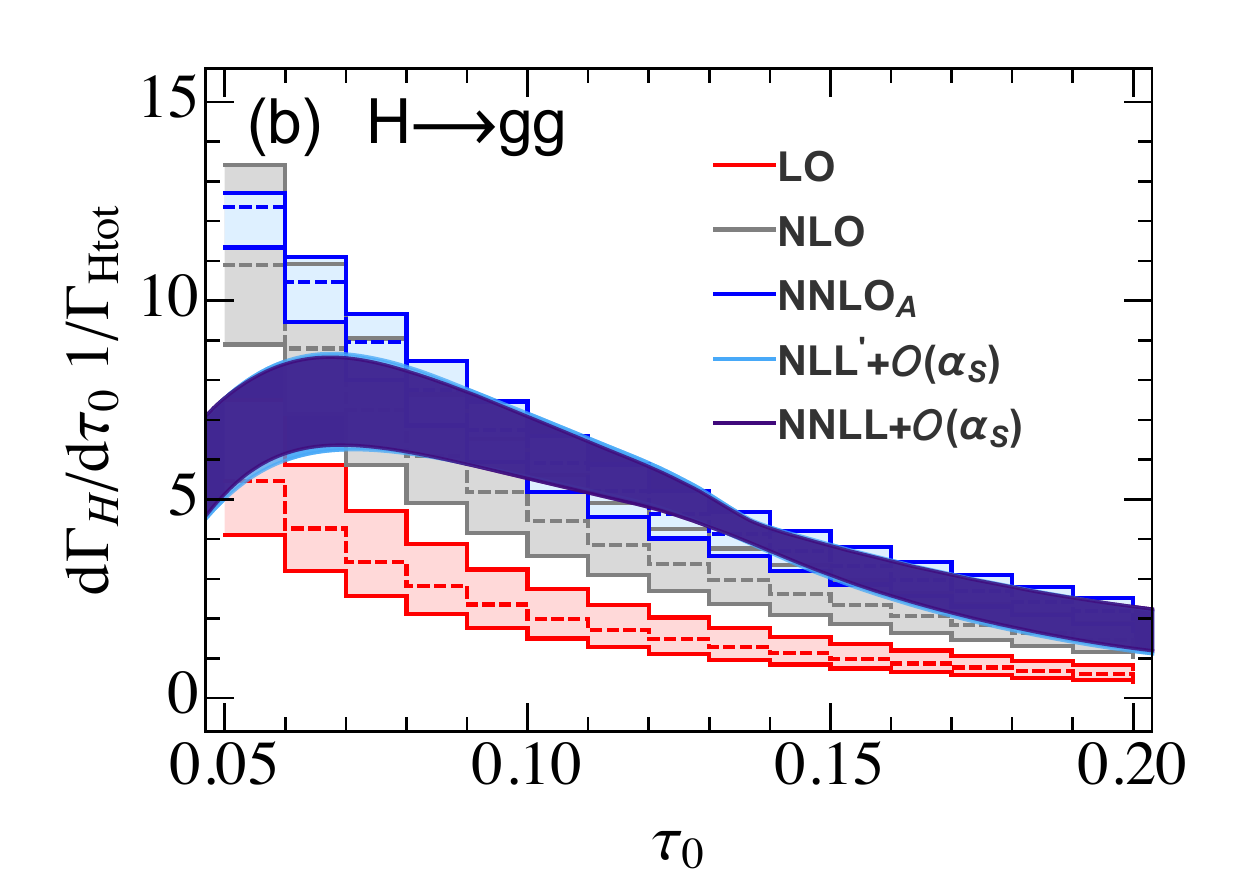}
 \vspace{-1em}
	\caption{The normalized thrust distributions for Higgs decaying to quark and gluons at $\rm NLL^\prime+\mathcal{O}(\alpha_s)$ and $\rm NNLL+\mathcal{O}(\alpha_s)$ accuracy (green and orange for $H\to q\bar{q}$ and light blue and purple for $H\to gg$), including a nonperturbative shape function, compared to LO (red), NLO (gray) and approximate NNLO (blue). The fixed order prediction is from~\cite{Gao:2019mlt}. Note that the normalization factor $\Gamma_{\rm Htot}$ is defined in NLO.}
	\label{fig:thrust}
\end{figure}
The differential distribution of angularities can be obtained by taking the derivative of \eq{res}; see Fig.~\ref{fig:sig} for the numerical results.
We note that the gluon distributions peak at much larger values compared to the quark case. It could be understood from the Casimir scaling $C_A/C_F$ in Sudakov factor.  The distributions from  gluon are also much broader compared to the quark cases due to the stronger QCD radiation. We also find that the larger $a$ would shift the peak to a larger value. This is because varying $a$ will change the proportions of two-jet-like events and three-or-more-jet-like events and  further to change the peak position of the $\tau_a$ distributions. 

\begin{figure}
	\centering
	\includegraphics[width=0.32\textwidth]{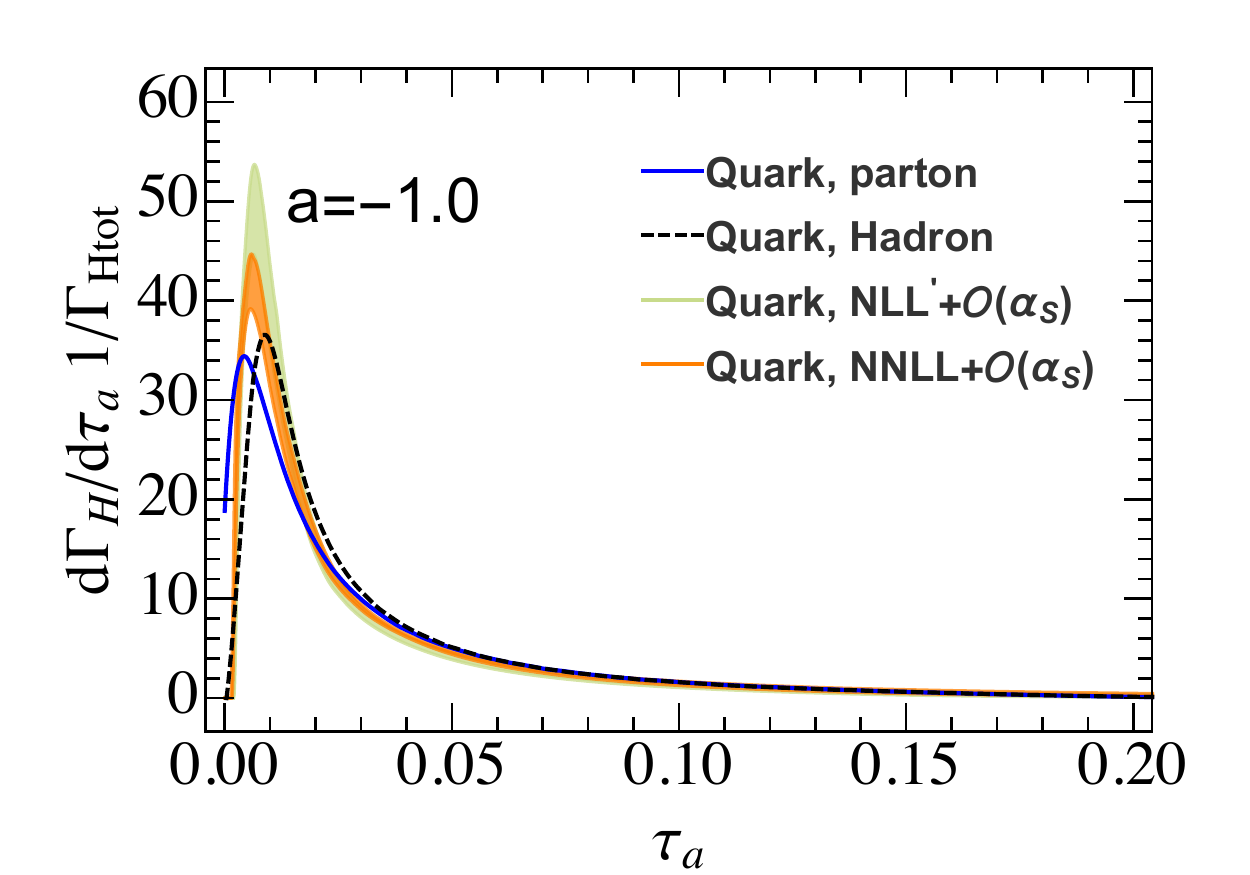}
	\includegraphics[width=0.32\textwidth]{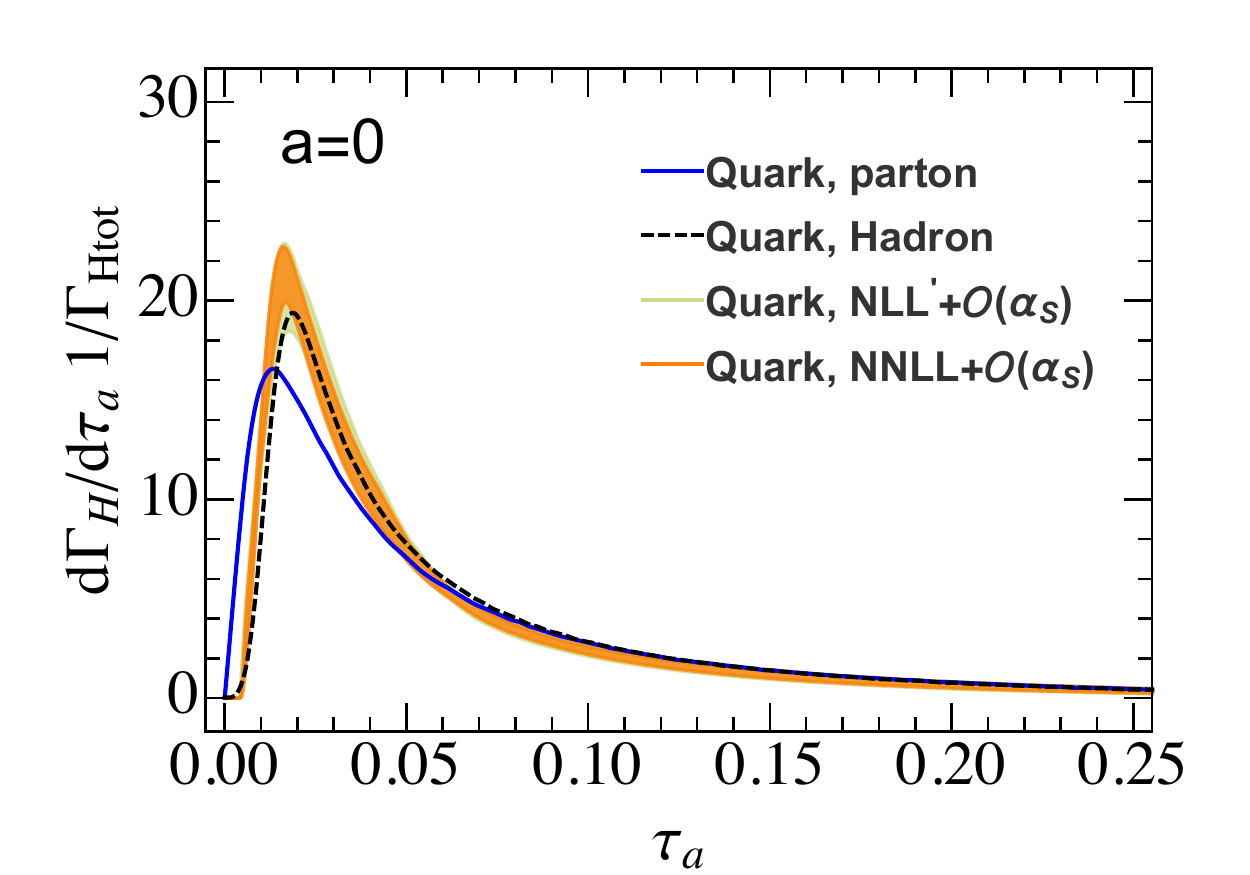}
	\includegraphics[width=0.32\textwidth]{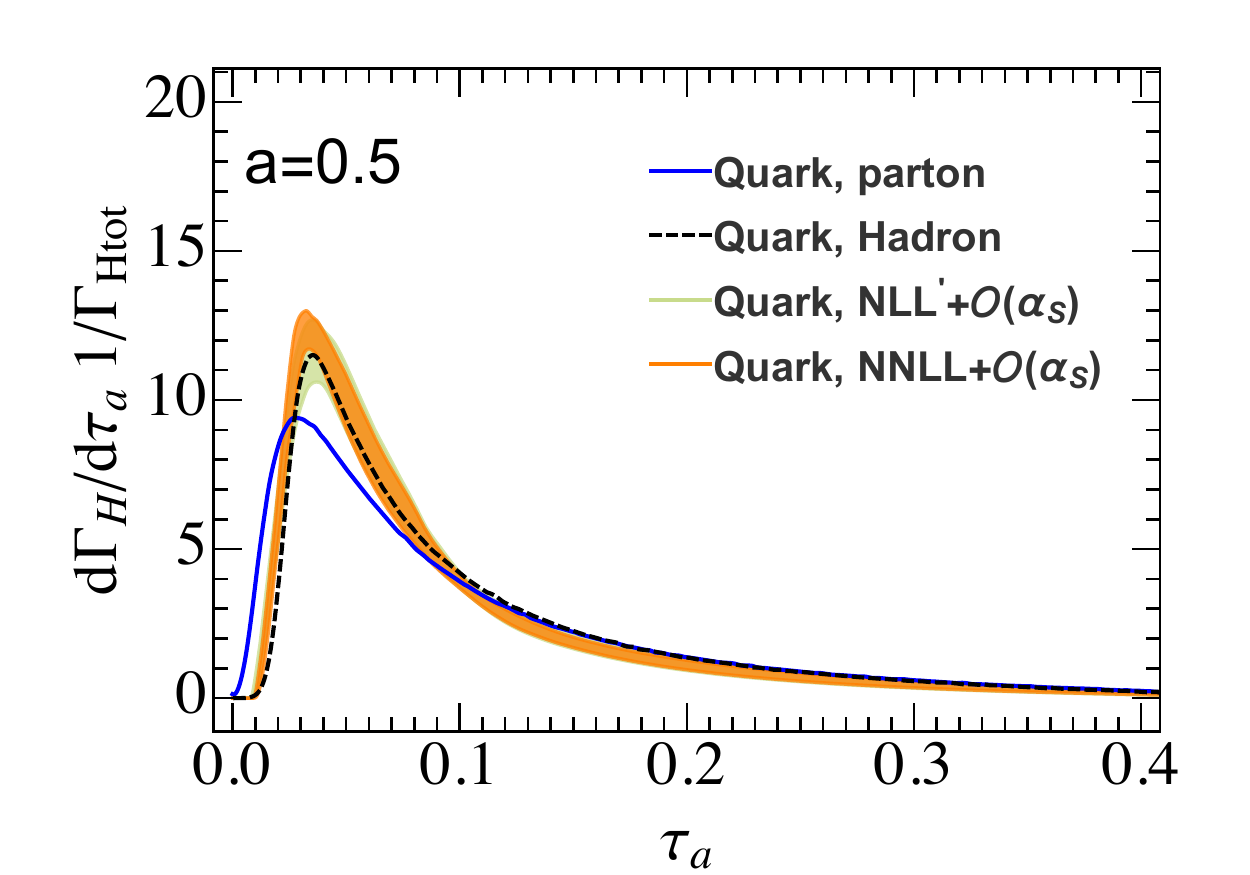}
	\includegraphics[width=0.32\textwidth]{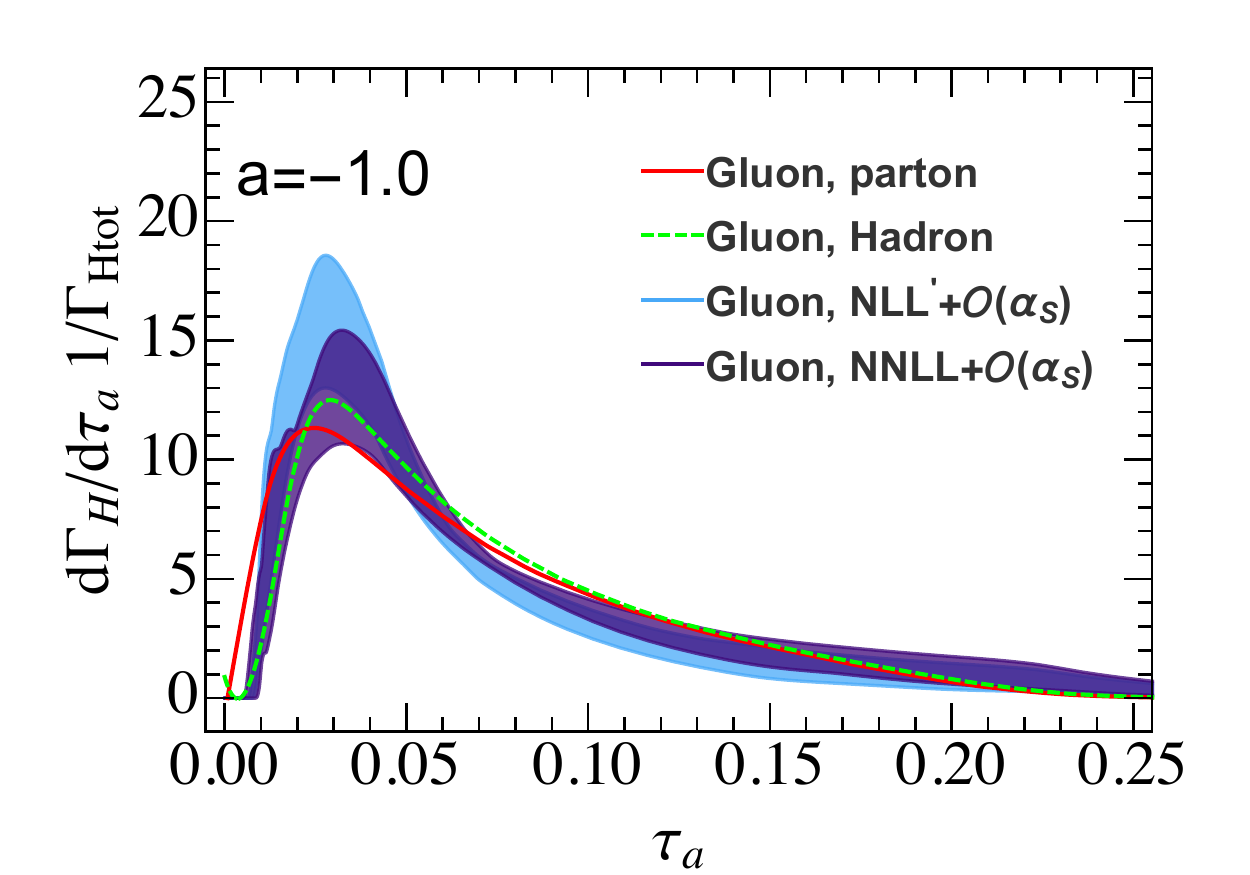}
	\includegraphics[width=0.32\textwidth]{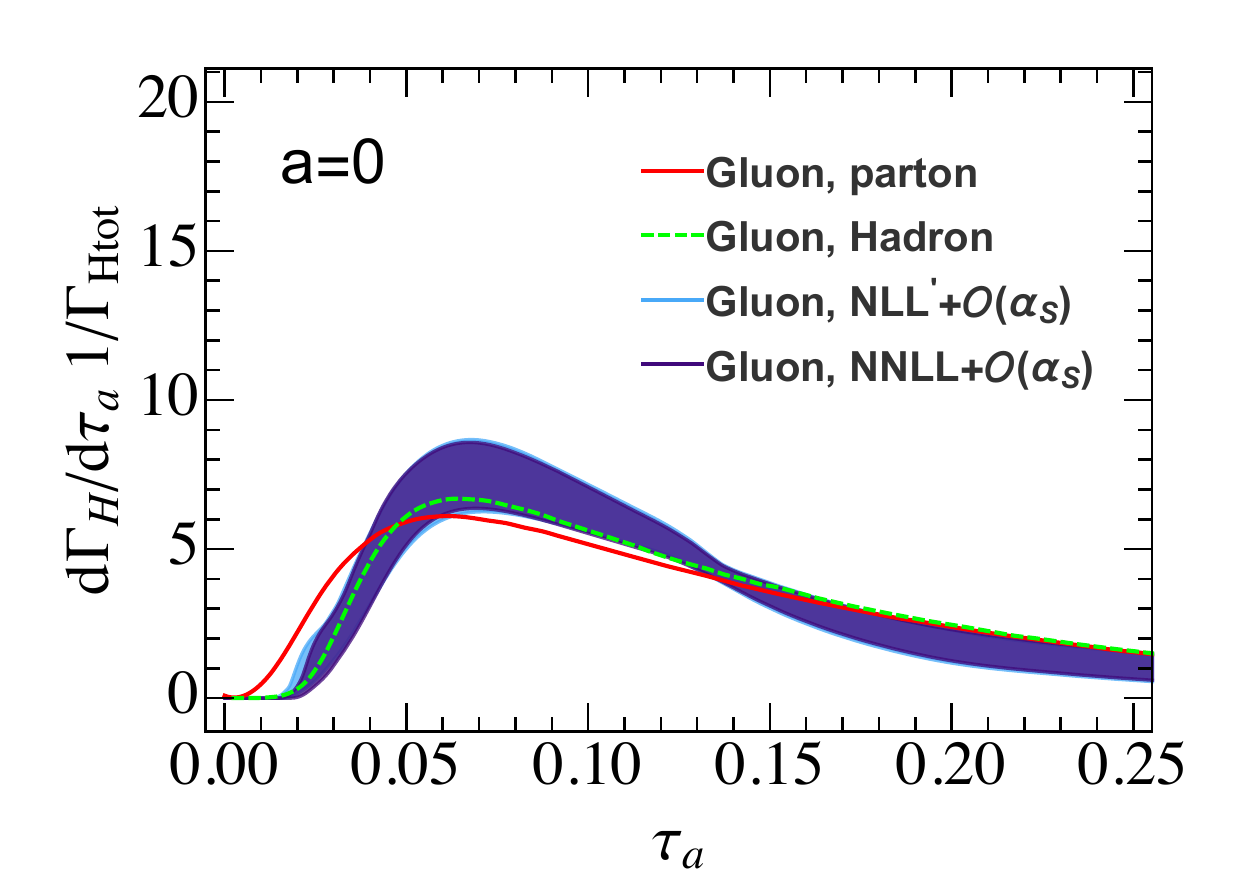}
	\includegraphics[width=0.32\textwidth]{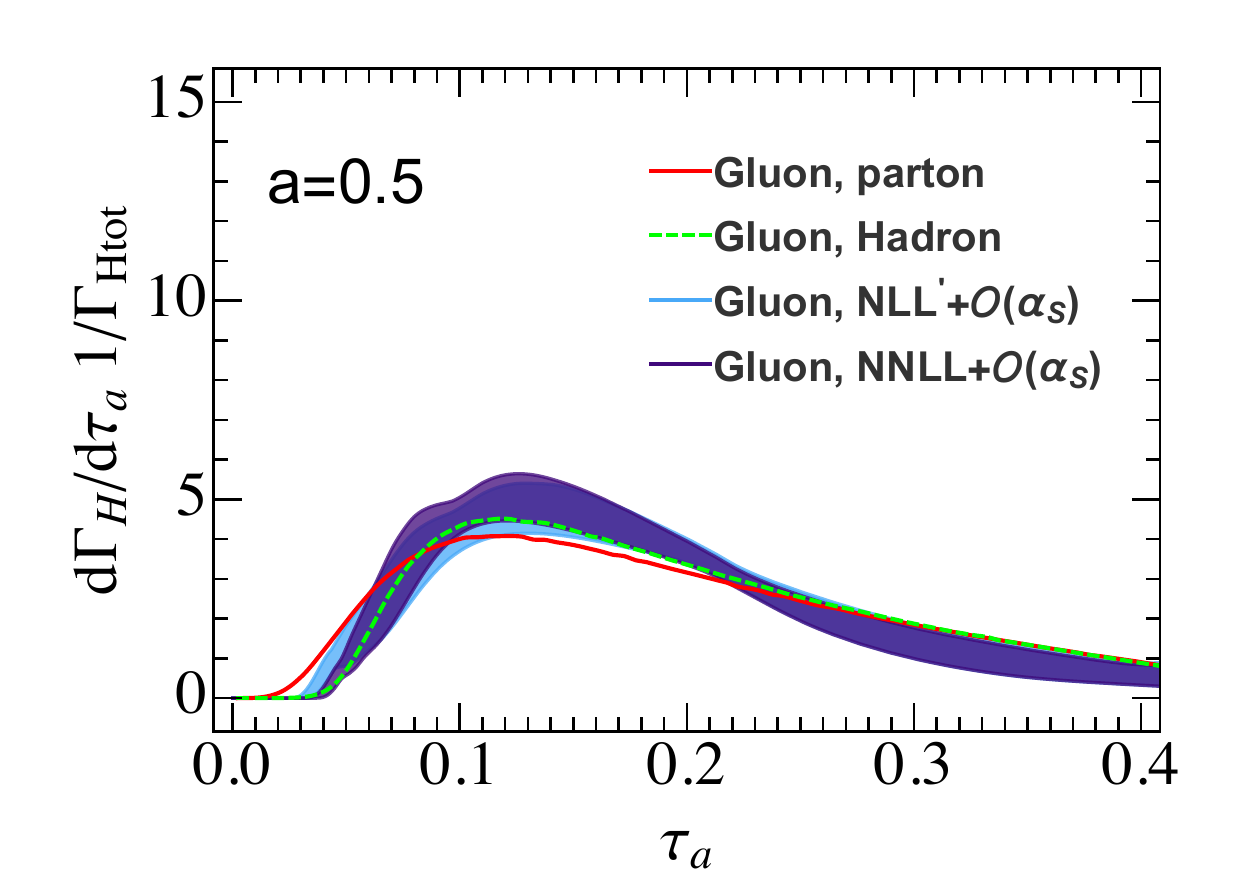}
 \vspace{-1em}
	\caption{The normalized angularity distributions for Higgs decaying to quark and gluons at $\rm NLL^\prime+\mathcal{O}(\alpha_s)$ and $\rm NNLL+\mathcal{O}(\alpha_s)$ accuracy, including shape function effects, compared to PYTHIA at parton level (solid lines) and hadron level (dashed lines) for $a=(-1.0,0,0.5)$.}
	\label{fig:py}
\end{figure}
In Fig.~\ref{fig:py}, we  compare our theoretical calculation of Higgs decaying to quarks and gluons to PYTHIA~\cite{Sjostrand:2007gs} at parton and hadron levels. It shows that PYTHIA  predictions at hadron level agree with our theoretical calculation very well, but the parton level 
results do not in the nonperturbative region at very small $\tau_a$.  
We should note that the design of profile functions is somewhat of an art. The choices of the parameters are based on obtaining properties of the theoretical predictions such as smoothness and convergence of uncertainty bands.

Since the thrust from Higgs boson decay has been calculated up to approximate $\rm{NNLO}$ (i.e. full NLO and singular NNLO) accuracy, it is also useful to compare our resummed results with the fixed order prediction in Ref.~\cite{Gao:2019mlt}; see Fig.~\ref{fig:thrust}. It shows that that our resummed prediction (which includes a shape function) agrees with the $\rm{NNLO}$ prediction very well in Higgs decaying to quark or gluon states at sufficiently large $\tau_0$, but there are deviations in the small $\tau_0$ region, where especially for gluons, where the effects of resummation and of the nonperturbative shape function cause the distribution to peak at some value of $\tau_0$.  However, if we focus on the region $\tau_0\in[0.1,0.20]$, our results agree with the approximate $\rm{NNLO}$ very well for both quark and gluon final states.

Our theoretical predictions show in Figs.~\ref{fig:sigc}--\ref{fig:py} include a soft shape function \eq{softfree}, with parameters chosen simply as described there. No attempt has been made to tune this for Higgs decay, they were guided simply by similar values they take in event shapes in $e^+e^-\to$ hadrons (e.g. \cite{Abbate:2010xh,Hoang:2014wka}), for purposes of producing illustrative numerical predictions. When a serious comparison to data is performed, the data can be used to constrain these soft shape function model parameters and test properties such as the universal scaling properties of the leading nonperturbative shift governed by $\Omega_1^{q,g}$ \cite{Berger:2003pk,Lee:2006nr}.

\section{Probing light quark Yukawa couplings}
\label{sec:yukawa}
Next, we will use  angularity distributions to probe light quark Yukawa couplings at CEPC  
and our results are easy to generalize  to other $e^+e^-$ colliders.
At CEPC, the Higgs boson is dominantly produced through the Higgs and $Z$ boson associated production, i.e. $e^+e^-\to H Z$. The signal we are interested in is 
Higgs decaying into $q\bar{q}$ with $q=u,d,s$
and $Z$ boson decaying into lepton pairs. The major SM backgrounds are $H(\to gg, b\bar{b},c\bar{c})Z$, $Zq\bar{q}$ and  $H(\to VV^*\to 4j)Z$, with $V=W^\pm,Z$~\footnote{In case one is concerned about an additional $Zgg$ final state background, since it can only be generated by $ZH(H\to gg)$ production, it is precisely the one we are trying to suppress by measuring $\tau_a$ on the final state.}.
To suppress the backgrounds from heavy quarks, we could use flavor tagging techniques. Ref.~\cite{Gao:2016jcm} has shown in this case that the heavy flavor backgrounds can be removed mostly if we require two non-$b$ and $c$ jets in the final state. The background $Zq\bar{q}$ can be highly suppressed after we include the kinematical cuts, e.g. recoil mass~\cite{Chen:2016zpw,CEPCStudyGroup:2018ghi}. As shown in Ref.~\cite{Gao:2016jcm}, after the kinematical cuts and requiring two light jets in the final state, we could get the number of background events for the $H\to gg$ down to $N_b^g=3070$ at $\sqrt{s}=250~{\rm GeV}$ with an integrated luminosity of $5~{\rm ab}^{-1}$.  The background of Higgs decaying to heavy flavor quarks ($H\to b\bar{b},c\bar{c}$) is about $N_b^{\rm HF}=0.1N_b^g$, and the number of $Zq\bar{q}$ events is $N_b^{\rm ZZ}=0.2N_b^g$.  The background of $H\to VV^*$ will contribute to the tail region of the angularity distributions and the number of events after including above analysis  is $N_b^{\rm 4q}=0.6~N_b^g$~~\cite{Gao:2016jcm}. 
It is clear that the gluon background is the major obstacle for probing  light quark Yukawa couplings.  Therefore, we propose to use the hadronic angularity distributions of Higgs boson to separate the gluon background from the signal. We should note that the kinematical cuts in this section just modify the normalization but not the shape of the $\tau_a$ distributions. Thus, we could use the normalized angularity distributions to study the Yukawa couplings without input any kinematical cuts for the signal and backgrounds.

In this work, we assume NP effects only change the branching ratio (BR) of
Higgs decaying to light quarks, then we define the ratio,
\beq
R_q=\frac{{\rm BR}(H\to q\bar{q})}{{\rm BR}(H\to gg)}=\frac{9C_A\pi^2v^2}{2m_H^2}\frac{y_q^2(m_H)}{\alpha_s^2(m_H)}\frac{1+\alpha_s(m_H)/(2\pi)\Gamma_{H1}^q}{1+\alpha_s(m_H)/(2\pi)\Gamma_{H1}^g}.
\eeq
The 1-loop corrections $\Gamma_{H1}^{q,g}$ to the Higgs decay widths were given in \eq{GammaH} and Table~\ref{tbl:NLO}. The total number of signal events is $N_s=R_qN_b^g$.

In the SM, $R_q\simeq 0$ due to the smallness of the quark mass. We divide the angularity into $N_{\rm bin}=10$ bins and use the binned likelihood function to estimate the sensitivity for the hypothesis with $R_q$ against the hypothesis with $R_q=0$~\cite{Cowan:2010js},
\beq
L(R_q)=\prod_{i=1}^{N_{\rm bin}}\frac{(s_i(R_q)+b_i)^{n_i}}{n_i!}e^{-s_i(R_q)-b_i},
\eeq
where $b_i$ and $n_i$ are the number of the backgrounds and observed event in the $i$th bin, respectively, and $s_i(R_q)$ is the number of signal events in the $i$th bin for the parameter $R_q$. We set $n_i=s_i(0)+b_i$. 
The number of signal events in each bin is determined by our theoretical calculation at ${\rm NNLL}+\mathcal{O}(\alpha_s)$ accuracy.  The number signal events in the $i$th bin is,
\beq
s_i(R_q)=N_s\int_i d\tau_a\frac{1}{\Gamma_{Htot}^q}\frac{d\Gamma_H^q}{d\tau_a},
\eeq
\begin{figure}
	\centering
	\includegraphics[width=0.32\textwidth]{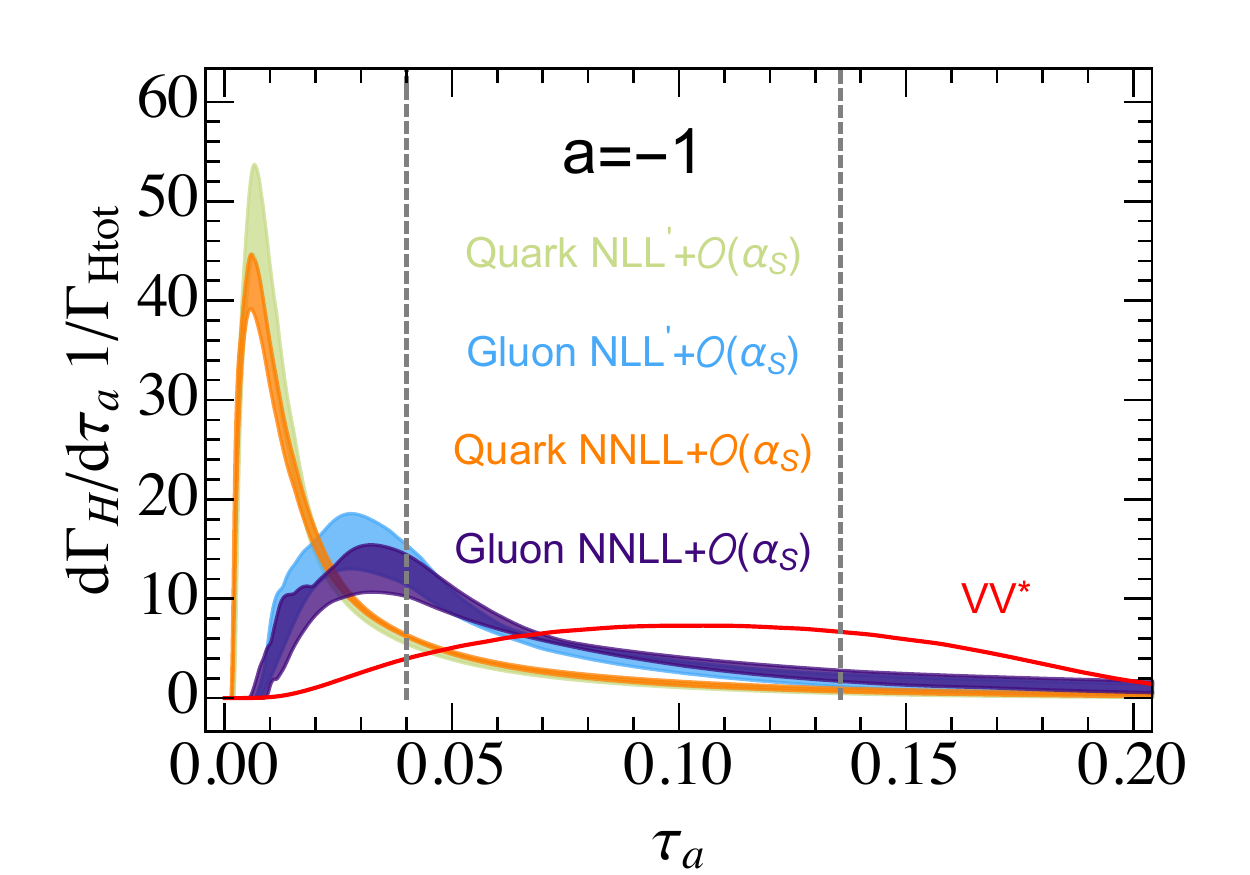}
	\includegraphics[width=0.32\textwidth]{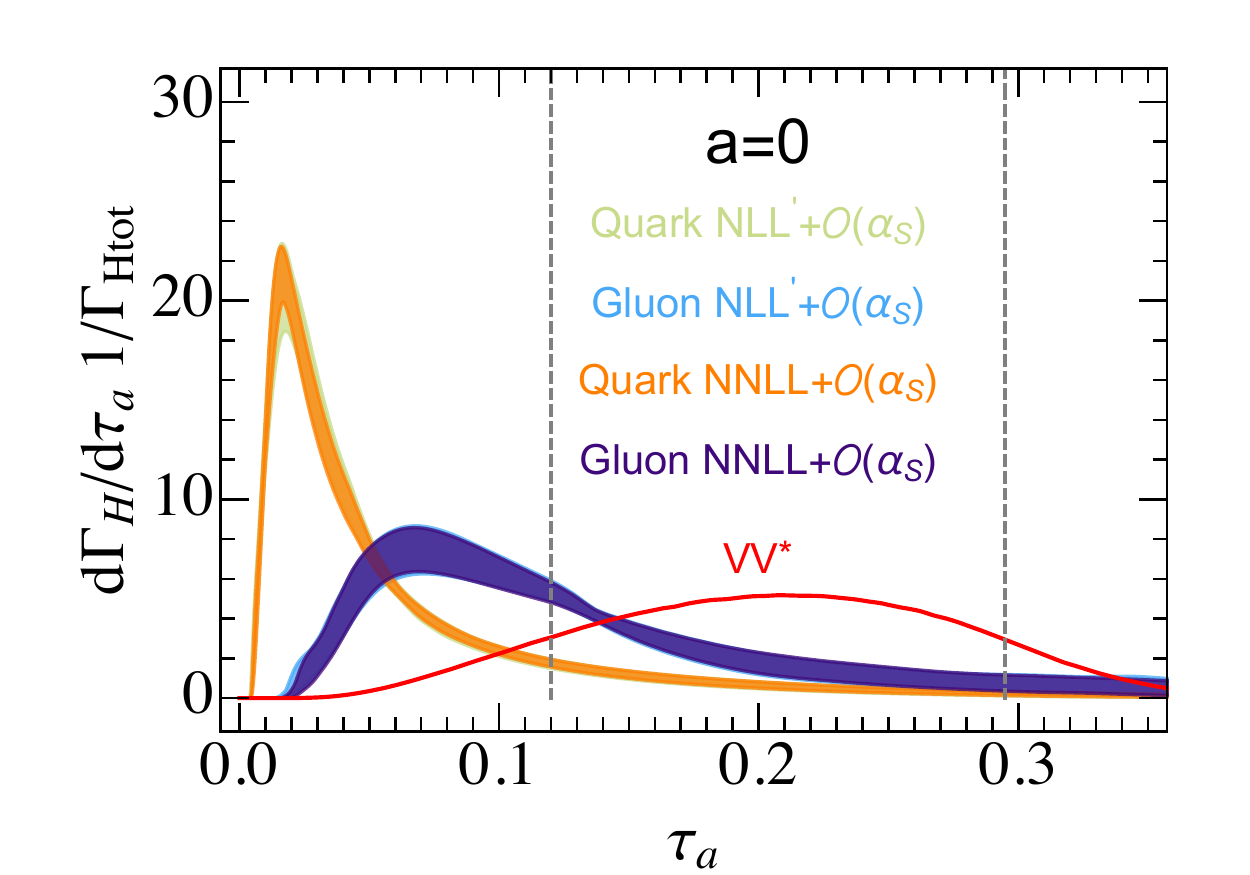}
        \includegraphics[width=0.32\textwidth]{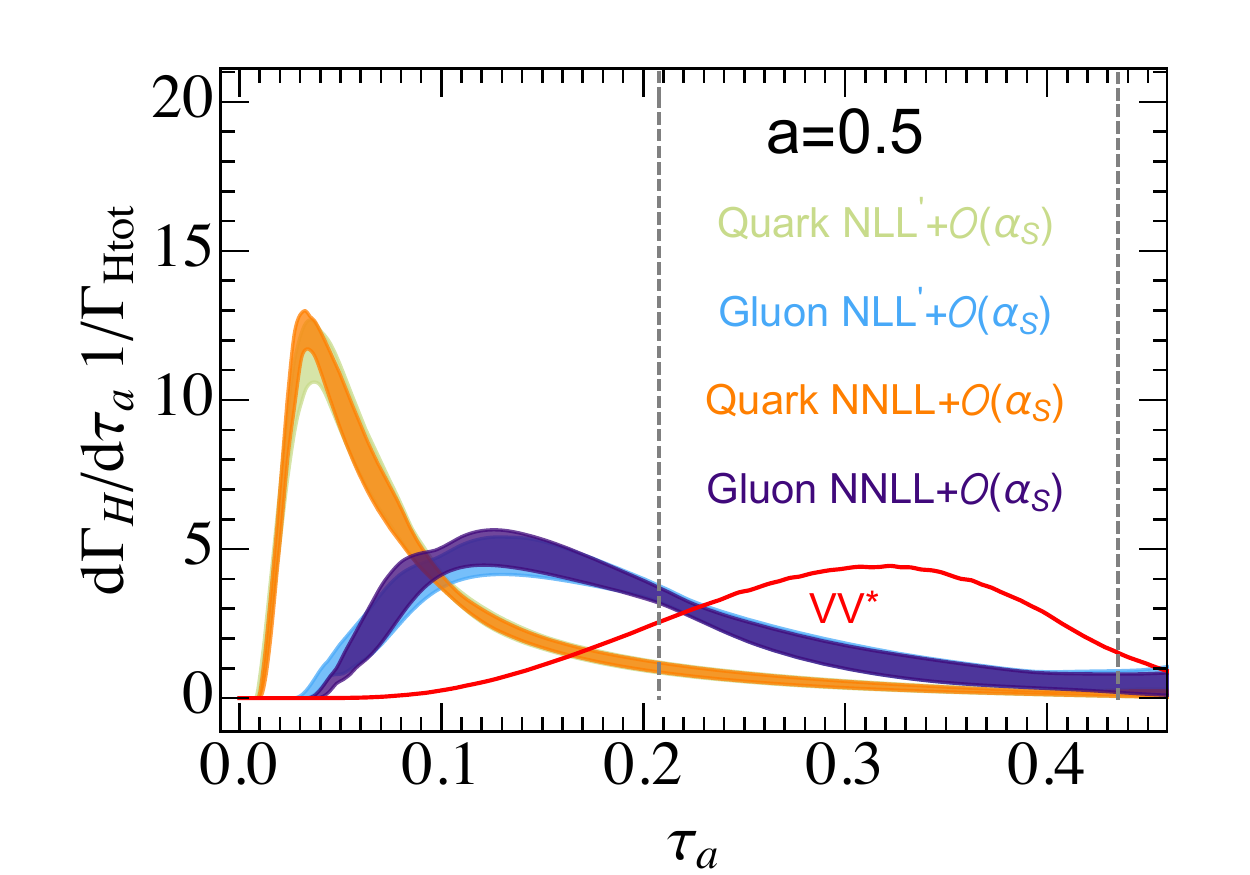}
        \vspace{-1em}
	\caption{The normalized distributions of angularity from PYTHIA at hadron level for $H\to VV^*\to 4j$  with  $a=(-1.0,0.0, 0.5)$ (red lines). The orange and purple lines are from our theoretical prediction for $H\to q\bar{q}$ and $H\to gg$. The perpendicular dashed lines denote the analysis regions [$\tau_{\rm min}(a),\tau_{\rm max}(a)$] for probing light quark Yukawa couplings. }
	\label{fig:hvv}
\end{figure}
The normalized angularity distributions from $H\to b\bar{b}, c\bar{c}$ are almost same with $H\to q\bar{q}$ except for very small $\tau_a$ region due to the mass effect of heavy quark. In order to avoid the possible heavy quark mass effect, the impact of non-perturbative function and the contamination of high tail backgrounds, we can very conservatively require $\tau_{\rm min}(a)<\tau_a<\tau_{\rm max}(a)$ with $\tau_{\rm min}(a)=\frac{15}{125}\times3^a$ and $\tau_{\rm max}(a)=0.295^{1-0.637a}$ (see the vertical dashed lines in Fig.~\ref{fig:sig}). Therefore, we expect the angularity distributions from heavy quarks should be same with the light quarks in this region. The distributions from $Zq\bar{q}$ should have the same shape as those predicted  in Ref.~\cite{Bell:2018gce} at ${\rm NNLL}^\prime+\mathcal{O}(\alpha_s^2)$ accuracy.  The background $H\to VV^*$ can be estimated by event generator Madgraph~\cite{Alwall:2014hca} and PYTHIA~\cite{Sjostrand:2007gs} at the leading order; see Fig.~\ref{fig:hvv}.  It is clear that the four-quark background from gauge boson pair is only sensitive to the tail region and far away from the peak of the signal. The
total backgrounds in $i$th bin is,
\beq
b_i=\sum_j N_b^j\int_i d\tau_a\frac{1}{\Gamma_{Htot}^j}\frac{d\Gamma_H^j}{d\tau_a}.
\eeq
Here $j=g,\rm{HF}, ZZ, 4q$ denotes the different background processes.

\begin{figure}
	\centering
	\includegraphics[width=0.45\textwidth]{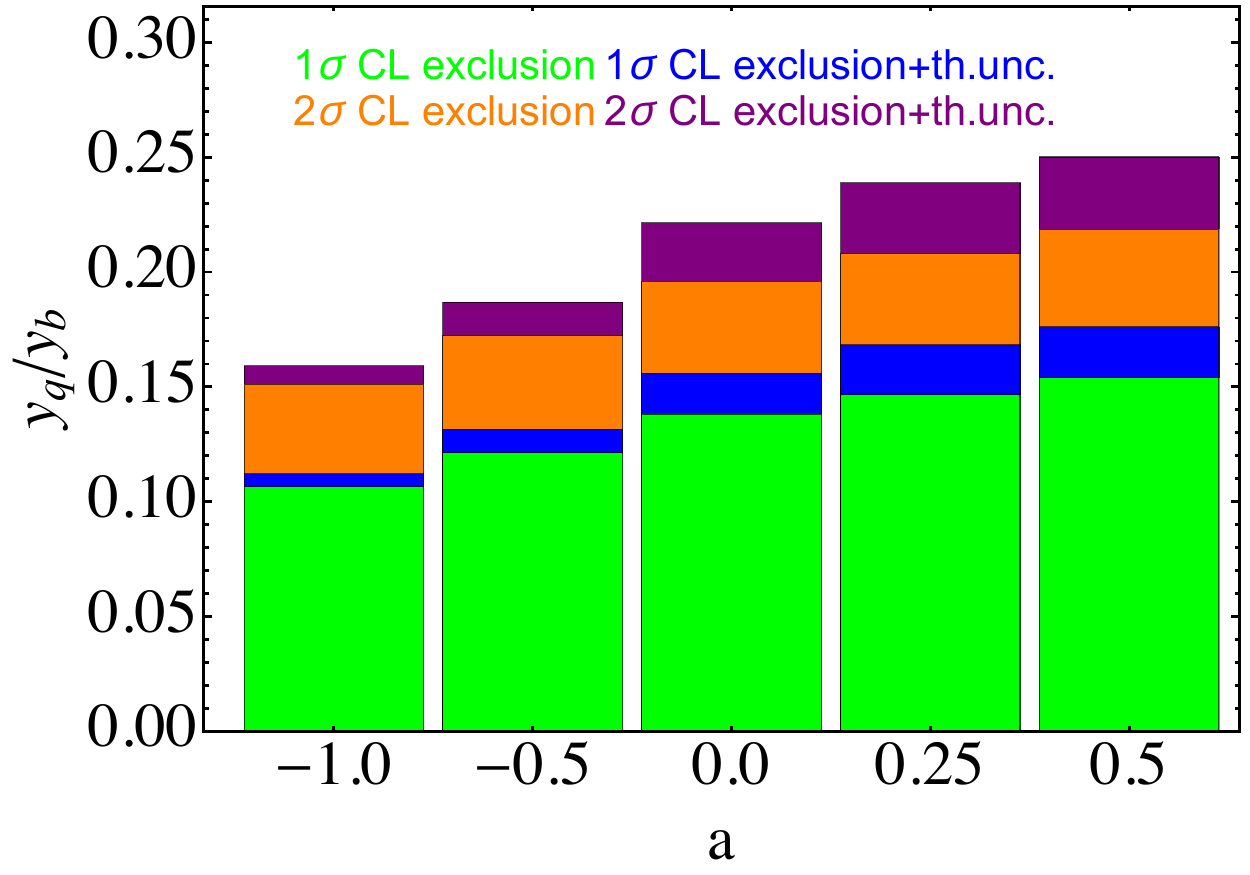}
    \includegraphics[width=0.46\textwidth]{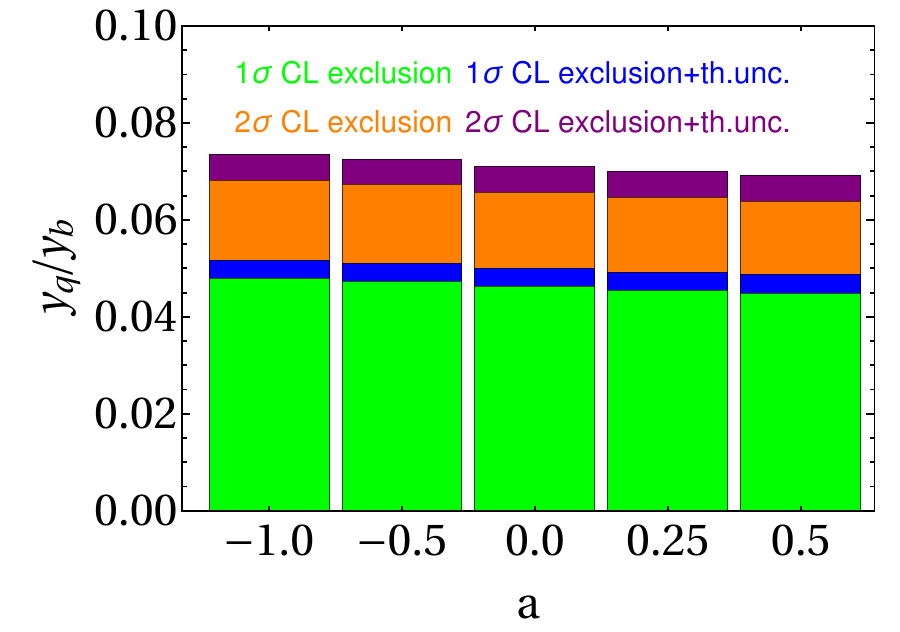}
  \vspace{-1em}
	\caption{Expected $1\sigma$ (green) and $2\sigma$ (orange) exclusion limit on Yukawa coupling $y_q(m_H)$ for values of  $a=(-1.0,-0.5,0.0,0.25,0.5)$ at $\rm {NNLL}+\mathcal{O}(\alpha_s)$.  $y_b$ is the bottom quark Yukawa coupling in the SM. The impact from theoretical uncertainties  are shown with blue ($1\sigma$) and purple ($2\sigma$) colors. We use the conservative analysis window $\tau_{\rm min}(a)<\tau_a<\tau_{\rm max}(a)$ in the left figure, while the more aggressive window $t_0^q(a)<\tau_a<\tau_{\rm max}(a)$ is used in the right figure.}
	\label{fig:Rlimit}
\end{figure}

We define the test ratio of likelihood function,
\beq
q^2=-2\ln\frac{L(R_q)}{L(0)}.
\eeq
The  parameter $q$ yields the exclusion of the hypothesis 1 with $R_q\neq 0$ versus the hypothesis 0 with $R_q=0$ at the $q\sigma$ confidence level. Thus,
\beq
q^2=2\left[\sum_{i=1}^{N_{\rm bin}} n_i\ln\frac{n_i}{n_i(R_q)}+n_i(R_q)-n_i\right],
\eeq
where $n_i(R_q)=s_i(R_q)+b_i$.  For simplicity, we normalize the light quark Yukawa couplings to the SM bottom quark $y_b\equiv y_b(m_H)$. 
Figure~\ref{fig:Rlimit} displays the expected $1\sigma$ (green) and $2\sigma$ (orange) confidence level exclusion limit on Yukawa coupling $y_q\equiv y_q(m_H)$ from the angularity distributions.
It shows that the angularity event shapes, in the conservative $\tau_a$ window we considered above, could give a fairly strong constraint
for the light quark Yukawa couplings, i.e. $y_q\lesssim 0.15 y_b$ (for $a=-1$) to $y_q\lesssim 0.22 y_b$ (for $a=0.5)$, at the $2\sigma$ confidence level.  The theoretical uncertainties will change the upper limit of light quark Yukawa couplings from 5\% to 14\%.  We note that the limit for $y_q$ obtained in this way is considerably larger than the results in Ref.~\cite{Gao:2016jcm}.
It could be understood from our analysis strategy, i.e. in this window, we have $\tau_a>\tau_{\min}(a)$, which is far away from the peak region of the signal, while it is not in Ref.~\cite{Gao:2016jcm}. 
From the PYTHIA prediction (see Fig.~\ref{fig:py}), we know the typical nonperturbative hadronization effects will shift the peak of angularities by about $\Delta\tau_a\sim 0.01$. Furthermore, the nonperturbative corrections to our predictions for light quark angularities remain small (or within the universal shift region) to considerably smaller values of $\tau_a$. If we push our analysis region a bit more aggressively into the peak region, e.g. to a lower limit of $\tau_a>t_0^g(a) \sim 3.5/125\times 3^a$ , then we obtain the upper limit $y_q\lesssim 0.09 y_b$  at $2\sigma$ confidence level for all our choices of $a$. We can try to be even more aggressive than this, and push the lower limit for $\tau_a$ left of the peak of the quark angularity distributions, even though nonperturbative and heavy-quark mass corrections are larger there---this is for illustrative, motivational purposes only. Choosing the analysis region $\tau_a>t_0^q(a)=1/125\times 3^a$, we obtain the stronger limit $y_q\lesssim 0.07 y_b$. We summarize the various limits we obtain on $y_q$ for the different analysis windows in Table~\ref{tbl:yq}. We stress that in this work we have used a simple educated guesses for the quark and gluon shape function parameters, to which this analysis region will be sensitive. If our guesses are close to accurate, though, the above limit is indicative of the power of a single angularity distribution to put a limit on $y_q$. A definitive limit would require further control of the shape function parameters through universality/scaling arguments and/or extraction from an independent dataset. Alternatively, to further improve the measurements and reduce the impact of the non-perturbative effects, we could use the  soft-drop grooming technique~\cite{Larkoski:2014wba,Frye:2016aiz,Lee:2019lge}, and  this could be considered in a future work. Nevertheless, the potential limits indicated in Table~\ref{tbl:yq} show the promise of angularities in separating quark signal from gluon background to obtain strong limits on $y_q$.

\begin{table}
\begin{center}
\begin{tabular}{c|c|c|c|c|c}
\hline 
$a$ & $-1.0$ & $-0.5$  & 0.0 & 0.25 & 0.5  \\ 
\hline 
$[\tau_{\rm min}(a),\tau_{\rm max}(a)]$ & 0.15 &  0.17 & 0.20 & 0.21 & 0.22  \\ 
\hline 
$[t_0^g(a),\tau_{\rm max}(a)]$ & 0.085 & 0.089 & 0.090 & 0.088 & 0.086  \\ 
\hline
$[t_0^q(a),\tau_{\rm max}(a)]$ & 0.068 & 0.067 & 0.066 & 0.065 & 0.064  \\ 
\hline 
\end{tabular} 
\end{center}
\vspace{-1em}
\caption{Expected $2\sigma$ exclusion limit on Yukawa coupling $y_q/y_b$ for values of $a=\{-1.0,-0.5,0.0,0.25,0.5\}$ at $\rm {NNLL}+\mathcal{O}(\alpha_s)$ accuracy with different analysis region, with $t_0^g(a)=3.5/125\times3^a$ and $t_0^q(a)=1/125\times 3^a$.}
\label{tbl:yq}
\end{table}

\begin{figure}
	\centering
	\includegraphics[width=0.7\textwidth]{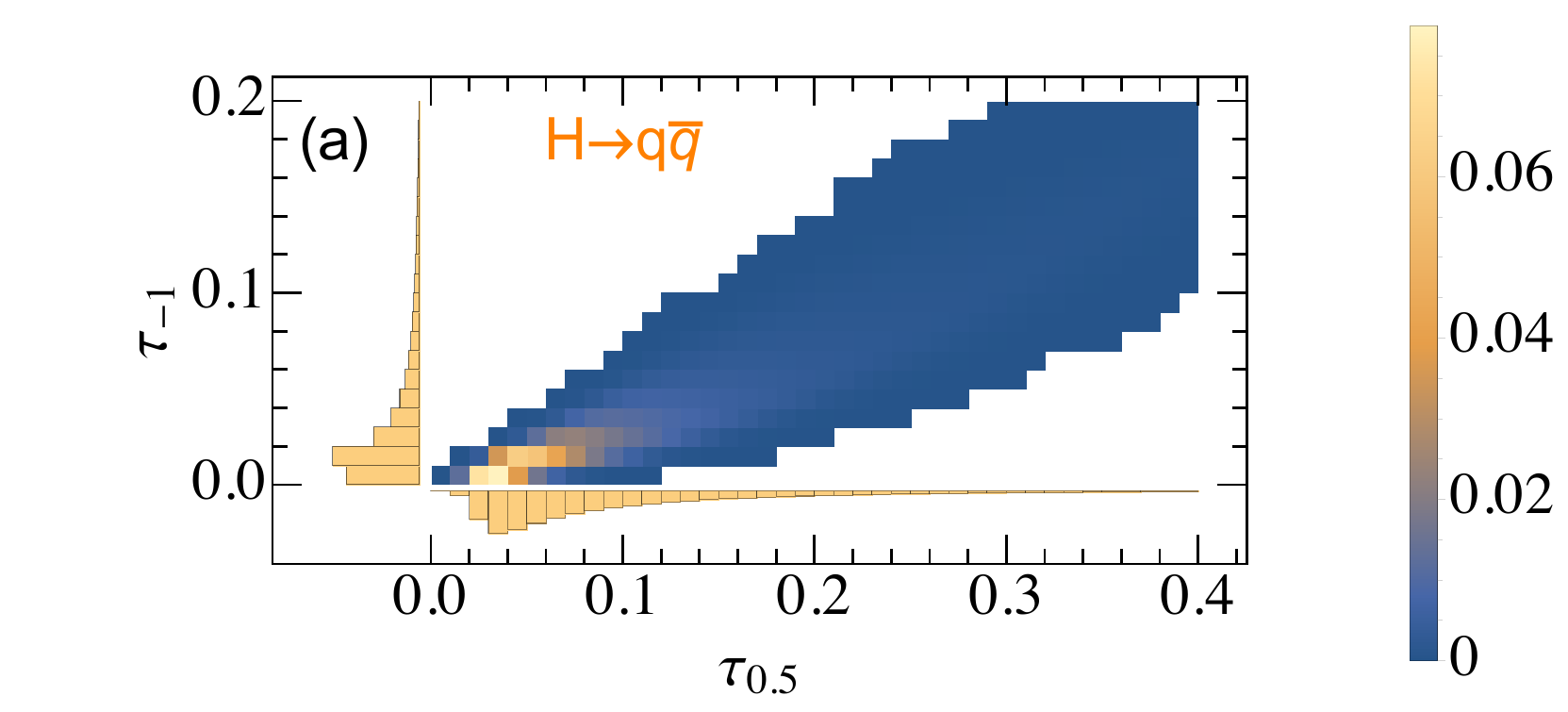}
		\includegraphics[width=0.7\textwidth]{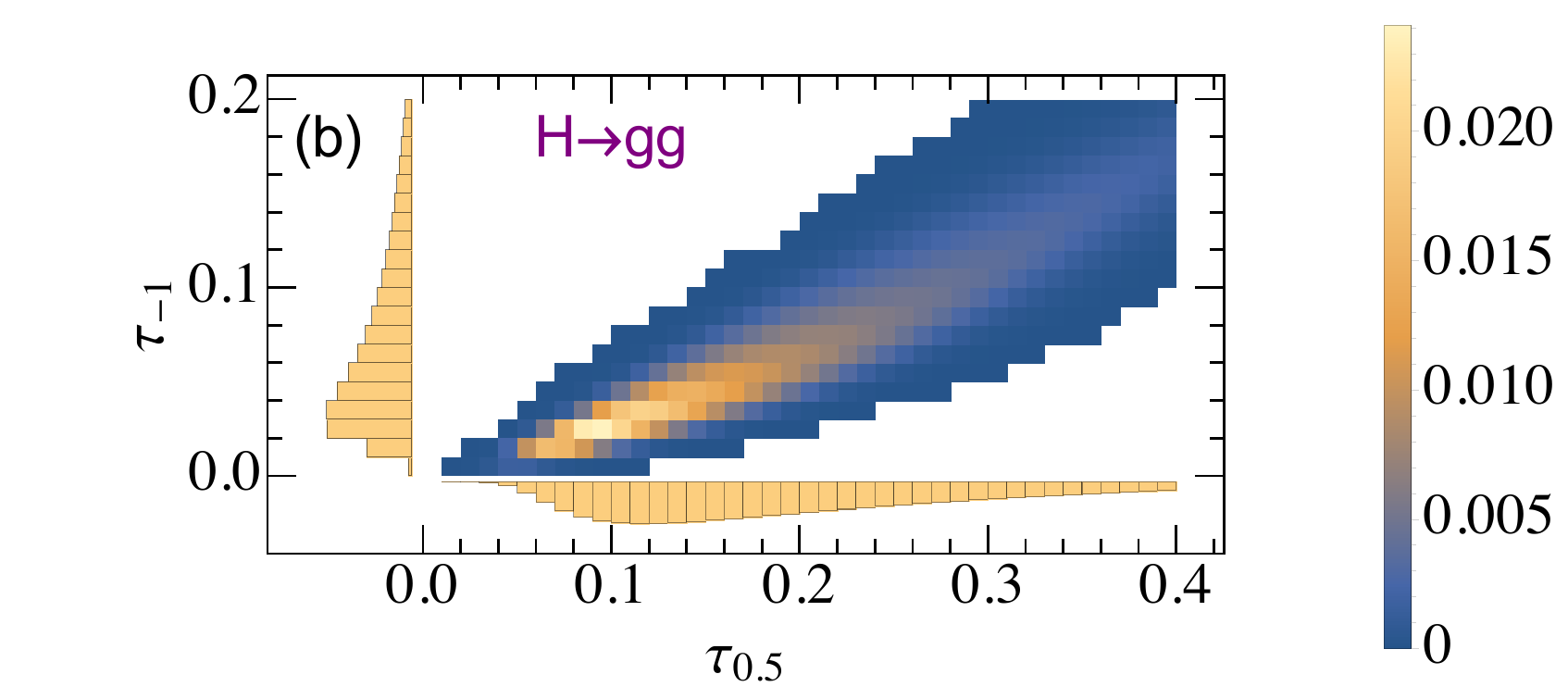}
  \vspace{-1em}
	\caption{The normalized double differential distributions of two jet angularities ($\tau_{-1.0},\tau_{0.5}$) from PYTHIA at hadron level for $H\to q\bar{q}$ and $H\to gg$. On the two side axes we show the corresponding single differential spectra. We analyse regions with the same conservative cuts $\tau_\text{min}(a)<\tau_a<\tau_\text{max}(a)$ for each $\tau_a$ as in the one-dimensional analyses used in Table~\ref{tbl:yq} to estimate the sensitivity for the $y_q$.}
	\label{fig:d2}
\end{figure}

Since the angularities depend on the continuous parameter $a$, it is also useful to combine the multiple angularity distributions in the analysis. Here we build upon ideas previously developed in \cite{ELVW}.
As an example, we show the normalized double differential distributions of angularities with $a=-1.0$ and $a=0.5$ from PYTHIA at hadron level for $H\to q\bar{q}$ and $H\to gg$  in Fig.~\ref{fig:d2}. Performing a likelihood analysis in the regions $\tau_\text{min}(a)<\tau_a<\tau_\text{max}(a)$ with the same min and max limits for each $a$ as in the single-angularity analysis above, we find that limits on the Yukawa couplings could be further improved about $2\%\sim3\%$  compared to the single angularity analysis at $2\sigma$ confidence level. Although the  two different angularities will improve the power of distinguishing the quark and gluon jets, the backgrounds from $H\to b\bar{b}, c\bar{c}$ and $Zq\bar{q}$ will become important since both of them have a similar angularity distributions as the signal. Therefore, the next step to further constraint the light quark Yukawa couplings could be from  improving the b-tagging efficiency, detector resolution and so on. If both $H\to b\bar{b}, c\bar{c}$ and $Zq\bar{q}$ could be further reduced about 10 times than current analysis, we estimate that the results from two different angularities could be improved about $12\%$ compared to the single angularity analysis. (Note that these estimates are obtained in our most conservative analysis windows.)
Methods for resummation of two angularities have been developed and applied to predictions in~\cite{Larkoski:2014tva,Procura:2018zpn}. 
Further development of such predictions for double differential distributions of angularities in Higgs decays and their applications to phenomenology would seem worthwhile.

\section{Summary}
\label{sec:sum}
In this paper,  we studied a class of event shape variables angularities from Higgs boson decay at the $e^+e^-$ colliders based on soft-collinear effective theory. Both the quark and gluon final states are calculated to ${\rm NLL}^\prime+\mathcal{O}(\alpha_s)$ and ${\rm NNLL}+\mathcal{O}(\alpha_s)$ accuracy.  From ${\rm NLL}^\prime+\mathcal{O}(\alpha_s)$ to ${\rm NNLL}+\mathcal{O}(\alpha_s)$ accuracy, we found a sizable correction for small $a$ angularties. The differential angularity distributions  also show a Casimir scaling $C_A/C_F$ from quark to gluon. We compared the predictions resulting from this analysis to PYTHIA  at both parton and hadron level, and find good agreement for hadron level results.  

Based on the difference of angularity distributions between quark and gluon final state, we  proposed to test the light quark Yukawa couplings through angularity distributions at lepton colliders.  As  an example, we show that the CEPC with $\sqrt{s}=250~{\rm GeV}$ and an integrated luminosity of $5~{\rm ab}^{-1}$ could give a constraint $y_q\lesssim 0.15-0.22 y_b$ for $q=u,d,s$, for different values of the angularity parameter $a$, at $2\sigma$ confidence level using a conservative analysis region $[\tau_{\rm min}(a),\tau_{\rm max}(a)]$ far away from small $\tau_a$ where hadronization and $b$-mass effects are larger. The theoretical uncertainty for this upper limit is around 10\%. In order to further improve the results, it is important to extend the analysis region to small $\tau_a$. It shows that the upper limit could be reduced to $y_q\lesssim 0.09 y_b$, similar to \cite{Gao:2016jcm}, if we push the analysis region down to $[t_0^g(a),\tau_{\rm max}(a)]$, or even smaller $y_q\lesssim 0.07 y_b$ in the region $[t_0^q(a),\tau_{\rm max}(a)]$. However, the theoretical prediction in the small $\tau_a$ region is strongly dependent on the non-perturbative model assumptions and this issue could be overcome by gaining better knowledge of the gluon soft shape function, in particular, or by utilizing soft-drop grooming techniques. Utilizing multiple angularities at once also shows promise in improving the potential limit on $y_q$ further.

{\it Note:} As this paper was being finalized, Ref.~\cite{Zhu:2023oka} appeared, computing Higgs angularity distributions to NNLL$'$ resummed and NLO fixed-order accuracy. We have not yet compared to the technical results of this paper; here we have focused more extensively on the phenomenological application of the results to the determination of Yukawa couplings.

\begin{acknowledgments}
BY would like to thank Zhongbo Kang and C.-P. Yuan for useful discussions, and Wan-Li Ju for discussions about the calculation of thrust in Higgs decay. This material is based upon work supported by the U.S. Department of Energy, Office of Science, Office of Nuclear Physics, and through an Early Career Research Award. Portions of the work were also supported by the Laboratory Directed Research and Development program of Los Alamos National Laboratory under project numbers 20190033ER and 20200775PRD4. Los Alamos National Laboratory is operated by Triad National Security, LLC, for the National Nuclear Security Administration of U.S. Department of Energy (Contract No. 89233218CNA000001). BY is also supported by the IHEP under Contract No. E25153U1. 
\end{acknowledgments}

\appendix
\section{Angularity distributions in Higgs decay to $\mathcal{O}(\alpha_s)$}
\label{app:nlo}
In this section, we give details of the calculation of the full QCD prediction for angularity $\tau_a$ distribution away from $\tau_a=0$. This follows closely the similar calculation given in \cite{Hornig:2009vb}. Both the virtual  and real diagrams can contribute to the angularity $\tau_a$ distribution at $\mathcal{O}(\alpha_s)$. However, the virtual diagram is proportional to $\delta(\tau_a)$, which is fully accounted for in the SCET prediction for the singular terms given by \eq{singulardist}. Here we only need to consider the terms contributing the difference between full QCD and SCET given by \eq{remainder}. So we only need to
consider the real diagram's contribution,
\beq
\frac{1}{\Gamma_{H0}^i}\frac{d\Gamma_H^i}{d\tau_a}\bigg \vert_{\rm QCD}=\left(\frac{\alpha_s}{2\pi}\right)A_a^i(\tau_a).
\eeq
For Higgs decaying to quark state, the coefficient $A_a^q(\tau_a)$ is,
\beq
\label{eq:Aaq}
A_a^q(\tau_a)=C_F\int dx_1 dx_2\frac{(1-x_1)^2+(1-x_2)^2+2x_1x_2}{(1-x_1)(1-x_2)}\delta(\tau_a-\tau_a(x_1,x_2)),
\eeq
where $x_{1,2}\equiv 2E_{1,2}/m_H$ are the energy fractions of any two of the three final state partons.

There are two subprocesses $H\to gg(\to q\bar{q})$ and $H\to ggg$ that can contribute to $A_a^g(\tau_a)$. For $H\to gg(\to q\bar{q})$ channel we have,
\beq
\label{eq:Agqq}
A_a^{gq\bar{q}}(\tau_a)=N_f\int dx_1 dx_2\frac{x_1^2+x_2^2-2x_1-2x_2+2}{x_1+x_2-1}\delta(\tau_a-\tau_a(x_1,x_2)).
\eeq
For $H\to ggg$,
\beq
\label{eq:Aggg}
A_a^{ggg}(\tau_a)=C_A\int dx_1 dx_2\frac{1+(1-x_1)^4+(1-x_2)^4+(1-x_1-x_2)^4}{3(1-x_1)(1-x_2)(x_1+x_2-1)}\delta(\tau_a-\tau_a(x_1,x_2)).
\eeq
The coefficient $A_a^g(\tau_a)=A_a^{gq\bar{q}}(\tau_a)+A_a^{ggg}(\tau_a)$.

\begin{figure}
	\centering
	\includegraphics[width=0.4\textwidth]{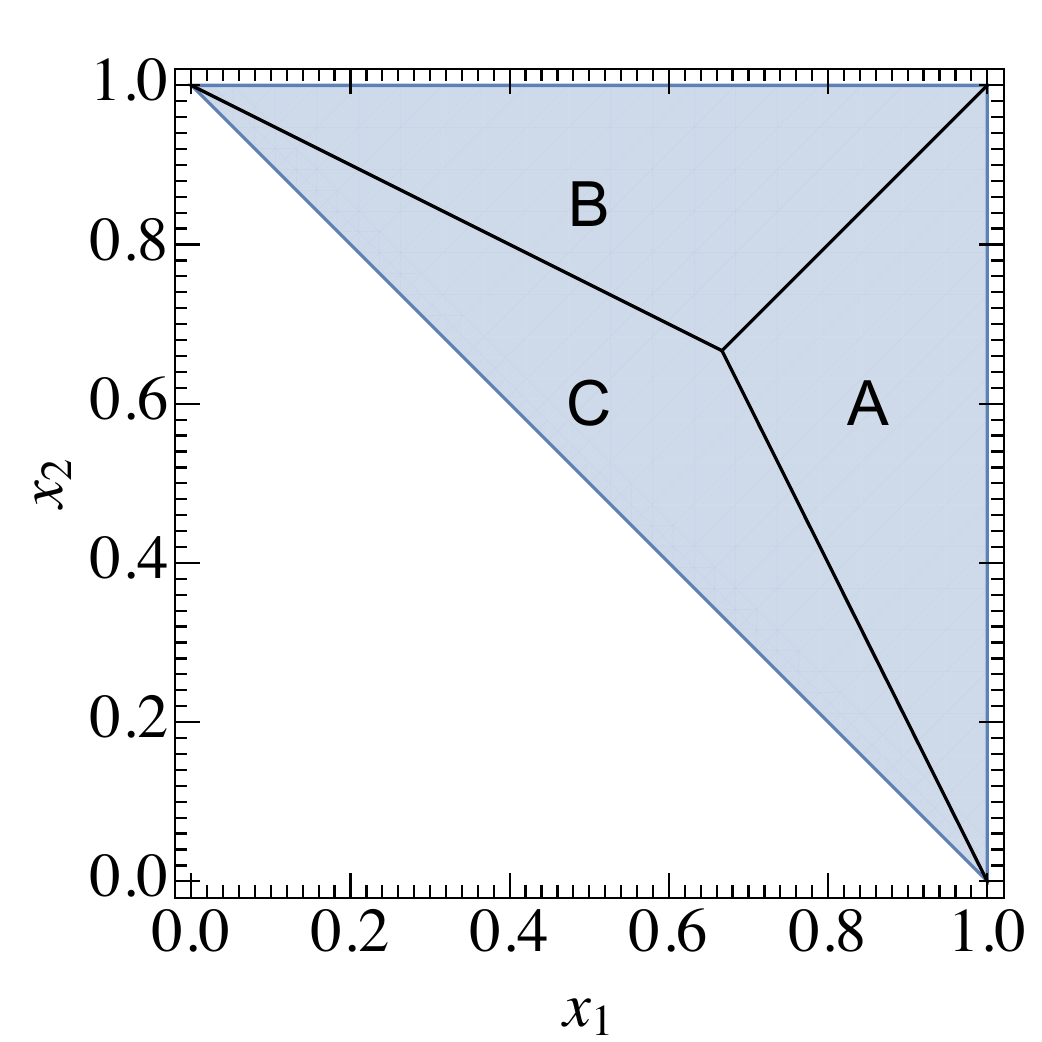}
 \vspace{-1em}
	\caption{The three body phase space.  The parameter $x_i$ is defined as $x_i=2E_i/m_H$ and $x_1+x_2+x_3=2$.}
	\label{fig:phase}
\end{figure}

The thrust axis in these 3-body final states is always along the direction of the particle with the largest energy. Therefore, the phase space in the $x_1,x_2$ plane can divided into three regions, as shown in Fig.~\ref{fig:phase}. The value of the angularity $\tau_a(x_1,x_2)$ is dependent on which parton has the largest energy.
In a region where $x_i>x_{j,k}$, the angularity $\tau_a(x_1,x_2)$ is given by,
\beq
\tau_a(x_1,x_2)\bigg \vert_{x_i>x_{j,k}}=\frac{1}{x_i}(1-x_i)^{1-a/2}\left[(1-x_j)^{1-a/2}(1-x_k)^{a/2}+(1-x_k)^{1-a/2}(1-x_j)^{a/2}\right].
\eeq
In the following, we will use phase space region $C$ where $x_3>x_{1,2}$ as an example to discuss the calculation of $A_a^i(\tau_a)$. The integration in phase space $A$ and $B$  can be related to the integration over region $C$ by the appropriate change of variables. The angularity in phase space $C$ can be written as,
\beq
\tau_a=\frac{1}{2-x_1-x_2}(x_1+x_2-1)^{1-a/2}\left[(1-x_1)^{1-a/2}(1-x_2)^{a/2}+(1-x_2)^{1-a/2}(1-x_1)^{a/2}\right].
\eeq
It turns out to be convenient to change the integration variables from $x_1$, $x_2$ to $\tau_a$ and $w$, where
\beq
w=\frac{1-x_1}{2-x_1-x_2}.
\eeq
Therefore, $x_1$ and $x_2$ are given by,
\begin{align}
\label{eq:x12w}
x_1(w,\tau_a)&=1-w+w\left[\frac{\tau_a}{w^{1-a/2}(1-w)^{a/2}+w^{a/2}(1-w)^{1-a/2}}\right]^{\frac{1}{1-a/2}},\nn\\
x_2(w,\tau_a)&=x_1(1-w,\tau_a).
\end{align}
In the phase space region $C$, we have
\beq
 x_1(w,\tau_a)\leq 2-x_1(w,\tau_a)-x_2(w,\tau_a),\quad x_2(w,\tau_a)\leq 2-x_1(w,\tau_a)-x_2(w,\tau_a)
 \eeq
 The integration region of $w$ is determined by solving these inequalities, which are equivalent to the conditions,
 \beq
 \label{eq:wminmax}
\tau_a=F_a(w_{\rm min}),\quad \tau_a=F_a(1-w_{\rm max}), 
 \eeq
whose solutions give the lower and upper limits of the integral over $w$, $w_{\rm min}$ and $w_{\rm max}$, respectively, where the function $F_a$ is given by
\beq
F_a(w)=\frac{w(1-w)^{a/2}}{(1+w)^{1-a/2}}(w^{1-a}+(1-w)^{1-a}).
\eeq 
These limits are themselves functions of $\tau_a$, and cease to have a solution above the maximally allowed value of $\tau_a=\tau_a^\text{max}$, which is $\tau_a^\text{max}= (1/3)^{1-a/2}$ at $\cO(\as)$.\footnote{This is true at least for values of $a\gtrsim -2$. See \cite{Hornig:2009vb} for subtleties for smaller values of $a$.} Then, for example, the integral for $A_a^q$ in \eq{Aaq} is expressed:
\beq
\label{eq:Aqintegral}
A_a^q(\tau_a) = C_F\int_{w_\text{min}(\tau_a)}^{w_\text{max}(\tau_a)} dw\,J(w,\tau_a) \biggl[ \frac{x_1^2 + x_2^2}{(1-x_1)(1-x_2)} + 2\frac{x_1^2+x_3^2}{(1-x_1)(1-x_3)}\biggr]_{x_{1,2} = x_{1,2}(w,\tau_a)}\,,
\eeq
where $x_{1,2}$ are expressed in the form \eq{x12w}, and $x_3=2-x_1-x_2$. The factor $J$ is the Jacobian associated with the variable transformation \eq{x12w},
\beq
\label{eq:Jacobian}
J(w,\tau_a) = \begin{vmatrix}
\frac{\partial x_1}{\partial w} & \frac{\partial x_1}{\partial \tau}  \\
\frac{\partial x_2}{\partial w} & \frac{\partial x_2}{\partial \tau} 
\end{vmatrix}\,.
\eeq
The endpoints $w_{\text{min,max}}$ in \eq{wminmax}, the Jacobian \eq{Jacobian}, and the integral \eq{Aqintegral} are all easily evaluated numerically, which is how we have computed the $A_a^i$'s and the resulting remainder functions illustrated in, e.g. \fig{ra}. The gluon channel contributions $A_a^g$ given by \eqs{Agqq}{Aggg} are expressed and evaluated similarly to \eq{Aqintegral}.

\section{Ingredients for NNLL resummation}
\label{app:NNLL}

In this Appendix we collect results needed to compute angularity distributions to NNLL accuracy.  The coefficients of  $\beta$ function  in the $\overline{\rm {MS}}$ scheme are given by~\cite{Tarasov:1980au,Larin:1993tp,vanRitbergen:1997va},
\begin{align}
\label{eq:betan}
\beta_0&=\frac{11}{3}C_A-\frac{4}{3}T_Fn_f, \\
\beta_1&=\frac{34}{3}C_A^2-\left(\frac{20}{3}C_A+4C_F\right)T_F n_f,\nn\\
\beta_2 &=\frac{2857}{54}C_A^3+\left(C_F^2-\frac{205}{18}C_FC_A-\frac{1415}{54}C_A^2\right)2T_Fn_f+\left(\frac{11}{9}C_F+\frac{79}{54}C_A\right)4T_F^2n_f^2.\,, \nn
\end{align}
and the coefficients of the anomalous dimension $\gamma_y$ of the Yukawa coupling $y_q$ in \eq{yqas_expansion} are given by~\cite{Gehrmann:2014vha}:
\begin{align}
\label{eq:gammayn}
\gamma_y^0 &=6C_F, \\
\gamma_y^1 &=3C_F^2+\frac{97}{3}C_AC_F-\frac{10}{3}C_Fn_f. \nn
\end{align}
The cusp anomalous dimensions up to 3-loop order~\cite{Korchemsky:1987wg,Moch:2004pa}
\begin{align}
\Gamma_0^i&=4C_i,\nn\\
\Gamma_1^i&=\left[\left(\frac{67}{9}-\frac{\pi^2}{3}\right)C_A-\frac{20}{9}T_FN_f\right]\Gamma_0^i,\nn\\
\Gamma_2^i &= \left[\Bigl( \frac{245}{6} - \frac{134\pi^2}{27} + \frac{11\pi^4}{45} + \frac{22\zeta_3}{3}\Bigr) C_A^2 + \Bigl(-\frac{418}{27} + \frac{40\pi^2}{27} - \frac{56\zeta_3}{3}\Bigr) C_A T_F n_f \right.\\
&\left.\quad + \Bigl(-\frac{55}{3} + 16\zeta_3\Bigr) C_F T_F n_f - \frac{16}{27} T_F^2 n_f^2\,\right]\Gamma_0^i. \nn
\end{align}
The non-cusp anomalous dimension for the hard function up to 2-loop level~\cite{Moch:2005id,Becher:2006mr,Idilbi:2006dg,Harlander:2009bw,Pak:2009bx,Berger:2010xi},
\begin{align}
\gamma_{H0}^q&=-12C_F,\nn\\
\gamma_{H1}^q &= - 2C_F 
\Bigl[
  \Bigl(\frac{82}{9} - 52 \zeta_3\Bigr) C_A
+ (3 - 4 \pi^2 + 48 \zeta_3) C_F
+ \Bigl(\frac{65}{9} + \pi^2 \Bigr) \beta_0 \Bigr]
\,,\nn\\
\gamma_{H0}^g&=-4\beta_0,\nn\\
\gamma_{H1}^g&=\left(-\frac{236}{9}+2\zeta_3\right)C_A^2+\left(-\frac{76}{9}+\frac{2\pi^2}{3}\right)C_A\beta_0-4\beta_1.
\end{align}
The 1-loop soft non-cusp anomalous dimension is zero, e.g. $\gamma_{S0}^i(a)=0$.
The 2-loop jet and soft anomalous dimensions are not known analytically, but are related by $\gamma_H^i + \gamma_J^i(a) + \gamma_S^i(a) =0$, and $\gamma_S^i(a)$ is known in numerically integrable form to 2-loop order, thanks to \cite{Bell:2018vaa}. The 2-loop soft anomalous dimension can be written in the form
\beq
\label{eq:gammaS1a}
\gamma_{S1}^i(a) = \frac{2C_i}{1-a} \Bigl[ \gamma_1^{CA} C_A + \gamma_1^{nf}(a) T_F n_f\Bigr] \,,
\eeq
where
\begin{align}
\gamma_1^{CA}(a) &= -\frac{808}{27} + \frac{11\pi^2}{9} + 28\zeta_3 - \Delta\gamma_1^{CA}(a) \nn\\
\gamma_1^{nf}(a) &= \frac{224}{27} - \frac{4\pi^2}{9} - \Delta\gamma_1^{nf}(a)\,,
\label{eq:gammaS1colors}
\end{align}
where the deviations from the $a=0$ values shown are given by the integrals:
\begin{align}
\label{eq:DeltagammaS}
\Delta\gamma_1^{CA}(a) \!&=\! \int_0^1 \!\! dx\int_0^1 \!\! dy\frac{32 x^2(1\plus xy\plus y^2)\bigl[ x(1\plus y^2) + (x\plus y)(1\plus xy)\bigr]}{y(1-x^2)(x+y)^2(1+xy)^2}
\ln\Bigl[\frac{(x^a+ xy) (x+ x^a y)}{x^a(1+ xy)(x+y)} \Bigr], \nn \\
\Delta\gamma_1^{nf}(a) \!&=\!  \int_0^1 \!\! dx\int_0^1 \!\! dy\frac{64 x^2(1 + y^2)}{(1-x^2)(x+y)^2(1+xy)^2}
\ln\Bigl[\frac{(x^a+ xy) (x+ x^a y)}{x^a(1+ xy)(x+y)} \Bigr],
\end{align}
which vanish for $a=0$. The integral representations can easily be evaluated numerically to high accuracy 
for any value of $a$, and the relevant values for our work are given in Table~\ref{tab:gamma1}.

\begin{table}[t]
\centering
\begin{tabular}{|c||c|c|c|c|c|}
\hline 
$a$ & $-1.0$ & $-0.5$ & 0.0 & 0.25 & 0.5 
\\ \hline \hline
$\gamma_1^{CA}$ & $1.0417$  & $9.8976$  & $15.795$ & $17.761$ & $19.132$ 
\\ \hline
$\gamma_1^{nf}$ & $-0.9571$  &  $1.8440$  & $3.9098$ & $4.6398$ & $5.1613$
\\ \hline
\end{tabular}
\caption{Coefficients of the two-loop soft anomalous dimension as defined in
Eq.~\ref{eq:gammaS1a}.}
\label{tab:gamma1}
\end{table}

\bibliography{reference}

\end{document}